\documentclass[aps,eqsecnum,preprint,floats,epsf,epsfig,nofootinbib,letter]{revtex4}
\usepackage{epsfig}

\begin{document}
\def\be{\begin{eqnarray}}
\def\en{\end{eqnarray}}
\def\non{\nonumber}
\def\la{\langle}
\def\ra{\rangle}
\def\nc{N_c^{\rm eff}}
\def\vp{\varepsilon}
\def\drho{\bar\rho}
\def\deta{\bar\eta}
\def\A{{\cal A}}
\def\B{{\cal B}}
\def\c{{\cal C}}
\def\d{{\cal D}}
\def\e{{\cal E}}
\def\p{{\cal P}}
\def\t{{\cal T}}
\def\up{\uparrow}
\def\dw{\downarrow}
\def\vma{{_{V-A}}}
\def\vpa{{_{V+A}}}
\def\smp{{_{S-P}}}
\def\spp{{_{S+P}}}
\def\J{{J/\psi}}
\def\ov{\overline}
\def\Lqcd{{\Lambda_{\rm QCD}}}
\def\pr{{Phys. Rev.}~}
\def\prl{{Phys. Rev. Lett.}~}
\def\pl{{Phys. Lett.}~}
\def\np{{Nucl. Phys.}~}
\def\zp{{Z. Phys.}~}
\def\lsim{ {\ \lower-1.2pt\vbox{\hbox{\rlap{$<$}\lower5pt\vbox{\hbox{$\sim$}
}}}\ } }
\def\gsim{ {\ \lower-1.2pt\vbox{\hbox{\rlap{$>$}\lower5pt\vbox{\hbox{$\sim$}
}}}\ } }

\font\el=cmbx10 scaled \magstep2{\obeylines\hfill May, 2008}
\vskip 0.7 cm
%\title{\vspace*{1cm} Polarizations in $B\to V A$ and $B\to AA$ Decays}

\centerline{\large\bf  Branching Ratios and Polarization in $B\to
VV,VA,AA$ Decays}
\bigskip
\centerline{\bf Hai-Yang Cheng$^{1}$ and Kwei-Chou Yang$^{2}$}
\medskip
\centerline{$^1$ Institute of Physics, Academia Sinica}
\centerline{Taipei, Taiwan 115, Republic of China}
\medskip
\centerline{$^2$ Department of Physics, Chung Yuan Christian
University} \centerline{Chung-Li, Taiwan 320, Republic of China}
\bigskip
%\bigskip

 \small
\begin{abstract}
\medskip
We present a detailed study of charmless two-body $B$ decays into final states
involving two vector mesons ($VV$) or two axial-vector mesons ($AA$) or one
vector and one axial-vector meson ($VA$),  within the framework of QCD
factorization, where $A$ is either a $^3P_1$ or $^1P_1$ axial-vector meson. The
main results are as follows. (i) In the presence of NLO nonfactorizable
corrections, effective Wilson coefficients $a_i^h$ are helicity dependent. For some penguin-dominated modes, the constructive (destructive) interference in the
negative-helicity (longitudinal-helicity) amplitude of the $\ov B\to VV$ decay
will render the former comparable to the latter and push up the transverse
polarization. (ii) In QCD factorization, the transverse polarization
fraction can be large for penguin-dominated charmless $VV$ modes by allowing for sizable penguin
annihilation contributions. (iii)
Using the measured $\bar K^{*0}\rho^-$ channel as an input, we predict the
branching ratios and polarization fractions for other $\ov B\to \bar K^*\rho$
decays.   (iv)  The smallness of the axial-vector decay constant
of the $^1P_1$ axial vector meson can be tested by measuring various $b_1\rho$
modes to see if $\Gamma(\ov B^0\to b_1^-\rho^+)\ll \Gamma(\ov B^0\to
b_1^+\rho^-)$ and $\Gamma(B^-\to b_1^-\rho^0)\ll \Gamma(B^-\to b_1^0\rho^-)$.
(v) For the penguin-dominated modes $a_1K^*$ and $b_1K^*$, it is found that
the former are dominated by transverse polarization amplitudes, whereas the
latter are governed by longitudinal polarization states.  (vi)  The rates
of $B\to K_1(1270)K^*$ and $K_1(1400)K^*$ are generally very small. The decay
modes $K_1^-K^{*+}$ and $K_1^+K^{*-}$ are of particular interest as they are
the only $AV$ modes which receive contributions solely from weak annihilation.
(vii) For tree-dominated $B\to AA$ decays, the $a_1^+a_1^-$, $a_1^-a_1^0$,
$a_1^-b_1^+$, $a_1^-b_1^0$, $b_1^+\rho^-$ and $b_1^0\rho^-$  modes have sizable
branching ratios, of order $(20\sim 40)\times 10^{-6}$. (viii) There are many
penguin-dominated $B\to AA$ decays within the reach of $B$ factories:
$K_1(1270)(a_1,b_1^\pm)$, $K_1(1400)(b_1,a_1^\pm)$,
$K_1(1270)(f_1(1285),f_1(1420))$ and $K_1(1400)(f_1(1420),h_1(1170))$.

\end{abstract}
\small

\maketitle

\section{Introduction}
Recently we have studied the charmless two-body $B$ decays involving an
axial-vector meson $A$ and a pseudoscalar meson $P$ in the final state
\cite{CY:AP,Yanga1}. There are two distinct types of axial-vector mesons,
namely, $^3P_1$ and $^1P_1$. We have studied their light-cone distribution
amplitudes  using the QCD sum rule method. Owing to the $G$-parity, the
chiral-even two-parton light-cone distribution amplitudes of the $^3P_1$
($^1P_1$) mesons are symmetric (antisymmetric) under the exchange of quark and
anti-quark momentum fractions in the SU(3) limit. For chiral-odd light-cone
distribution amplitudes, it is the other way around. In this work, we will
generalize our previous study to charmless $VA$ and $AA$ modes. Moreover, we
will use this chance to re-examine $B\to VV$ decays.

The charmless decays $B\to VV,VA,AA$ are expected to have rich physics as they
have three polarization states. Through polarization studies, these channels
can shed light on the underlying helicity structure of the decay mechanism.
Experimentally, $B\to K^*\phi$ decays have been studied with full angular
analysis and hence can provide information on polarization fractions and
relative strong phases among various helicity amplitudes. Historically, it was
the observation of large transverse polarization in $B\to K^*\phi$ decays that
had triggered a burst of theoretical and experimental interest in the study of
charmless $B\to VV$ decays. BaBar and Belle have observed that $f_L\sim 1/2$
and $f_\parallel\sim f_\bot\sim 1/4$ in the $K^*\phi$ channels
\cite{BaBar:KVV,Belle:KVphib}, where $f_L,f_\bot$ and $f_\parallel$ are the
longitudinal, perpendicular, and parallel polarization fractions, respectively.
The transverse polarization fraction $f_T=f_\parallel+f_\bot\sim 1/2$ is found
to be of the same order magnitude as the longitudinal one $f_L$ in the
penguin-dominated $K^*\phi$ and $K^*\rho$ modes (except the decay $B^-\to
K^{*-}\rho^0$). While the naive expectation of $f_\parallel\sim f_\bot$ is
borne out by experiment, the observed large $f_T$ is in contradiction to the
naive anticipation of a small transverse polarization of order $f_T\sim
m_V^2/m_B^2$. This has promoted many to explore the possibility of new physics
in penguin-dominated $B\to VV$ decays. If so, the new physics effects should
also manifest themselves in penguin-dominated $VA$ and $AA$ modes.

The analysis of charmless $B\to VV$ decays within the framework of QCD
factorization \cite{BBNS,BN} was first performed by us \cite{CY:VV} followed by
many others \cite{YDYang,YDYangnew,VV:QCDF,Kagan,Yang&Das,BenekeVV}. In these
studies, NLO corrections to the helicity-dependent coefficients $a_i^h$ such as
vertex corrections, penguin contributions and hard spectator scattering were
calculated. However, most of the early results do not agree with each other due
to the incorrect projection  on the polarization states. Recently, Beneke,
Rohrer and Yang \cite{BenekeVV} have used the correct light-cone projection
operators and computed complete NLO corrections to $a_i^h$ and weak
annihilation amplitudes. We will follow their work closely in the study of
$B\to VV$ decays.

The generalization of the analysis of $B\to VV$ decays to $VA$ and $AA$ modes
is  highly nontrivial. First of all, while the $^3P_1$ meson behaves similarly
to the vector meson, this is not the case for the $^1P_1$ meson. For the
latter, its decay constant vanishes in the SU(3) limit and its chiral-even
two-parton light-cone distribution amplitude (LCDA) is anti-symmetric under the
exchange of quark and anti-quark momentum fractions in the SU(3) limit due to
the $G$ parity, contrary to the symmetric behavior for the $^3P_1$ meson.
Second,  there are two mixing effects for axial-vector mesons: one is the
mixing between $^3P_1$ and $^1P_1$ states, e.g., $K_{1A}$ and $K_{1B}$ and the
other is the mixing among $^3P_1$ or $^1P_1$ states themselves. In this work we
will derive the longitudinal and transverse projectors for axial-vector mesons
and work out the hard spectator scattering and annihilation contributions to
$VA$ and $AA$ decays.

Since the resolution of the $K^*\phi$ polarization anomaly may call for new
physics beyond the standard model, this issue has received much attention in
the past years. However, there are two crucial points that have been often
overlooked in the literature. First, a reliable estimate of polarization
fractions cannot be achieved unless the decay rate is correctly reproduced.
Second, all the existing calculations except
\cite{CY:VV,YDYang,Yang&Das,BenekeVV} assume that the effective Wilson
coefficients $a_i^h$ are helicity independent. This leads to the scaling law:
$f_T\sim {\cal O}(m_V^2/m_B^2)$. Calculations based on naive factorization
often predict too small $B\to K^*\phi$ and $B\to K^*\rho$ rates by a factor of
$2\sim 3$. Obviously, it does not make sense at all to compare theory with
experiment for $f_{L,T}$ at this stage as the definition of polarization
fractions depends on the partial rate and hence the prediction can be easily
off by a factor of $2\sim 3$. The first task is to have some mechanism to bring
up the rates. While the QCD factorization and pQCD \cite{Mishima} approaches
rely on penguin annihilation, soft-collinear effective theory invokes charming
penguin \cite{SCET} and the final-state interaction model considers final-state
rescattering of intermediate charm states \cite{Colangelo,Ladisa,CCSfsi}. Once
the measured rate is reproduced, then it becomes sensible to ask what is the
effect of this mechanism on polarization fractions. Next, it is important to
consider NLO corrections to various helicity coefficients  $a_i^h$, such as
vertex corrections, penguin and hard spectator scattering contributions. It
turns out that in some of $\ov B\to VV$ decays, e.g. $\ov B\to \bar
K^*\phi,\bar K^{*0}\rho^0$,  NLO nonfactorizable corrections will render
negative-helicity amplitude comparable to the longitudinal one and hence will
bring up the transverse polarization. Therefore, any serious solution to the
polarization puzzle should take into account NLO effects on $a_i^h$.

There have been a few studies of charmless $B\to AV$ and $B\to AA$ decays in
the literature \cite{Yang1P1,ChenK1,Calderon}. Except for \cite{Yang1P1} done
in QCD factorization, the analysis in other two references was carried out in
the framework of  generalized factorization in which the nonfactorizable
effects are described by the parameter $N_c^{\rm eff}$, the effective number of
colors. It has been claimed in \cite{Calderon} that most of $B\to AV$ decays
are suppressed and $\Gamma(B\to AV)<\Gamma(B\to AP)$. This seems to be in
contradiction to the naive anticipation that $AV$ modes will have larger rates
because of the existence of three polarization states for the vector meson. One
of the main motivations for this work is to examine if the claim of
\cite{Calderon} holds.

The present paper is organized as follows. In Sec. II we summarize all the
input parameters relevant to the present work, such as the mixing angles, decay
constants, form factors and light-cone distribution amplitudes for $^3P_1$ and
$^1P_1$ axial-vector mesons and their Gegenbauer moments. We then apply QCD
factorization in Sec.  III to study $B\to VV,VA,AA$ decays and derive the
relevant spectator interaction and annihilation terms. Results and discussions
are presented in Sec. IV. Sec. V contains our conclusions. Flavor operators and
the factorizable amplitudes of selective $B\to AV$ and $AA$ decays are
summarized in Appendices A and B, respectively. In Appendix C we give an
explicit evaluation of the annihilation amplitude for the decay $B\to VA$.
Since annihilation and hard spectator scattering amplitudes involve end-point
divergences $X_A^h$, we give explicit expressions of them for various $VV,VA$
and $AA$ modes in terms of $X_A^h$ in Appendices D and E.

\section{Input parameters}
In this section we shall briefly discuss and summarize
all the input parameters relevant to the present work, such as the
mixing angles, decay constants, form factors and light-cone
distribution amplitudes for vector and axial-vector
mesons.

\subsection{Mixing angles}
Mixing angles of the axial-vector mesons have been discussed in \cite{YangNP}
and \cite{CY:AP}. Here we recapitulate the main points. For axial-vector mesons
there are two mixing angles of interest: one is the mixing between $^3P_1$ and
$^1P_1$ states, e.g., $K_{1A}$ and $K_{1B}$ and the other is the mixing among
$^3P_1$ or $^1P_1$ states themselves, for example, the $^3P_1$ states
$f_1(1285)$ and $f_1(1420)$ have mixing due to SU(3) breaking effects.

The non-strange axial vector mesons, for example, the neutral $a_1(1260)$ and
$b_1(1235)$ cannot have mixing because of the opposite $C$-parities. In the
isospin limit, charged $a_1(1260)$ and  $b_1(1235)$ also cannot have mixing
because of the opposite $G$-parities. On the contrary, the strange partners of
$a_1(1260)$ and $b_1(1235)$, namely, $K_{1A}$ and $K_{1B}$, respectively, are
not mass eigenstates and they are mixed together due to the strange and
non-strange light quark mass difference. We write
 \be \label{eq:mixing}
 K_1(1270) &=& K_{1A}\sin\theta_{K_1}+K_{1B}\cos\theta_{K_1}, \non \\
 K_1(1400) &=& K_{1A}\cos\theta_{K_1}-K_{1B}\sin\theta_{K_1}.
 \en
Various experimental information yields $\theta_{K_1}\approx \pm37^\circ$ and
$\pm58^\circ$ (see e.g. \cite{ChengDAP}). The sign of $\theta_{K_1}$ is
intimately related to the relative phase of the $K_{1A}$ and $K_{1B}$ states.
We choose the phase convention such that the decay constants of $K_{1A}$ and
$K_{1B}$ are of the same sign, while the $B\to K_{1A}$ and $B\to K_{1B}$ form
factors are opposite in sign. In this convention for $K_{1A}$ and $K_{1B}$, the
mixing angle $\theta_{K_1}$ is favored to be negative as implied by the
experimental measurement of the ratio of $K_1\gamma$ production in $B$ decays
\cite{CY:AP,Hatanaka:2008xj}.

Just like the $\eta-\eta'$ mixing in the pseudoscalar sector, the $1^1P_1$
states  $h_1(1170)$ and $h_1(1380)$ may be mixed in terms of the pure octet
$h_8$ and singlet $h_1$,
 \begin{eqnarray}
 |h_1(1170)\rangle = |h_1\rangle\cos\theta_{^1P_1}+|h_8\rangle\sin\theta_{^1P_1},
 \quad
 |h_1(1380)\rangle = -|h_1\rangle\sin\theta_{^1P_1} +|h_8\rangle\cos\theta_{^1P_1} \,,
 \end{eqnarray}
and likewise  the $1^3P_1$ states $f_1(1285)$ and $f_1(1420)$ have mixing via
\begin{eqnarray}
 |f_1(1285)\rangle = |f_1\rangle\cos\theta_{^3P_1}+|f_8\rangle\sin\theta_{^3P_1},
 \quad |f_1(1420)\rangle =
 -|f_1\rangle\sin\theta_{^3P_1} +|f_8\rangle\cos\theta_{^3P_1} \,.
 \end{eqnarray}
Using the Gell-Mann-Okubo mass formula \cite{CloseBook,PDG}, we found that the
mixing angles $\theta_{^1P_1}$ and $\theta_{^3P_1}$ depend on the angle
$\theta_{K_1}$ and are given by \cite{CY:AP} \be \label{eq:mixingangle}
 \theta_{^1P_1}=25.2^\circ, \quad && \theta_{^3P_1}=27.9^\circ, \quad {\rm for}~\theta_{K_1}=-37^\circ, \non \\
  \theta_{^1P_1}\simeq 0^\circ, \quad && \theta_{^3P_1}=53.2^\circ, \quad {\rm for}~\theta_{K_1}=-58^\circ.
 \en

\subsection{Decay constants and form factors}
 Decay constants of vector and axial-vector mesons are defined as
 \be \label{eq:decayc}
 \la V(p,\epsilon)|\bar q_2\gamma_\mu q_1|0\ra &=& -if_V m_V\epsilon^*_\mu, \non \\
 \la ^{3(1)}P_1(p,\epsilon)|\bar q_2\gamma_\mu\gamma_5 q_1|0\ra &=& if_{^3P_1(^1P_1)} m_{^3P_1(^1P_1)}\epsilon^*_\mu.
 \en
Transverse decay constants are defined via the tensor current
by
\begin{eqnarray}
   \langle ^{3(1)}P_1(p,\epsilon) |\bar q_2 \sigma_{\mu\nu}\gamma_5q_1 |0\rangle
 &=& f_{^{3(1)}P_1}^\perp (\epsilon_{\mu}^* p^\nu -
\epsilon_{\nu}^* p^\mu)\,,   \non \\
   \langle V(p,\epsilon) |\bar q_2 \sigma_{\mu\nu}q_1 |0\rangle
 &=& -f_V^\perp (\epsilon_{\mu}^* p^\nu -
\epsilon_{\nu}^* p^\mu)\,.
   \label{eq:tensor-1p1-2}
\end{eqnarray}
The decay constants $f_{^1P_1}$ of the $^1P_1$ non-strange neutral mesons
$b_1^0(1235),h_1(1170),h_1(1380)$ vanish due to charge conjugation invariance.
Likewise, the decay constant $f_{b_1}$ of the charged $b_1$ vanishes owing to
its even $G$-parity valid in the isospin limit. In general, the decay constants
$f_{^1P_1}$ and $f_{^3P_1}^\perp$ are zero in the SU(3) limit. As discussed in
\cite{CY:AP}, they are related to $f_{^1P_1}^\perp$ and $f_{^3P_1}$,
respectively, via
 \be \label{eq:f1p1}
 f_{^1P_1}=f_{^1P_1}^\perp(\mu) a_0^{\parallel,^1P_1}(\mu), \qquad
  f_{^3P_1}^\perp(\mu)=f_{^3P_1}a_0^{\perp,^3P_1}(\mu),
 \en
where $a_0^{\parallel,^{1,3}P_1}$ are the zeroth Gegenbauer moment of
$\Phi_\parallel^{^{1,3}P_1}$ to be defined later. Since we will assume isospin
symmetry in practical calculations, this means that $f_{^1P_1}=0$ for the $b_1$
and $h_1$ mesons and $f^\perp_{^3P_1}=0$ for $a_1$ and $f_1$ mesons. Note that
since $f_{^1P_1}$ and $f^\perp_{^3P_1}$ are $G$-parity violating quantities,
their signs have to be flipped from particle to antiparticle due to the
$G$-parity, for example, $f_{K^+_{1B}}=-f_{K^-_{1B}}$. In the present work, the
$G$-parity violating parameters, e.g. $a_1^K, a_1^{\parallel,K_{1A}},
a_{0,2}^{\perp,K_{1A}}, a_1^{\perp,K_{1B}}$ and $a_{0,2}^{\parallel,K_{1B}}$,
are considered for mesons containing a strange quark.

For the decay constants $f_{f_1(1285)}^q$ and
$f_{f_1(1420)}^q$ for $1^3P_1$ states defined by
\begin{eqnarray}
\langle 0| \bar q\gamma_\mu \gamma_5 q |
f_1(1285)(P,\lambda)\rangle
 &=& -i m_{f_1(1285)} f_{f_1(1285)}^q \epsilon_\mu^{(\lambda)}\,,
 \label{eq:decay-def1}\\ \non
\langle 0| \bar q\gamma_\mu \gamma_5 q |
f_1(1420)(P,\lambda)\rangle
 &=& -i m_{f_1(1420)} f_{f_1(1420)}^q \epsilon_\mu^{(\lambda)}\,,
  \label{eq:decay-def2}
\end{eqnarray}
and the tensor decay constants for $1^1P_1$ states defined by
  \begin{eqnarray}
\langle 0| \bar q\sigma_{\mu\nu} q | h_1(1170)(P,\lambda)\rangle
 &=& i f_{h_1(1170)}^{\perp,q}\,\epsilon_{\mu\nu\alpha\beta}
  \epsilon_{(\lambda)}^\alpha P^\beta\,, \label{eq:decay-def3}\\ \non
\langle 0| \bar q\sigma_{\mu\nu} q | h_1(1380)(P,\lambda)\rangle
 &=& i f_{h_1(1380)}^{\perp,q}\,\epsilon_{\mu\nu\alpha\beta}
  \epsilon_{(\lambda)}^\alpha P^\beta\,, \label{eq:decay-def4}
\end{eqnarray}
the reader is referred to \cite{YangNP,CY:AP} for details.

Form factors for the $\ov B\to A$ and $\ov B\to V$ transitions
read as
\begin{eqnarray} \label{eq:FF}
\langle{A}(p, \lambda)|A_\mu|{\overline B} (p_B)\rangle
 &=& i \frac{2}{m_B - m_{A}} \varepsilon_{\mu\nu\alpha\beta}
 \epsilon_{(\lambda)}^{*\nu}
 p_B^\alpha p^{\beta} A^{BA}(q^2),
\nonumber \\
 \langle A (p,\lambda)|V_\mu|{\overline B}(p_B)\rangle
 &=& - \Bigg\{ (m_B - m_{A}) \epsilon^{(\lambda)*}_{\mu} V_1^{BA}(q^2)
 - (\epsilon^{(\lambda)*} \cdot p_B)
(p_B + p)_\mu \frac{V_2^{BA}(q^2)}{m_B - m_{A}}
\nonumber \\
&& - 2 m_{A} \frac{\epsilon^{(\lambda)*}\cdot p_B}{q^2} q^\mu
\left[V_3^{BA}(q^2) - V_0^{BA}(q^2)\right]\Bigg\}, \non \\
\langle{V}(p, \lambda)|V_\mu|{\overline B} (p_B)\rangle
 &=& -i \frac{2}{m_B + m_{V}} \varepsilon_{\mu\nu\alpha\beta}
 \epsilon_{(\lambda)}^{*\nu}
 p_B^\alpha p^{\beta} V^{BV}(q^2), \nonumber \\
 \langle V (p,\lambda)|A_\mu|{\overline B}(p_B)\rangle
 &=& (m_B + m_{V}) \epsilon^{(\lambda)*}_{\mu} A_1^{BV}(q^2)
 - (\epsilon^{(\lambda)*} \cdot p_B)
(p_B + p)_\mu \frac{A_2^{BV}(q^2)}{m_B + m_{V}}
\nonumber \\
&& - 2 m_{V} \frac{\epsilon^{(\lambda)*}\cdot p_B}{q^2} q^\mu
\left[A_3^{BV}(q^2) - A_0^{BV}(q^2)\right],
 \end{eqnarray}
where $q = p_B - p$, $V_3^{BA}(0) = V_0^{BA}(0)$ and
 \begin{eqnarray}
 V_3^{BA}(q^2) &=&  \frac{m_B - m_{A}}{2 m_{A}} V_1^{BA}(q^2) - \frac{m_B
 + m_{A}}{2 m_{A}} V_2^{BA}(q^2), \non \\
  A_3^{BV}(q^2) &=&  \frac{m_B + m_{V}}{2 m_{V}} A_1^{BV}(q^2) - \frac{m_B
 - m_{V}}{2 m_{V}} A_2^{BV}(q^2).
 \end{eqnarray}
Form factors for $B\to a_1(1260),b_1(1235),K_{1A},K_{1B}$
transitions have been calculated in  the relativistic covariant
light-front quark model (LFQM) (Table \ref{tab:FFinLF})  \cite{CCH},
the light-cone sum rule (LCSR) method (Table \ref{tab:FFinLCSR})
\cite{YangFF}, and the pQCD approach \cite{pQCD}.
Various $B\to A$ form factors also can be obatined in the
Isgur-Scora-Grinstein-Wise  (ISGW) model \cite{ISGW,ISGW2} based on the
nonrelativistic constituent quark picture. However, as pointed out in
\cite{CY:AP}, the predicted form factor $V_0^{Ba_1}(0)\approx 1.0$ in the ISGW2
model \cite{ISGW2} is too big and will lead to too large rates for $\ov B^0\to
a_1^\pm\pi^\mp$ and the wrong pattern $\B(\ov B^0\to a_1^+\pi^-)\gg\B(\ov
B^0\to a_1^-\pi^+)$, in contradiction to the experimental result $\B(\ov B^0\to
a_1^+\pi^-)\sim {1\over 2}\B(\ov B^0\to a_1^-\pi^+)$. This may imply that
relativistic effects in heavy-to-light transitions at maximum recoil that have
been neglected in the ISGW model should be taken into account in order to get
realistic form factors.

%%%%%%%%%%%%%%%%%%%%%%%%%%%%%%%%%%%%%%%%%%%%%%%%%%%%%
\begin{table}[t]
\caption{Form factors for
$B\to a_1,b_1,K_{1A},K_{1B}$ transitions obtained
in the covariant light-front model \cite{CCH} are fitted to the
3-parameter form Eq. (\ref{eq:FFpara}) except for the form factor
$V_2$ denoted by $^{*}$ for which the fit formula Eq.
(\ref{eq:FFpara1}) is used.} \label{tab:FFinLF}
\begin{ruledtabular}
\begin{tabular}{| c c c c || c c c c |}
~~~$F$~~~~~
    & $F(0)$
    &$a$
    & $b$
& ~~~ $F$~~~~~
    & $F(0)$
    & $a$
    & $b$
 \\
\hline
$A^{Ba_1}$
    & $0.25$
    & 1.51
    & 0.64
&$A^{Bb_1}$
    & $-0.10$
    & 1.92
    & 1.62
    \\
$V^{Ba_1}_0$
    & $0.13$
    &  1.71
    &  1.23
&$V^{Bb_1}_0$
    & $-0.39$
    & 1.41
    & 0.66
    \\
$V^{Ba_1}_1$
    & $0.37$
    & $0.29$
    & 0.14
&$V^{Bb_1}_1$
    & $0.18$
    & $1.03$
    & 0.32
    \\
$V^{Ba_1}_2$
    & $0.18$
    & $1.14$
    & $0.49$
&$V^{Bb_1}_2$
    & $0.03^*$
    & $2.13^*$
    & $2.39^*$
    \\
$A^{BK_{1A}}$
    & $0.26$
    & 1.47
    & 0.59
&$A^{BK_{1B}}$
    & $-0.11$
    & 1.88
    & 1.53
    \\
$V^{BK_{1A}}_0$
    & $0.14$
    &  1.62
    &  1.14
&$V^{BK_{1B}}_0$
    & $-0.41$
    & $1.40$
    & $0.64$
    \\
$V^{BK_{1A}}_1$
    & $0.39$
    & $0.21$
    & 0.16
& $V^{BK_{1B}}_1$
    & $-0.19$
    & 0.96
    & 0.30
    \\
$V^{BK_{1A}}_2$
    & $0.17$
    & $1.02$
    & $0.45$
&$V^{BK_{1B}}_2$
    & $0.05^*$
    & $1.78^*$
    & $2.12^*$
    \\
\end{tabular}
\end{ruledtabular}
\end{table}

%%%%%%%%%%%%%%%%%%%%%%%%%%%%%%%%%%%%%%%%%%%%%%%%%%%%%
\begin{table}[t]
\caption{Same as Table \ref{tab:FFinLF} but
in the light-cone sum rule model \cite{YangFF}.} \label{tab:FFinLCSR}
\begin{ruledtabular}
\begin{tabular}{| c c c c || c c c c |}
~~~$F$~~~~~
    & $F(0)$~~~~~
    &$a$~~~~~
    & $b$~~~~~~
& ~~~ $F$~~~~~
    & $F(0)$~~~~~
    & $a$~~~~~
    & $b$~~~~~~
 \\
    \hline
$A^{Ba_1}$
    & $0.30\pm0.05$
    & $1.64$
    & $0.986$
&$A^{Bb_1}$
    & $-0.16\pm0.03$
    & $1.69$
    & ~~$0.910$
    \\
$V_0^{Ba_1}$
    & $0.30\pm 0.05$
    & $1.77$
    & $0.926$
&$V_0^{Bb_1}$
    & $-0.39\pm0.07$
    & $1.22$
    & ~~$0.426$
    \\
$V_1^{Ba_1}$
    & $0.60\pm0.11$
    & ~~$0.645$
    & $0.250$
&$V_1^{Bb_1}$
    & $-0.32\pm 0.06$
    & ~~$0.748$
    & ~~$0.063$
    \\
$V_2^{Ba_1}$
    & $0.26\pm0.05$
    & $1.48$
    & $1.00~~$
&$V_2^{Bb_1}$
    & $-0.06\pm0.01$
    & ~~$0.539$
    & $1.76$
    \\
$A^{BK_{1A}}$
    & $0.27\pm0.05$
    & $1.60$
    & $0.974$
&$A^{BK_{1B}}$
    & $-0.22^{+0.06}_{-0.04}$
    & $1.72$
    & ~~$0.912$
    \\
$V_0^{BK_{1A}}$
    & $0.22\pm0.04$
    & $2.40$
    & $1.78~~$
&$V_0^{BK_{1B}}$
    & $-0.45^{+0.12}_{-0.08}$
    & $1.34$
    & ~~$0.690$
    \\
$V_1^{BK_{1A}}$
    & $0.56\pm0.11$
    & ~~$0.635$
    & $0.211$
&$V_1^{BK_{1B}}$
    & $-0.48^{+0.13}_{-0.08}$
    & ~~$0.729$
    & ~~$0.074$
    \\
$V_2^{BK_{1A}}$
    & $0.25\pm 0.05$
    & $1.51$
    & $1.18~~$
&$V_2^{BK_{1B}}$
    & $-0.10^{+0.03}_{-0.02}$
    & ~~$0.919$
    & ~~$0.855$
    \\
$A^{Bf_1}$
    & $0.18\pm0.03$
    & $1.63$
    & $0.900$
&$A^{Bh_1}$
    & $-0.10\pm0.02$
    & $1.54$
    & ~~$0.848$
    \\
$V_0^{Bf_1}$
    & $0.18\pm 0.03$
    & $1.81$
    & $0.880$
&$V_0^{Bh_1}$
    & $-0.24\pm0.04$
    & $1.16$
    & ~~$0.294$
    \\
$V_1^{Bf_1}$
    & $0.37\pm0.07$
    & ~~$0.640$
    & $0.153$
&$V_1^{Bh_1}$
    & $-0.21\pm 0.04$
    & ~~$0.612$
    & ~~$0.078$
    \\
$V_2^{Bf_1}$
    & $0.16\pm0.03$
    & $1.47$
    & $0.956$
&$V_2^{Bh_1}$
    & $-0.04\pm0.01$
    & ~~$0.500$
    & $1.63$
    \\
$A^{Bf_8}$
    & $0.13\pm0.02$
    & $1.64$
    & $0.919$
&$A^{Bh_8}$
    & $-0.08\pm0.02$
    & $1.56$
    & ~~$0.827$
    \\
$V_0^{Bf_8}$
    & $0.12\pm 0.02$
    & $1.84$
    & $0.749$
&$V_0^{Bh_8}$
    & $-0.18\pm0.03$
    & $1.22$
    & ~~$0.609$
    \\
$V_1^{Bf_8}$
    & $0.26\pm0.05$
    & ~~$0.644$
    & $0.209$
&$V_1^{Bh_8}$
    & $-0.18\pm 0.03$
    & ~~$0.623$
    & ~~$0.094$
    \\
$V_2^{Bf_8}$
    & $0.11\pm0.02$
    & $1.49$
    & $1.09~~$
&$V_2^{Bh_8}$
    & $-0.03\pm0.01$
    & ~~$0.529$
    & $1.53$
    \\
\end{tabular}
\end{ruledtabular}
\end{table}

It should be stressed that in the convention of the present work and LCSR, the
decay constants of $^1P_1$ and $^3P_1$ axial-vector mesons are of the same
sign, while form factors $V_i^{B\to ^1P_1}$ and $V_i^{B\to ^3P_1}$ have
opposite signs. The sign convention is the other way around in the LFQM and
pQCD calculations. Therefore, as explained in \cite{CY:AP}, we put additional
minus signs to the $B\to\, ^1P_1$ form factors in Table \ref{tab:FFinLF}.

The momentum dependence of the form factors calculated in the light-front quark
model and the LCSR approach is parametrized in the three-parameter form:
 \be \label{eq:FFpara}
 F(q^2)=\,{F(0)\over 1-a\,q^2/m_{B}^2+b\,q^4/m_{B}^4}\,
 \en
In the LFQM we use a different parametrization for the form factor $V_2(q^2)$ in
some transitions  \cite{CCH}
 \be \label{eq:FFpara1}
 F(q^2)=\,{F(0)\over (1-q^2/m_{B}^2)[1-a\,q^2/m_{B}^2+b\,q^4/m_{B}^4]}.
 \en

For $B\to \rho,K^*,\omega$ form factors, we shall use the results in \cite{Ball:BV}
obtained from light-cone sum rules.

\subsection{Light-cone distribution amplitudes}
The light-cone distribution amplitudes (LCDAs) relevant for the
present study are defined as~\cite{Ball,YangNP}
\begin{eqnarray}
  \langle V(P,\lambda)|\bar q_1(y) \gamma_\mu q_2(x)|0\rangle
  && = -if_V m_V \, \int_0^1
      du \,  e^{i (u \, p y +
    \bar u p x)}
   \left\{p_\mu \,
    \frac{\epsilon^{*(\lambda)} z}{p z} \, \Phi_\parallel(u)
         +\varepsilon_{\perp\mu}^{*(\lambda)} \, g_\perp^{(v)}(u) \right.
    \nonumber\\
    &&\ \ \left. - \frac{1}{2}z_{\mu}
 \frac{\epsilon^{*(\lambda)} z }{(p  z)^{2}} m_{V}^{2} g_{3}(u)
 \right\}, \label{app:vec-evendef1} \\
  \langle V(P,\lambda)|\bar q_1(y) \gamma_\mu \gamma_5 q_2(x)|0\rangle
  && = i  f_V \Bigg(1-{f_V^{\perp} \over f_V}{m_{q_1}+m_{q_2}\over m_V}\Bigg)
   m_V \,\varepsilon_{\mu\nu\rho\sigma} \,
      \epsilon^{*\nu}_{(\lambda)} p^{\rho} z^\sigma \nonumber\\
  && \ \ \ \times \int_0^1 du \, e^{i (u \, p y + \bar u\, p  x)} \,
       \frac{g_\perp^{(a)}(u)}{4}\,, \label{app:vec-evendef2}
\end{eqnarray}

\begin{eqnarray}
  &&\langle V(P,\lambda)|\bar q_1(y) \sigma_{\mu\nu} q_2(x) |0\rangle
  =  - f_V^{\perp} \,\int_0^1 du \, e^{i (u \, p y +
    \bar u\, p x)} \,
\Bigg\{(\varepsilon^{*(\lambda)}_{\perp\mu} p_{\nu} -
  \varepsilon_{\perp\nu}^{*(\lambda)}  p_{\mu})
  \Phi_\perp(u)\nonumber\\
&& \hspace*{+5cm}
  + \,\frac{m_A^2\,\epsilon^{*(\lambda)} z}{(p z)^2} \,
   (p_\mu z_\nu -
    p_\nu  z_\mu) \, h_\parallel^{(t)}(u)\nonumber\\
 && \hspace*{+5cm} + \frac{1}{2}
(\varepsilon^{*(\lambda)}_{\perp \mu} z_\nu
-\varepsilon^{*(\lambda)}_{\perp \nu} z_\mu) \frac{m_{V}^{2}}{p\cdot
z}
 h_{3}(u) \Bigg\}\,,\label{app:vec-odddef1}\\
&&\langle V(P,\lambda)|\bar q_1(y) q_2(x) |0\rangle
  =  - f_V^\perp \Bigg(1-{f_V \over f_V^\perp}{m_{q_1}+m_{q_2}\over m_V}\Bigg)
 m_{V}^2 \, (\epsilon^{*(\lambda)}z) \,\int_0^1 du \, e^{i (u \, p y +
 \bar u\, p x)}  \, \frac{h_\parallel^{(s)}(u)}{2}\,,{} \nonumber\\
 \label{app:vec-odddef2}
\end{eqnarray}
for the vector meson, and
\begin{eqnarray}
  \langle A(P,\lambda)|\bar q_1(y) \gamma_\mu \gamma_5 q_2(x)|0\rangle
  && = if_A m_A \, \int_0^1
      du \,  e^{i (u \, p y +
    \bar u p x)}
   \left\{p_\mu \,
    \frac{\epsilon^{*(\lambda)} z}{p z} \, \Phi_\parallel(u)
         +\varepsilon_{\perp\mu}^{*(\lambda)} \, g_\perp^{(a)}(u) \right.
    \nonumber\\
    &&\ \ \left. - \frac{1}{2}z_{\mu}
 \frac{\epsilon^{*(\lambda)} z }{(p  z)^{2}} m_{A}^{2} g_{3}(u)
 \right\}, \label{app:axial-evendef1} \\
  \langle A(P,\lambda)|\bar q_1(y) \gamma_\mu q_2(x)|0\rangle
  && = - i f_A m_A \,\varepsilon_{\mu\nu\rho\sigma} \,
      \epsilon^{*\nu}_{(\lambda)} p^{\rho} z^\sigma \, \int_0^1 du \,
   e^{i (u \, p y + \bar u\, p  x)} \,
       \frac{g_\perp^{(v)}(u)}{4}, \label{app:axial-evendef2}
\end{eqnarray}
\begin{eqnarray}
  &&\langle A(P,\lambda)|\bar q_1(y) \sigma_{\mu\nu}\gamma_5 q_2(x)
            |0\rangle
  =  f_A^{\perp} \,\int_0^1 du \, e^{i (u \, p y +
    \bar u\, p x)} \,
\Bigg\{(\varepsilon^{*(\lambda)}_{\perp\mu} p_{\nu} -
  \varepsilon_{\perp\nu}^{*(\lambda)}  p_{\mu})
  \Phi_\perp(u)\nonumber\\
&& \hspace*{+5cm}
  + \,\frac{m_A^2\,\epsilon^{*(\lambda)} z}{(p z)^2} \,
   (p_\mu z_\nu -
    p_\nu  z_\mu) \, h_\parallel^{(t)}(u)\nonumber\\
 && \hspace*{+5cm} + \frac{1}{2}
(\varepsilon^{*(\lambda)}_{\perp \mu} z_\nu
 -\varepsilon^{*(\lambda)}_{\perp \nu} z_\mu) \frac{m_{A}^{2}}{pz}
 h_{3}(u)
\Bigg\},\label{app:axial-odddef1}\\
&&\langle A(P,\lambda)|\bar q_1(y) \gamma_5 q_2(x) |0\rangle
  =  f_A^\perp
 m_{A}^2 (\epsilon^{*(\lambda)} z)\,\int_0^1 du \, e^{i (u \, p y +
    \bar u\, p x)}  \, \frac{h_\parallel^{(p)}(u)}{2}\,, \label{app:axial-odddef2}
\end{eqnarray}
for the axial-vector meson, where $z=y-x$ with $z^2=0$ and we have
introduced the light-like vector $p_\mu=P_\mu-m_{V(A)}^2 z_\mu/(2
P z)$ with the meson momentum ${P}^2=m_{V(A)}^2$. Here
$\Phi_\parallel, \Phi_\perp$ are twist-2 LCDAs, $g_\perp^{(a)},
g_\perp^{(v)}, h_\parallel^{(t)}, h_\parallel^{(p)}$  twist-3 ones,
and $g_3, h_3$  twist-4. In the definitions of LCDAs, the
longitudinal and transverse {\it projections} of polarization
vectors $\epsilon^{*(\lambda)}_\mu$ along the $z-$direction for the (axial-)vector meson
are given by \cite{Ball}
\begin{eqnarray}\label{eq:polprojectiors}
 && \varepsilon^{*(\lambda)}_{\parallel\, \mu} \equiv
     \frac{\epsilon^{*(\lambda)} z}{p z} \left(
      p_\mu-\frac{m_{V(A)}^2}{2 p z} \,z_\mu\right), \qquad
 \varepsilon^{*(\lambda)}_{\perp\, \mu}
        = \epsilon^{*(\lambda)}_\mu -\varepsilon^{*(\lambda)}_{\parallel\,
        \mu}\,.
\end{eqnarray}
One should distinguish the above projectors from the {\it exactly}
longitudinal and transverse polarization vectors of the
(axial-)vector meson, which are independent of the coordinate variable $z=y-x$, defined as
\begin{equation}
\epsilon^{*(0)\mu}= \frac{E}{m_{V(A)}}
\Bigg[ \Bigg(1-\frac{m_{V(A)}^2}{4E^2}\Bigg) n_-^\mu
 - \frac{m_{V(A)}^2}{4 E^2} n_+^\mu\Bigg], \quad
\epsilon^{*(\lambda)\mu}_\perp \equiv \Bigg(\epsilon^{*(\lambda)\mu} - \frac{\epsilon^{*(\lambda)}
n_+}{2}\,n_-^\mu- \frac{\epsilon^{*(\lambda)} n_-}{2}\,n_+^\mu \Bigg)\delta_{\lambda,\pm1}\,,
\end{equation}
where we have defined two light-like
vectors $n_\pm^\mu$ with $n_-^\mu\equiv (1,0,0,-1)$, and $n_+^\mu\equiv
(1,0,0,1)$ and assumed
that the meson moves along the $n_-^\mu$ direction.

In the QCDF calculation, the LCDAs of the vector meson appear in
the following way \cite{Beneke:2000wa}
  \begin{eqnarray}
  &&  \langle V(P,\lambda)|\bar q_{1\,\alpha}(y) \, q_{2\, \delta}(x)|0\rangle
= -\frac{i}{4} \, \int_0^1 du \,  e^{i (u \, p y +
    \bar u p x)}
\nonumber\\[0.1cm]
  && \quad \times\,\Bigg\{ f_V m_V \left(
    \not\!p \, \frac{\epsilon^{*(\lambda)} z}{p z} \,
    \Phi_\parallel(u) +  \not\! \varepsilon^{*(\lambda)}_\perp \,
    g_\perp^{(v)}(u) +  \varepsilon_{\mu\nu\rho\sigma} \,
    \epsilon^{*\mu}_{(\lambda)}  p^{\rho} z^\sigma \, \gamma^\nu\gamma_5
    \, \frac{g_\perp^{(a)}(u)}{4}\right)
\nonumber \\[0.1em]
  && \qquad \,\,\,
  + \,f^{\perp}_V \left(\not\!\varepsilon_\perp^{*(\lambda)} \not\! p \,
  \Phi_\perp(u)-i{m_V^2 (\epsilon^{*(\lambda)} z) \over (p\cdot z)^2}\sigma_{\mu\nu}p^\mu
  z^\nu h_\parallel^{(t)}(u)-im_V^2 (\epsilon^{*(\lambda)} z) {h_\parallel^{(s)}
  (u)\over 2}\right) \non \\
  && \qquad \,\,\, +{\cal O}[(x-y)^2]\Bigg\}_{\delta\alpha}\,. \label{eq:DAs}
 \end{eqnarray}
Here, all the components of the parton should be taken
into account in the calculation before the collinear approximation
is applied, so that one can assign the momenta
 \begin{eqnarray}
 k_1^\mu = u E n_-^\mu
+k_\perp^\mu + \frac{k_\perp^2}{4 uE}n_+^\mu\,,
 \qquad
 k_2^\mu = \bar u E n_-^\mu
- k_\perp^\mu + \frac{k_\perp^2}{4 \bar uE}n_+^\mu\,,
\end{eqnarray}
to the quark and antiquark, respectively, in an energetic light final-state meson
with the momentum $P^\mu$ and mass $m$, satisfying the relation
$P^\mu =En_-^\mu + m^2 n_+^\mu/(4E) \simeq E n_-^\mu$. To obtain
the light-cone projection operator of the meson in the momentum
space, we apply the following substitution in the calculation
\begin{equation}
z^\mu \to -i \frac{\partial}{\partial k_{1\, \mu}}\simeq -i \Bigg(
\frac{n_+^\mu}{2E}\frac{\partial}{\partial u} +
\frac{\partial}{\partial k_{\perp\, \mu}}\Bigg)\,,
\end{equation}
where terms of order $k_\perp^2$ have been omitted. Moreover, to
perform the calculation in the momentum space, we need to express
Eq.~(\ref{eq:DAs}) in terms of $z$-independent variables, $P$ and
$\epsilon^{*(\lambda)}$, instead of $p$ and
$\varepsilon^{*(\lambda)}$. Consequently, the light-cone projection operator of
the meson in the momentum space, including twist-3 two-parton
distribution amplitudes, reads
\begin{equation}
  M_{\delta\alpha} =  M_{\delta\alpha}{}_\parallel +
   M_{\delta\alpha}{}_\perp\,,
\label{app:rhomeson2}
\end{equation}
where $M_{\delta\alpha}{}_\parallel$ and
$M_{\delta\alpha}{}_\perp$ are the longitudinal and transverse
projectors, respectively.

For the vector meson, the longitudinal projector reads
\cite{Beneke:2000wa}
 \begin{eqnarray} \label{eq:VproL}
M^V_\parallel &=& -i\frac{f_V}{4} \, \frac{m_V(\epsilon^*_{(\lambda)}
  n_+)}{2}
 \not\! n_- \,\Phi_\parallel(u)
- i\frac{f_V^\perp m_V}{4}  \,\frac{m_V(\epsilon^*_{(\lambda)} n_+)}{2E}
 \, \Bigg\{ -\frac{i}{2}\,\sigma_{\mu\nu} \,  n_-^\mu  n_+^\nu \,
 h_\parallel^{(t)}(u)
\nonumber\\
&& \hspace*{-0.0cm} - \,i E\int_0^u dv \,(\Phi_\perp(v) -
 h_\parallel^{(t)}(v)) \
 \sigma_{\mu\nu} n_-^\mu \, \frac{\partial}{\partial k_\perp{}_\nu}
  +  \frac{h_\parallel'{}^{(s)}(u)}{2} \Bigg\}\, \Bigg|_{k=u p}+{\cal O}
  \Bigg[\bigg(\frac{m_V}{E}\bigg)^2\Bigg]\,,
\end{eqnarray}
and the transverse projector has the form
 \begin{eqnarray} \label{eq:VproT}
 M^V_\perp &=& -i\frac{f^{\perp}_V}{4} E\not\! \epsilon^{*(\lambda)}_\perp\not\! n_-  \,
   \Phi_\perp(u)\nonumber\\
&&  -i \frac{f_V m_V}{4} \,\Bigg\{\not\! \epsilon^{*(\lambda)}_\perp \,
g_\perp^{(v)}(u) -  \, E\int_0^u dv\, (\Phi_\parallel(v) -
g_\perp^{(v)}(v))
       \not\! n_- \, \epsilon^{*(\lambda)}_{\perp\mu} \,\frac{\partial}{\partial
         k_{\perp\mu}}
\cr && + \,i \varepsilon_{\mu\nu\rho\sigma} \,
        \gamma^\mu\epsilon_\perp^{*(\lambda)\nu} \,  n_-^\rho \gamma_5
         \left[n_+^\sigma \,{g_\perp'^{(a)}(u)\over 8}-
          E\,\frac{g_\perp^{(a)}(u)}{4} \, \frac{\partial}{\partial
         k_\perp{}_\sigma}\right]
 \Bigg\} \, \Bigg|_{k=up}  +{\cal O}\Bigg[\bigg(\frac{m_V}{E}\bigg)^2\Bigg],
\end{eqnarray}
where $k_\perp$ is the transverse momentum of the $q_1$ quark in
the vector meson. For the axial-vector meson, the longitudinal
projector is given by
 \begin{eqnarray} \label{eq:AproL}
M^A_\parallel &=& -i\frac{f_A}{4} \, \frac{m_A(\epsilon^*_{(\lambda)} n_+)}{2}
 \not\! n_- \gamma_5\,\Phi_\parallel(u)
 +i\frac{f_A^\perp m_A}{4}  \,\frac{m_A(\epsilon^*_{(\lambda)} n_+)}{2E}
 \, \Bigg\{-\frac{i}{2}\,\sigma_{\mu\nu}\gamma_5 \,  n_-^\mu  n_+^\nu \,
 h_\parallel^{(t)}(u)
\nonumber\\
&& \hspace*{-0.0cm} - \,i E\int_0^u dv \,(\Phi_\perp(v) -
h_\parallel^{(t)}(v)) \
     \sigma_{\mu\nu} \gamma_5 n_-^\mu
     \, \frac{\partial}{\partial k_\perp{}_\nu}
  +\gamma_5\frac{h_\parallel'^{(p)}(u)}{2}\Bigg\}\, \Bigg|_{k=u p}
  +{\cal O}\Bigg[\bigg(\frac{m_A}{E}\bigg)^2\Bigg]\,, \ \ \
\end{eqnarray}
 and the transverse projector  given by
 \begin{eqnarray} \label{eq:AproT}
M^A_\perp &=& i\frac{f^{\perp}_A}{4} E\not\!
\epsilon^{*(\lambda)}_\perp\not\! n_- \gamma_5 \,
   \Phi_\perp(u)\nonumber\\
&&  -i \frac{f_Am_A}{4} \,\Bigg\{\not\! \epsilon^{*(\lambda)}_\perp\gamma_5 \,
g_\perp^{(a)}(u) -  \, E\int_0^u dv\, (\Phi_\parallel(v) - g_\perp^{(a)}(v))
 \not\! n_-\gamma_5 \, \epsilon^{*(\lambda)}_{\perp\mu}
\,\frac{\partial}{\partial
         k_{\perp\mu}}
\cr && + \,i \varepsilon_{\mu\nu\rho\sigma} \,
       \gamma^\mu \epsilon_\perp^{*(\lambda)\nu} \,  n_-^\rho
         \left[n_+^\sigma \,{g_\perp'^{(v)}(u)\over 8}-
          E\,\frac{g_\perp^{(v)}(u)}{4} \, \frac{\partial}{\partial
         k_\perp{}_\sigma}\right]
 \Bigg\}
 \, \Bigg|_{k=up} +{\cal O}\Bigg[\bigg(\frac{m_A}{E}\bigg)^2\Bigg]\,.
\end{eqnarray}
In the present study, we choose the
coordinate systems in the Jackson convention;  that is, in the
$\overline B$ rest frame, one of the vector or axial-vector mesons
is moving along the $z$ axis of the coordinate system and the
other along the $-z$ axis, while the $x$ axes of both daughter
particles are parallel \cite{T'Jampens}
 \begin{eqnarray}
 &&\epsilon_{1}^{\mu(0)}=(p_c, 0, 0, E_1)/m_1,\ \ \ \
   \epsilon_{2}^{\mu(0)}=(p_c, 0, 0, -E_{2})/m_{2}, \nonumber\\
 &&\epsilon_{1}^{\mu (\pm 1)}=\frac{1}{\sqrt{2}}(0, \mp 1, -i, 0),
 \ \ \ \
   \epsilon_{2}^{\mu (\pm 1)}=\frac{1}{\sqrt{2}}(0, \mp 1, +i,
   0),
 \end{eqnarray}
where $p_c$ is the center mass momentum of the final state meson
and $\epsilon_1^{*(\pm1)}\cdot
\epsilon_2^{*(\pm1)}=-\delta_{\pm1,\pm1}$. In the large energy
limit, if the $A$ meson moves along the $n_-^\mu$ direction, we will
have $\epsilon_A^{*(\lambda)}\cdot n_+ =2E_A/m_A\,
\delta_{\lambda,0}$ and $\epsilon_A^{*(\lambda)}\cdot n_- =0$. Note
that if the coordinate systems are in the Jacob-Wick convention
where the $y$ axes of both decay particles are parallel, the
transverse polarization vectors of the second meson will become
$\epsilon_{2}^{\mu}(\pm 1)=(0, \pm 1, -i, 0)/\sqrt{2}$ and
$\epsilon_1^{*(\pm1)}\cdot
\epsilon_2^{*(\pm1)}=\delta_{\pm1,\pm1}$. In general, the QCDF
amplitudes can be reduced to the form of $\int_0^1 du \, {\rm tr}
(M^A\dots)$.

To obtain the projector on the transverse polarization states in
the helicity basis, one can insert
$\epsilon^*_\perp=\epsilon^*_\mp$ to obtain
 \be \label{eq:MV-+}
 M^V_\mp(u) &=& -i{f_V^\perp\over 4}E\not\!
\epsilon^{*(\lambda)}_\mp\not\! n_-\Phi_\perp^V(u) \non \\
 &&-i{f_V m_V\over 8}\Bigg\{ \not\!
\epsilon^{*(\lambda)}_\mp(1-\gamma_5)
\left(g_\perp^{(v)}(u)\pm {g_\perp'^{(a)}(u)\over 4}\right)+\not\!
\epsilon^{*(\lambda)}_\mp(1+\gamma_5)
\left(g_\perp^{(v)}(u)\mp {g_\perp'^{(a)}(u)\over 4}\right) \non \\
&&\ \ \ -E\not\! n_-(1-\gamma_5)\left(\int^u_0
dv(\Phi_\parallel(v)-g_\perp^{(v)}(v)) \mp{g_\perp^{(a)}(u)\over
4}\right)\epsilon^*_{\mp\nu}{\partial\over \partial k_{\perp\nu}}
 \non \\
&&\ \ \ -E\not\! n_-(1+\gamma_5)\left(\int^u_0 dv(\Phi_\parallel(v)-g_\perp^{(v)}(v))
\pm{g_\perp^{(a)}(u)\over 4}\right)\epsilon^*_{\mp\nu}{\partial\over \partial k_{\perp\nu}}
\Bigg\} \Bigg|_{k_\perp=0}+{\cal O}\left[\left({m_V\over E}\right)^2\right], \non \\
 \en
and
 \be \label{eq:MA-+}
 M^A_\mp(u) &=& i{f_A^\perp\over 4}E\not\!
\epsilon^{*(\lambda)}_\mp\not\! n_-\gamma_5\Phi_\perp^A(u) \non \\
 &&-i{f_A m_A\over 8}\Bigg\{ -\not\!
\epsilon^{*(\lambda)}_\mp(1-\gamma_5)\left(g_\perp^{(a)}(u)\pm {g_\perp'^{(v)}(u)\over 4}\right)+\not\!
\epsilon^{*(\lambda)}_\mp(1+\gamma_5)\left(g_\perp^{(a)}(u)\mp {g_\perp'^{(v)}(u)\over 4}\right) \non \\
&&\ \ \ +E\not\! n_-(1-\gamma_5)\left(\int^u_0
dv(\Phi_\parallel(v)-g_\perp^{(a)}(v)) \mp{g_\perp^{(v)}(u)\over
4}\right)\epsilon^*_{\mp\nu}{\partial\over \partial k_{\perp\nu}}
 \non \\
&&\ \ \ -E\not\! n_-(1+\gamma_5)\left(\int^u_0
dv(\Phi_\parallel(v)-g_\perp^{(a)}(v)) \pm{g_\perp^{(v)}(u)\over
4}\right)\epsilon^*_{\mp\nu}{\partial\over \partial k_{\perp\nu}}
\Bigg\} \Bigg|_{k_\perp=0}+{\cal O}\left[\left({m_A\over E}\right)^2\right]. \non \\
 \en

Applying equations of motion to LCDAs, one can obtain the
following Wandzura-Wilczek relations in which twist-3 LCDAs are related to
the twist-2 ones \cite{Ball} via
 \begin{eqnarray}
  g_\perp^{(v)}(u) &=& \frac12 \left[ \,\int_0^u
    \frac{\Phi_\parallel(v)}{\bar v} \, dv  + \int_u^1
    \frac{\Phi_\parallel(v)}{v}\, dv  \right] +\ldots \,,\label{eq:v-ww1}
 \\
  g_\perp^{(a)}(u) &=& 2 \left[ \bar u \int_0^u
    \frac{\Phi_\parallel(v)}{\bar v}\, dv  + u \, \int_u^1
    \frac{\Phi_\parallel(v)}{v}\, dv  \right] +\ldots \,,
\label{app:v-ww2}\\
h_\parallel^{(t)}(u) &=& (2u-1) \left[ \,\int_0^u
    \frac{\Phi_\perp(v)}{\bar v} \, dv  - \int_u^1
    \frac{\Phi_\perp(v)}{v}\, dv  \right] +\ldots \,,\label{app:v-ww3}
 \\
  h_\parallel^{(s)}(u) &=& 2 \left[ \bar u \int_0^u
    \frac{\Phi_\perp(v)}{\bar v}\, dv  + u \, \int_u^1
    \frac{\Phi_\perp(v)}{v}\, dv  \right] +\ldots \,,
\label{app:v-ww4}
 \end{eqnarray}
for vector mesons, and
 \begin{eqnarray}
  g_\perp^{(a)}(u) &=& \frac12 \left[ \,\int_0^u
    \frac{\Phi_\parallel(v)}{\bar v} \, dv  + \int_u^1
    \frac{\Phi_\parallel(v)}{v}\, dv  \right] +\ldots \,,\label{app:a-ww1}
 \\
  g_\perp^{(v)}(u) &=& 2 \left[ \bar u \int_0^u
    \frac{\Phi_\parallel(v)}{\bar v}\, dv  + u \, \int_u^1
    \frac{\Phi_\parallel(v)}{v}\, dv  \right] +\ldots \,,
\label{app:a-ww2}\\
h_\parallel^{(t)}(u) &=& (2u-1) \left[ \,\int_0^u
    \frac{\Phi_\perp(v)}{\bar v} \, dv  - \int_u^1
    \frac{\Phi_\perp(v)}{v}\, dv  \right] +\ldots \,,\label{app:a-ww3}
 \\
  h_\parallel^{(p)}(u) &=& 2 \left[ \bar u \int_0^u
    \frac{\Phi_\perp(v)}{\bar v}\, dv  + u \, \int_u^1
    \frac{\Phi_\perp(v)}{v}\, dv  \right] +\ldots \,,
\label{app:a-ww4}
 \end{eqnarray}
for axial-vector mesons, where the ellipses denote additional
contributions from three-particle distribution amplitudes
containing gluons and terms proportional to light quark masses,
which we do not consider here and below.
Eqs.~(\ref{eq:v-ww1})-(\ref{app:a-ww4}) further give us
\begin{eqnarray}\label{eq:v-ww}
  && \frac{g_\perp^{\prime(a)}(v)}{4}+g_\perp^{(v)}(v)= \int_v^1
\frac{\Phi_\parallel (u)}{u}du \equiv \Phi_+(v)\,,\nonumber\\
&& \frac{g_\perp^{\prime(a)}(v)}{4}-g_\perp^{(v)}(v)=- \int_0^v
\frac{\Phi_\parallel (u)}{\bar u}du \equiv -\Phi_-(v)\,,\nonumber\\
 &&
h_\parallel^{\prime(s)}(v)= -2\Bigg[ \int_0^v
\frac{\Phi_\perp(u)}{\bar u}du -\int_v^1
\frac{\Phi_\perp(u)}{u}du \Bigg] \equiv -2 \Phi_v(v), \\
 &&\int_0^v du \big( \Phi_\perp (u) -h^{(t)}_\parallel
(u))= v\bar v\Bigg[ \int_0^v \frac{\Phi_\perp(u)}{\bar u}du
-\int_v^1
\frac{\Phi_\perp(u)}{u}du\Bigg] = v \bar v\Phi_v(v), \nonumber\\
&&\int_0^v du \big( \Phi_\parallel (u) -g^{(v)}_\perp (u))=
 \frac{1}{2}\Bigg[ \bar v\int_0^v \frac{\Phi_\parallel(u)}{\bar u}du
 -v \int_v^1 \frac{\Phi_\parallel(u)}{u}du\Bigg]
 =\frac{1}{2}\bigg(\bar v \Phi_-(v) -v \Phi_+(v)\bigg), \nonumber
\end{eqnarray}
for vector mesons, and
\begin{eqnarray}\label{eq:a-ww}
  && \frac{g_\perp^{\prime(v)}(v)}{4}+g_\perp^{(a)}(v)= \int_v^1
\frac{\Phi_\parallel (u)}{u}du  \equiv \Phi_+ (v)\,,\nonumber\\
&& \frac{g_\perp^{\prime(v)}(v)}{4}-g_\perp^{(a)}(v)=- \int_0^v
\frac{\Phi_\parallel (u)}{\bar u}du \equiv -\Phi_-(v)\,,\nonumber\\
 &&
h_\parallel^{\prime(p)}(v)= -2\Bigg[ \int_0^v
\frac{\Phi_\perp(u)}{\bar u}du -\int_v^1
\frac{\Phi_\perp(u)}{u}du \Bigg] \equiv -2 \Phi_a(v), \nonumber\\
 &&\int_0^v du \big( \Phi_\perp (u) -h^{(t)}_\parallel
(u))= v\bar v\Bigg[ \int_0^v \frac{\Phi_\perp(u)}{\bar u}du
-\int_v^1
\frac{\Phi_\perp(u)}{u}du\Bigg] = v \bar v\Phi_a(v), \\
&&\int_0^v du \big( \Phi_\parallel (u) -g^{(a)}_\perp (u))=
 \frac{1}{2}\Bigg[ \bar v\int_0^v \frac{\Phi_\parallel(u)}{\bar u}du
 -v \int_v^1 \frac{\Phi_\parallel(u)}{u}du\Bigg]
 =\frac{1}{2}\bigg(\bar v \Phi_-(v) -v \Phi_+(v)\bigg),
 \nonumber
\end{eqnarray}
for axial-vector mesons.

After applying the Wandzura-Wilczek relations, the transverse
helicity projectors (\ref{eq:MV-+}) and (\ref{eq:MA-+}) can be
simplified to
  \be \label{eq:MV-+simple}
 M^V_\mp(u) &=& -i{f_V^\perp\over 4}E\not\!
\epsilon^{*(\lambda)}_\mp\not\! n_-\Phi_\perp^V(u) \non \\
 &&-i{f_V m_V\over 8}\Bigg\{
\epsilon^{*(\lambda)}_{\mp\nu}\Phi_+(u)\left[\gamma^\nu(1\mp\gamma_5) +
 uE\not\! n_-(1\mp\gamma_5){\partial\over \partial k_{\perp\nu}}\right]
 \non \\
&& \quad
 +\epsilon^{*(\lambda)}_{\mp\nu}\Phi_-(u)\left[\gamma^\nu(1\pm\gamma_5) -
 \bar{u}E\not\! n_-(1\pm\gamma_5){\partial\over \partial k_{\perp\nu}}\right]
 \Bigg\}\Bigg|_{k_\perp=0}+{\cal O}\left[\left({m_V\over E}\right)^2\right], \non \\
 \en
and
  \be \label{eq:MA-+simple}
 M^A_\mp(u) &=& i{f_A^\perp\over 4}E\not\!
\epsilon^{*(\lambda)}_\mp\not\! n_-\gamma_5\Phi_\perp^A(u) \non \\
 &&-i{f_A m_A\over 8}\Bigg\{
\epsilon^{*(\lambda)}_{\mp\nu}\Phi_+(u)\left[\gamma^\nu(1\mp\gamma_5) +
 uE\not\! n_-(1\mp\gamma_5){\partial\over \partial k_{\perp\nu}}\right]\gamma_5
 \non \\
&& \qquad
 +\epsilon^{*(\lambda)}_{\mp\nu}\Phi_-(u)\left[\gamma^\nu(1\pm\gamma_5) -
 \bar{u}E\not\! n_-(1\pm\gamma_5){\partial\over \partial k_{\perp\nu}}\right]
 \gamma_5\Bigg\}\Bigg|_{k_\perp=0}+{\cal O}\left[\left({m_A\over E}\right)^2\right]. \non \\
 \en

From Eqs. (\ref{eq:VproL})-(\ref{eq:AproT}) and
(\ref{eq:v-ww1})-(\ref{eq:a-ww}), we see that $\Phi_+$ and $\Phi_-$
project onto transversely polarized vector or axial-vector mesons in
which quark and antiquark flips helicity, respectively, while
$\Phi_{v(a)}$ projects onto longitudinally polarized vector
(axial-vector) mesons in which either the quark or antiquark flips
helicity.

We next specify the LCDAs for vector and axial-vector mesons. The
general expressions of LCDAs are
 \be
 \Phi_V(x,\mu)=6x(1-x)\left[1+\sum_{n=1}^\infty
 a_n^V(\mu)C_n^{3/2}(2x-1)\right],
 \en
and
 \be
 \Phi_v(x,\mu)=3\left[2x-1+\sum_{n=1}^\infty
 a_{n}^{\bot,V}(\mu)P_{n+1}(2x-1)\right],
 \en
for the vector meson, where $P_n(x)$ are the Legendre polynomials.
The normalization of LCDAs is
 \be
 \int^1_0dx \Phi_V(x)=1, \qquad \int^1_0dx \Phi_v(x)=0.
 \en

The explicit expressions of the LCDAs of axial-vector mesons have
been discussed in details in \cite{YangNP,CY:AP}. We use
\begin{eqnarray}
\Phi_\perp^{^1P_1} (u) &=&  6 u \bar u
 \left\{ 1 +3a_1^{\perp,^1P_1}(2u-1)+a_2^{\perp,^1P_1}\, \frac{3}{2} \bigg[ 5 (2u-1)^2  - 1 \bigg]
 \right\}, \non \\
 \Phi_\parallel^{^1P_1} (u) & = &  6 u \bar u \left\{a_0^{\parallel,^1P_1} +
 3a_1^{\parallel,^1P_1}(2u-1)+a_2^{\parallel,^1P_1}\,
 \frac{3}{2} \bigg[ 5 (2u-1)^2  - 1 \bigg]\right\}, \non \\
 \Phi_a^{^1P_1}(u) &=& 3\left[(2u-1)+\sum_{n=1}^\infty
 a_{n}^{\perp,^1P_1}(\mu)P_{n+1}(2u-1)\right],
 \label{eq:LCDA1P1}
\end{eqnarray}
for $^1P_1$ mesons, and
\begin{eqnarray} \label{eq:LCDA3P1}
\Phi_\parallel^{^3P_1} (u)& = & 6 u \bar u
 \left\{ 1 +3a_1^{\parallel,^3P_1}(2u-1)+a_2^{\parallel,^3P_1}\, \frac{3}{2} \bigg[ 5 (2u-1)^2  - 1 \bigg]
 \right\}, \non \\
 \Phi_\perp^{^3P_1} (u) & = & 6 u \bar u\left\{a_0^{\perp,^3P_1}+3a_1^{\perp,^3P_1}(2u-1)+a_2^{\perp,^3P_1}\,
 \frac{3}{2} \bigg[ 5 (2u-1)^2  - 1 \bigg]\right\}, \non \\
 \Phi_a^{^3P_1}(u) &=& 3\left[a_0^{\perp,^3P_1}(2u-1)+\sum_{n=1}^\infty
  a_{n}^{\perp,^3P_1}(\mu)P_{n+1}(2u-1)\right],
\end{eqnarray}
for $^3P_1$ mesons. The normalization conditions are
 \be \label{eq:nor}
 && \int^1_0dx\Phi_\parallel^{^1P_1}(x)=a_0^{\parallel,^1P_1}, \qquad
 \int^1_0dx\Phi_\parallel^{^3P_1}(x)=1, \non \\
  && \int^1_0dx\Phi_\perp^{^1P_1}(x)=1, \qquad\qquad
 \int^1_0dx\Phi_\parallel^{^3P_1}(x)=a_0^{\perp,^3P_1}, \non \\
 && \int^1_0 dx\Phi_a^{^1P_1}(x)= \int^1_0 dx\Phi_a^{^3P_1}(x)=0.
 \en

It should be stressed that the LCDAs $\Phi_\parallel^{^1P_1}$ and
$\Phi_\perp^{^3P_1}$ are defined with the decay constants $f^\bot_{^1P_1}$ and
$f_{^3P_1}$, respectively, even though their corresponding normalizations are
$f_{^1P_1}$ and $f^\perp_{^3P_1}$. As stressed in \cite{CY:AP}, if we employ
the decay constants $f_{^1P_1}$ and $f^\perp_{^3P_1}$ to define the the LCDAs
$\Phi_\parallel^{^1P_1}$ and $\Phi_\perp^{^3P_1}$, they will have the form \be
 \Phi_\parallel^{^1P_1} (u) & = &  f_{^1P_1}6 u \bar u \left\{1+\mu_{^1P_1}
 \sum_{i=1}^2a_i^{\parallel,^1P_1}C_i^{3/2}(2u-1)
 \right\}, \non \\
  \Phi_\perp^{^3P_1} (u) & = & f_{^3P_1}^\perp6 u \bar u
  \left\{1+\mu_{^3P_1}\sum_{i=1}^2a_i^{\perp,^3P_1}C_i^{3/2}(2u-1) \right\},
 \en
where $\mu_{^1P_1}=1/a_0^{\parallel,^1P_1}$ and
$\mu_{^3P_1}=1/a_0^{\perp,^3P_1}$  which become infinite in the SU(3) limit.
Therefore, it is most convenient to use Eq. (\ref{eq:LCDA1P1}) for the LCDA
$\Phi_\parallel^{^1P_1}$ and (\ref{eq:LCDA3P1}) for the LCDA
$\Phi_\parallel^{^3P_1}$ which amount to treating the decay constant of $^1P_1$
as $f_{^1P_1}^\perp$ and the tensor decay constant of $^3P_1$ as $f_{^3P_1}$.
Of course, this does not mean that $f_{^1P_1}$ ($f_{^3P_1}^\perp$) is equal to
$f_{^1P_1}^\perp$ ($f_{^3P_1}$).

For the $B$ meson, we use the light-cone projector
\cite{Beneke:2000wa}
 \be
 M_{\beta\alpha}^B=-{if_Bm_B\over 4}\Bigg\{{1+v\!\!\!/\over 2}
 \left[\Phi_1^B(\omega)n\!\!\!/_+
+\Phi_2^B(\omega)\Big(n\!\!\!/_--l_+\gamma_\perp^\nu{\partial\over
 \partial l_\perp^\nu}\Big)\right]\gamma_5\Bigg\}_{\alpha\beta}.
 \en
The integral of the $B$
meson wave function is parameterized as \cite{BBNS}
 \begin{eqnarray}
 \int_0^1 \frac{d\rho}{1-\rho}\Phi_1^B(\rho) \equiv
 \frac{m_B}{\lambda_B}\,,
 \end{eqnarray}
where $1-\rho$ is the momentum fraction carried by the light
spectator quark in the $B$ meson.

\subsection{A summary of input quantities}
It is useful to summarize all the input quantities we have used in this work.

For the CKM matrix elements, we use the Wolfenstein parameters
$A=0.807\pm0.018$, $\lambda=0.2265\pm0.0008$, $\bar \rho=0.141^{+0.029}_{-0.017}$ and $\bar
\eta=0.343\pm0.016$ \cite{CKMfitter}. The corresponding three unitarity
angles are $\alpha=(90.7^{+4.5}_{-1.9})^\circ$, $\beta=(21.7\pm0.017)^\circ$ and
$\gamma=(67.6^{+2.8}_{-4.5})^\circ$.

For the running quark masses we shall use \cite{PDG,Xing}
 \be \label{eq:quarkmass}
 && m_b(m_b)=4.2\,{\rm GeV}, \qquad~~~~ m_b(2.1\,{\rm GeV})=4.94\,{\rm
 GeV}, \qquad m_b(1\,{\rm GeV})=6.34\,{\rm
 GeV}, \non \\
 && m_c(m_b)=0.91\,{\rm GeV}, \qquad~~~ m_c(2.1\,{\rm GeV})=1.06\,{\rm  GeV},
 \qquad m_c(1\,{\rm GeV})=1.32\,{\rm
 GeV}, \non \\
 && m_s(2.1\,{\rm GeV})=95\,{\rm MeV}, \quad~ m_s(1\,{\rm GeV})=118\,{\rm
 MeV}, \non\\
 && m_d(2.1\,{\rm GeV})=5.0\,{\rm  MeV}, \quad~ m_u(2.1\,{\rm GeV})=2.2\,{\rm
 MeV}.
 \en
Among the quarks, the strange quark gives the major theoretical uncertainty to
the decay amplitude. Hence, we will only consider the uncertainty in the
strange quark mass given by $m_s(2.1\,{\rm GeV})=95\pm20$ MeV.
Notice that for the one-loop penguin contribution, the relevant quark mass is the pole mass rather than the current one \cite{Li:2006jb}. Since the penguin loop correction is governed by the ratio of the pole masses squared $s_i\equiv(m_i^{\rm pole}/m_b^{\rm pole})^2$ [see Eqs. (\ref{eq:PK}) and (\ref{eq:P6}) below] and since the pole mass is meaningful only for heavy quarks, we only need to consider the ratio of $c$ and $b$ quark pole masses given by $s_c=(0.3)^2$.

 The strong
coupling constants employed in the present work are \be \label{eq:alphas}
 \alpha_s(4.2\,{\rm GeV})=0.221,\quad\alpha_s(2.1\,{\rm GeV})=0.293,
 \quad\alpha_s(1.45\,{\rm GeV})=0.359,\quad \alpha_s(1\,{\rm GeV})=0.495\,. \non \\
 \en
For longitudinal
and transverse decay constants of the vector mesons, we use (in
units of MeV)
 \be
 && f_\rho=216\pm3, \qquad f_\omega=187\pm5, \qquad~ f_{K^*}=220\pm5,
 \qquad~~ f_\phi=215\pm5, \non \\
 && f_\rho^\bot=165\pm9, \qquad f_\omega^\bot=151\pm9, \qquad f_{K^*}^\bot=185\pm10, \qquad
 f_\phi^\bot=186\pm9\,,
 \en
where the values of $f_V$ and $f_V^\bot(1{\rm GeV})$ are taken from \cite{BallfV}.

%%%%%%%%%%%%%%%%%%%%%%%%%%%%%%%%%%%%%%%%%%%%%%%%%%%%%%%%%%%%%%%%%%%%%%%%%%%%%%%
\begin{table}[t]
\caption[]{Gegenbauer moments of $\Phi_\perp$ and $\Phi_\parallel$ for $1^3P_1$
and $1^1P_1$ mesons, respectively, where $a_0^{\perp,K_{1A}}$ and
$a_0^{\parallel,K_{1B}}$ are updated from the $B\to K_1 \gamma$ analysis
\cite{Hatanaka:2008xj}, and $a_1^{\parallel,K_{1A}}, a_2^{\perp,K_{1A}},
a_2^{\parallel,K_{1B}}$, and $a_1^{\perp,K_{1B}}$ are then obtained from
Eq.~(141) in \cite{YangNP}. The scale dependence of Gegenbauer moments is
referred to Eq. (\ref{eq:alphas}).} \label{tab:Gegenbauer}
\renewcommand{\arraystretch}{1.8}
\addtolength{\arraycolsep}{0.4pt} {\small
$$
\begin{array}{|c|c|c|c|c|c|c|}\hline
 \mu & a_2^{\parallel, a_1(1260)}& a_2^{\parallel,f_1^{^3P_1}}
 & a_2^{\parallel,f_8^{^3P_1}} & a_2^{\parallel, K_{1A}}
 & \multicolumn{2}{|c|}{a_1^{\parallel, K_{1A}}}
 \\ \hline
\begin{array}{c} {\rm 1~GeV}   \\ {\rm 2.2~GeV} \end{array}&
\begin{array}{c} -0.02\pm 0.02 \\ -0.01\pm 0.01  \end{array}&
\begin{array}{c}  -0.04\pm 0.03 \\ -0.03 \pm 0.02 \end{array}&
\begin{array}{c} -0.07\pm 0.04\\ -0.05 \pm  0.03 \end{array}&
\begin{array}{c} -0.05\pm 0.03\\  -0.04 \pm 0.02 \end{array}&
\multicolumn{2}{c|}{\begin{array}{c} {-0.30^{+0.26}_{-0.00}}\\ {-0.24^{+0.21}_{-0.00}}\end{array}}
\\ \hline\hline
 \mu & a_1^{\perp, a_1(1260)}& a_1^{\perp,f_1^{^3P_1}}
 & a_1^{\perp,f_8^{^3P_1}} & a_1^{\perp, K_{1A}}
 & a_0^{\perp, K_{1A}}    & a_2^{\perp, K_{1A}}
 \\ \hline
\begin{array}{c} {\rm 1~GeV}   \\ {\rm 2.2~GeV} \end{array}&
\begin{array}{c} -1.04\pm 0.34 \\ -0.81\pm 0.26  \end{array}&
\begin{array}{c} -1.06\pm 0.36 \\ -0.82\pm 0.28   \end{array}&
\begin{array}{c} -1.11\pm 0.31 \\ -0.86\pm 0.24   \end{array}&
\begin{array}{c} -1.08\pm 0.48 \\ -0.84\pm 0.37   \end{array}&
\begin{array}{c}  0.26^{+0.03}_{-0.22}\\ 0.24^{+0.03}_{-0.21}  \end{array}&
\begin{array}{c}  0.02\pm 0.21\\   0.01\pm 0.15   \end{array}
\\ \hline\hline
\mu & a_1^{\parallel, b_1(1235)}& a_1^{\parallel,h_1^{^1P_1}}
 & a_1^{\parallel,h_8^{^1P_1}} & a_1^{\parallel, K_{1B}}
 & a_0^{\parallel, K_{1B}}    & a_2^{\parallel, K_{1B}}
 \\ \hline
\begin{array}{c} {\rm 1~GeV}   \\   {\rm 2.2~GeV} \end{array}&
\begin{array}{c} -1.95\pm 0.35 \\  -1.56\pm 0.28  \end{array}&
\begin{array}{c} -2.00\pm 0.35 \\  -1.60\pm 0.28  \end{array}&
\begin{array}{c} -1.95\pm 0.35 \\  -1.56\pm 0.28  \end{array}&
\begin{array}{c} -1.95\pm 0.45 \\  -1.56\pm 0.36  \end{array}&
\begin{array}{c}  -0.15\pm 0.15 \\   -0.15\pm 0.15  \end{array}&
\begin{array}{c}  0.09^{+0.16}_{-0.18} \\    0.06^{+0.11}_{-0.13}  \end{array}
\\ \hline\hline
 \mu & a_2^{\perp, b_1(1235)}& a_2^{\perp,h_1^{^1P_1}}
 & a_2^{\perp,h_8^{^1P_1}} & a_2^{\perp, K_{1B}}
 & \multicolumn{2}{|c|}{a_1^{\perp, K_{1B}}}
 \\ \hline
\begin{array}{c}  {\rm 1~GeV}  \\   {\rm 2.2~GeV} \end{array}&
\begin{array}{c}  0.03\pm 0.19 \\   0.02\pm 0.14  \end{array}&
\begin{array}{c}  0.18\pm 0.22 \\  0.14 \pm 0.17  \end{array}&
\begin{array}{c}   0.14\pm 0.22\\  0.11 \pm  0.17 \end{array}&
\begin{array}{c}  -0.02\pm 0.22\\ -0.02 \pm 0.17  \end{array}&
\multicolumn{2}{c|}{\begin{array}{c} 0.30^{+0.00}_{-0.31}\\
0.25^{+0.00}_{-0.26}\end{array}}
\\ \hline
\end{array}
$$}
\end{table}
%%%%%%%%%%%%%%%%%%%%%%%%%%%%%%%%%%%%%%%%%%%%%%%%%%%%%%%%%%%%%%%%%%%%%%%%%%%%%%%
%

The decay constants $f_{^3P_1}$ for $a_1$, $f_1$, $f_8$ \footnote{Recall that
$f_8$ and $f_1$ are SU(3)-octet and -singlet states.} and
$f^\perp_{^1P_1}(1\,{\rm GeV})$ for $b_1$, $h_1$, $h_8$ obtained from QCD sum
rule methods are listed in \cite{YangNP}. For the decay constants of $K_{1A}$
and $K_{1B}$ we use \be f_{K_{1A}}=250\pm13~{\rm MeV}, \qquad
f_{K_{1B}}=a_0^{\parallel,K_{1B}}f_{K_{1B}}^\bot\approx -28~{\rm MeV},
 \en
where uses of $f_{K_{1B}}^\bot=190\pm 10$ MeV \cite{YangNP} and the value of
$a_0^{\parallel,K_{1B}}$ from Table \ref{tab:Gegenbauer} have been made.
Therefore,
\begin{eqnarray} \label{eq:fK1}
f_{K_{1}(1270)} &=& -184  \pm 11~ {\rm MeV} \,, \quad
f_{K_{1}(1400)} = 177 \pm 12~ {\rm MeV} \,,
 \qquad {\rm for\ }\theta_{K_1}=-37^\circ, \non \\
f_{K_{1}(1270)} &=& -234 \pm 15~ {\rm MeV}, \quad f_{K_{1}(1400)} =
100 \pm 12~ {\rm MeV} \,,
 \qquad~ {\rm for\ }\theta_{K_1}=-58^\circ.
\end{eqnarray}

The Gegenbauer moments $a_i^{(\bot),V}$ and $a_i^{\parallel\, (\perp),A}$ have
been studied using the QCD sum rule method. Here we employ the most
recent updated values evaluated at $\mu=1$ GeV \cite{Ball2007}
 \be
 && a_1^{K^*}=0.03\pm0.02, \quad a_1^{\bot,K^*}=0.04\pm0.03, \quad
 a_2^{K^*}=0.11\pm0.09, \quad
 a_2^{\bot,K^*}=0.10\pm0.08,\non \\
 &&  a_2^{\rho,\omega}=0.15\pm0.07, \quad
 a_2^{\bot,\rho,\omega}=0.14\pm0.06,\quad a_2^\phi=0.18\pm0.08,
 \quad~~ a_2^{\bot,\phi}=0.14\pm0.07.
 \en
Note that $a_1^V=0$, $a_1^{\bot,V}$=0 for $V=\rho,\omega,\phi$. The Gegenbauer
moments $a_i^{\parallel\, (\perp),A}$ for axial-vector mesons are summarized in
Table \ref{tab:Gegenbauer}. This table is taken from \cite{YangNP} with some
updates on the Gegenbauer moments $a_{0}^{\perp,K_{1A}}$,
$a_{0}^{\parallel,K_{1B}}$, $a_1^{\parallel,K_{1A}}, a_2^{\perp,K_{1A}},
a_2^{\parallel,K_{1B}}$, $a_1^{\perp,K_{1B}}$,  $a_1^{\perp,a_1}$ and
$a_1^{\perp,f_1^{^3P_1}}$. As stressed before, the values of the $G$-parity
violating Gegenbauer moments (e.g. $a_1^K, a_1^{\parallel,K_{1A}},
a_{0,2}^{\perp,K_{1A}}, a_1^{\perp,K_{1B}}$ and $a_{0,2}^{\parallel,K_{1B}})$
are displayed for the mesons containing a strange quark. Their signs are
flipped for the mesons containing a $\bar s$ quark. In general,
$\Phi_K(1-x)=\Phi_{\bar K}(x)$.

For the $B$ meson, we shall use $\lambda_B ({\rm 1~GeV}) =(250\pm 100)$~MeV for
its wave function and $f_{B}=210\pm 20$ MeV for its decay constant.

The Wilson coefficients $c_i(\mu)$ at various scales, $\mu=4.4$ GeV, 2.1 GeV,
1.45 GeV and 1 GeV are taken from \cite{Groot}. For the renormalization scale
of the decay amplitude, we choose $\mu=m_b(m_b)$. However, as will be discussed
below, the hard spectator and annihilation contributions will be evaluated at
the hard-collinear scale $\mu_h=\sqrt{\mu\Lambda_h}$ with $\Lambda_h\approx 500
$ MeV.

\section{$B\to VA,AA$ decays in QCD factorization}
Within the framework of QCD factorization \cite{BBNS}, the effective
Hamiltonian matrix elements are written in the form
\begin{equation}\label{fac}
   \langle M_1M_2 |{\cal H}_{\rm eff}|\overline B\rangle
  \! =\! \frac{G_F}{\sqrt2}\sum_{p=u,c} \! \lambda_p\,
\!   \langle M_1M_2 |{\cal T_A}^{h,p}\!+\!{\cal
T_B}^{h,p}|\overline B\rangle \,,
\end{equation}
where $\lambda_p\equiv V_{pb}V_{pq}^*$ with $q=d,s$, and the
superscript $h $ denotes the helicity of the final-state meson.  ${\cal
T_A}^{h,p}$ describes contributions from naive factorization, vertex
corrections, penguin contractions and spectator scattering expressed
in terms of the flavor operators $a_i^{p,h}$, while ${\cal T_B}$
contains annihilation topology amplitudes characterized by  the
annihilation operators $b_i^{p,h}$.

The flavor operators $a_i^{p,h}$ are basically the Wilson coefficients
in conjunction with short-distance nonfactorizable corrections such
as vertex corrections and hard spectator interactions. In general,
they have the expressions \cite{BBNS,BN}
 \be \label{eq:ai}
  a_i^{p,h}(M_1M_2) &=&
 \left(c_i+{c_{i\pm1}\over N_c}\right)N_i^h(M_2) \int_0^1 \Phi^{M_2,h}(x)
 dx \nonumber\\
  && + {c_{i\pm1}\over N_c}\,{C_F\alpha_s\over
 4\pi}\Big[V_i^h(M_2)+{4\pi^2\over N_c}H_i^h(M_1M_2)\Big]+P_i^{h,p}(M_2),
 \en
where $i=1,\cdots,10$,  the upper (lower) signs apply when $i$ is
odd (even), $c_i$ are the Wilson coefficients,
$C_F=(N_c^2-1)/(2N_c)$ with $N_c=3$, $M_2$ is the emitted meson
and $M_1$ shares the same spectator quark with the $B$ meson. The
quantities $V_i^h(M_2)$ account for vertex corrections,
$H_i^h(M_1M_2)$ for hard spectator interactions with a hard gluon
exchange between the emitted meson and the spectator quark of the
$B$ meson and $P_i(M_2)$ for penguin contractions.  The LCDA
$\Phi^{M,h}$ in the first term of Eq. (\ref{eq:ai}) is
$\Phi_\parallel^M$ for $h=0$ and $\Phi_\perp^M$ for $h=\pm$. The expression
of the quantities $N_i^h(M_2)$ reads
 \be
 N_i^h(M_2)=\cases{0, & $i=6,8$, \cr
                 1, & else. \cr}
 \en

\vskip 0.2in \noindent {\it \underline{Vertex corrections}} \vskip 0.1in

The vertex corrections are given by
\begin{equation}\label{vertex0}
   V^0_i(M_{2}) = \left\{\,\,
   \begin{array}{ll}
    {\displaystyle \int_0^1\!dx\,\Phi_\parallel^{M_2}(x)\,
     \Big[ 12\ln\frac{m_b}{\mu} - 18 + g(x) \Big]} \,, & \qquad
     (i=\mbox{1--4},9,10) \\[0.4cm]
   {\displaystyle \int_0^1\!dx\,\Phi_\parallel^{M_2}(x)\,
     \Big[ - 12\ln\frac{m_b}{\mu} + 6 - g(1-x) \Big]} \,, & \qquad
     (i=5,7) \\[0.4cm]
   {\displaystyle \int_0^1\!dx\, \Phi_{m_2}(x)\,\Big[ -6 + h(x) \Big]}
    \,, & \qquad (i=6,8)
   \end{array}\right.
\end{equation}
\begin{equation}\label{vertexpm}
   V^\pm_i(M_2) = \left\{\,\,
   \begin{array}{ll}
    {\displaystyle \int_0^1\!dx\, \Phi^{M_2}_\pm(x) \,
     \Big[ 12\ln\frac{m_b}{\mu} - 18 + g_T(x) \Big]} \,, & \
     (i=\mbox{1--4},9,10) \\[0.4cm]
   {\displaystyle \int_0^1\!dx\, \Phi^{M_2}_\mp(x) \,
     \Big[ - 12\ln\frac{m_b}{\mu} + 6 - g_T(1-x) \Big]}\,, & \
     (i=5,7) \\[0.4cm]
  0, & \ (i=6,8)
   \end{array}\right.
\end{equation}
with
 \begin{eqnarray}
 g(x)&=& 3\Bigg( \frac{1-2x}{1-x} \ln x -i\pi\Bigg)\nonumber\\
  && + \Big[ 2{\rm Li}_2(x) -\ln^2x +\frac{2\ln x}{1-x} -(3+2i\pi)\ln x -
 (x \leftrightarrow 1-x)\Big]\,, \nonumber\\
 h(x)&=&  2{\rm Li}_2(x) -\ln^2x -(1+2i\pi)\ln x -
 (x \leftrightarrow 1-x)\,, \nonumber\\
 g_T(x)&=& g(x) + \frac{\ln x}{\bar x}\,,
 \end{eqnarray}
where $\bar x=1-x$, $\Phi^M_\parallel$ is a twist-2 light-cone distribution
amplitude of the meson $M$, $\Phi_m$ (for the longitudinal
component), and $\Phi_\pm$ (for transverse components) are twist-3
ones. Specifically, $\Phi_m=\Phi_v$ for $M=V$ and $\Phi_m=\Phi_a$
for $M=A$.

\vskip 0.2in \noindent{\it \underline{Hard spectator terms}} \vskip 0.1in

$H^h_i(M_1 M_2)$ arise from  hard spectator interactions with a hard
gluon exchange between the emitted  meson and the spectator quark of
the $\overline B$ meson. $H^0_i(M_1 M_2)$ have the expressions:
\begin{eqnarray}\label{eq:sepc01}
  H^0_i(M_1 M_2)= {if_B f_{M_1} f_{M_2} \over X^{(\overline{B} M_1,
  M_2)}_0}\,{m_B\over\lambda_B} \int^1_0 d u d v \,
 \Bigg( \frac{\Phi^{M_1}_\parallel(u) \Phi^{M_2}_\parallel(v)}{\bar u \bar v}
 \pm r_\chi^{M_1}
  \frac{\Phi_{m_1} (u) \Phi^{M_2}_\parallel(v)}{\bar u v}\Bigg),
 \hspace{0.5cm}
 \end{eqnarray}
for $i=1-4,9,10$,
\begin{eqnarray}\label{eq:spec02}
  H^0_i(M_1 M_2)= -{if_B f_{M_1} f_{M_2} \over X^{(\overline{B} M_1, M_2)}_0}
  \,{m_B\over\lambda_B}\int^1_0 d u d v \,
 \Bigg( \frac{\Phi^{M_1}_\parallel(u) \Phi^{M_2}_\parallel(v)}{\bar u  v}
 \pm r_\chi^{M_1}
  \frac{\Phi_{m_1} (u) \Phi^{M_2}_\parallel(v)}{\bar u \bar v}\Bigg),
 \hspace{0.5cm}
 \end{eqnarray}
for $i=5,7$, and $H^0_i(M_1 M_2)=0$ for $i=6,8$, where the upper (lower) signs
apply when $M_1=V$ ($M_1=A$). The transverse hard spectator terms $H^\pm_i(M_1
M_2)$ read
 \begin{eqnarray}
 H^-_i(M_1 M_2) &=&  \eta^-(M_1M_2) {2i f_B
 f^{\perp}_{M_1} f_{M_2} m_{M_2} \over m_B X^{({\overline B} {M_1}, M_2)}_-}
 \,{m_B\over\lambda_B}\int^1_0 dudv\,
 {\Phi^{M_{1}}_\perp(u)\Phi_-^{M_2}(v)\over \bar u^2 v}, \label{eq:H1m} \\
 H^+_i(M_1 M_2) &=& \eta^+(M_1M_2)
 \frac{2i f_B f_{M_1} f_{M_2} m_{M_1} m_{M_2}}{m_B^2
  X^{({\overline B} M_1, M_2)}_+}\,{m_B\over\lambda_B}\int^1_0 dudv\, {(\bar u-v)
  \Phi_+^{M_1}(u)\Phi_+^{M_2}(v)\over \bar u^2\bar v^2}, \label{eq:H1p}
  \en
 for  $i=1-4,9,10$,
and
 \begin{eqnarray}
 H^-_i(M_1 M_2) &=&  \sigma^-(M_1M_2) {2i f_B
 f^{\perp}_{M_1} f_{M_2} m_{M_2} \over m_B X^{({\overline B} {M_1}, M_2)}_-}
 \,{m_B\over\lambda_B}\int^1_0 dudv\,
 {\Phi^{M_{1}}_\perp(u)\Phi_+^{M_2}(v)\over \bar u^2\bar v},\label{eq:H5m}
  \\
 H^+_i(M_1 M_2) &=& \sigma^+(M_1M_2)
 \frac{2i f_B f_{M_1} f_{M_2} m_{M_1} m_{M_2}}{m_B^2
  X^{({\overline B} M_1, M_2)}_+}\,{m_B\over\lambda_B}\int^1_0 dudv\, {(u-v)
  \Phi_+^{M_1}(u)\Phi_-^{M_2}(v)\over \bar u^2 v^2}, \label{eq:H5p}
  \en
for $i=5,7$, and
 \begin{eqnarray} \label{eq:H6}
 H^-_i(M_1 M_2) &=&  \sigma^-(M_1M_2) {i f_B
 f_{M_1} f_{M_2} m_{M_2} \over m_B X^{({\overline B} {M_1}, M_2)}_-}
 {m_Bm_{M_1}\over m_{M_2}^2}\,
 \,{m_B\over\lambda_B}\int^1_0 dudv\,
 {\Phi^{M_{1}}_+(u)\Phi_\perp^{M_2}(v)\over v\bar u\bar v}, \label{eq:H6m} \\
 H^+_i(M_1 M_2) &=& 0,
 \en
for $i=6,8$,
where
 \be
 \eta^-(M_1M_2) &=&\cases{ +1; & for $M_1M_2=VV,VA$, \cr
                        -1; & for $M_1M_2=AV,AA$, \cr
                        } \non \\
  \sigma^-(M_1M_2)&=& \cases{ +1; & for $M_1M_2=VA,AV$, \cr
                        -1; & for $M_1M_2=VV,AA$,} \non \\
  \sigma^+(M_1M_2) &=& \cases{ +1; & for $M_1M_2=VA,AA$, \cr
                        -1; & for $M_1M_2=VV,AV$,}
 \en
and $\eta^+(M_1M_2)=-1$. To write down Eq. (\ref{eq:H6}), we have factored out the
$r_\chi^{M_2}$ term so that  $a_6^\pm$ will
contribute to the decay amplitude in the product of
$r_\chi^{M_2}a_6^\pm\propto  r_\chi^{M_2}H_6^\pm$. Two remarks are
in order: (i) We have checked explicitly that the hard spectator
terms depend on the $B$ meson wave function $\Phi^B_1(\rho)$, but
not on $\Phi_2^B(\rho)$. (ii) Since Beneke {\it et al.}
\cite{BenekeVV} adopted the Jacob convention for transverse polarization
states, they have $\epsilon_1^*\cdot\!\epsilon_2^*=1$. As a
consequence, their expressions for the parameters $\eta^\pm$,
$\sigma^\pm$ and the decay amplitudes $X_\pm^{(\ov BV_1,V_2)}$ defined
below have signs opposite to ours. Nevertheless, the expressions for
$H_i^\pm(M_1M_2)$ are independent of the choice for transverse
polarization vectors.

The helicity dependent factorizable amplitudes defined by
 \be \label{eq:Xamplitude}
 X^{(\ov BM_1,M_2)}=\la M_2(p_2,\epsilon_2^*)|J_\mu|0\ra\la M_1(p_1,\epsilon_1^*)|J^\mu|B\ra
 \en
have the expressions
 \be \label{eq:Xh}
 X_0^{(\ov BV_1,M_2)} &=& {if_{M_2}\over 2m_{V_1}}\left[
 (m_B^2-m_{V_1}^2-m_{M_2}^2)(m_B+m_{V_1})A_1^{BV_1}(q^2)-{4m_B^2p_c^2\over
 m_B+m_{V_1}}A_2^{BV_1}(q^2)\right], \non \\
 X_0^{(\ov BA_1,M_2)} &=& {if_{M_2}\over 2m_{A_1}}\left[
 (m_B^2-m_{A_1}^2-m_{M_2}^2)(m_B-m_{A_1})V_1^{BA_1}(q^2)-{4m_B^2p_c^2\over
 m_B-m_{A_1}}V_2^{BA_1}(q^2)\right], \non \\
 X_\pm^{(\ov BV_1,M_2)} &=& -if_{M_2}m_Bm_{M_2}\left[
 \left(1+{m_{V_1}\over m_B}\right)A_1^{BV_1}(q^2)\mp{2p_c\over
 m_B+m_{V_1}}V^{BV_1}(q^2)\right], \non \\
 X_\pm^{(\ov BA_1,M_2)} &=& -if_{M_2}m_Bm_{M_2}\left[
 \left(1-{m_{A_1}\over m_B}\right)V_1^{BA_1}(q^2)\mp{2p_c\over
 m_B-m_{A_1}}A^{BA_1}(q^2)\right],
 \en
where $M_2=V_2,A_2$ and $p_c$ is the c.m. momentum.

\vskip 0.2in \noindent{\it \underline{Penguin terms}} \vskip 0.1in

At order $\alpha_s$, corrections from penguin contractions are
present only for $i=4,6$. For $i=4$ we obtain
\begin{eqnarray}\label{eq:PK}
   P_4^{h,p}(M_2) &=& \frac{C_F\alpha_s}{4\pi N_c}\,\Bigg\{
    c_1  \bigg[ G^h_{M_2}(s_p)+g_{M_2}\bigg] \!
    + c_3 \!\bigg[  G^h_{M_2}(s_s) + G^h_{M_2}(1) +2g_{M_2}\bigg] \non \\ && + (c_4+c_6)\!
    \sum_{i=u}^b  \left[G^h_{M_2}(s_i)+g'_{M_2}\right]
%    && + \frac{3}{2}(c_8+c_{10})\!
%    \sum_{i=u}^b  e_i G^h_{M_2}(s_i)
%    + \frac{3}{2}c_9 [ e_{q'} G_{M_2}^h (s_{q'})- \frac{1}{3} %G_{M_2}^h (s_{b})]
     -2 c_{8g}^{\rm eff} G^h_g \Bigg\} \,,
\end{eqnarray}
where $s_i=m_i^2/m_b^2$ and the function $G^h_{M_2}(s)$ is given by
\begin{eqnarray}  \label{eq:GK}
 G^h_{M_2}(s) &=&
 4\int^1_0 du\,\Phi^{M_2,h}(u)\int^1_0 dx\,x\bar x
  \ln[s-\bar u x\bar x-i\epsilon], \nonumber \\
  g_{M_2} &=& \left( {4\over 3}\ln \frac{m_b}{\mu}+{2\over 3}\right)\int^1_0\Phi^{M_2,h}(x)dx, \non \\
   g'_{M_2} &=& {4\over 3}\ln \frac{m_b}{\mu}\int^1_0\Phi^{M_2,h}(x)dx,
 \end{eqnarray}
with $\Phi^{M_2,0}=\Phi^{M_2}_\|$, $\Phi^{M_2,\pm}=\Phi^{M_2}_\pm$.
For $i=6$, the result for the penguin contribution is
\begin{eqnarray} \label{eq:P6}
   P_6^{h,p}(M_2)=\frac{C_F\alpha_s}{4\pi N_c}\,\Bigg\{ \!
    c_1 \hat G^h_{M_2}(s_p)
    + c_3\bigg[ \hat G^h_{M_2}(s_s) + \hat G^h_{M_2}(1)
    \bigg]
    + (c_4+c_6)\sum_{i=u}^b  \hat G^h_{M_2}(s_i) \! \Bigg\}.
    \hspace{0.5cm}
\end{eqnarray}
In analogy with (\ref{eq:GK}), the function $\hat G_{M_2}(s)$ is
defined as
 \begin{eqnarray} \label{eq:P6.1}
 \hat G^0_{M_2}(s) &=&  4\int^1_0 du\, \Phi_{m_2}(u) \int^1_0 dx\,x\bar x
  \ln[s-\bar u x\bar x-i\epsilon], \nonumber \\
 \hat G^\pm_{M_2}(s) &=& 0\,.
 \end{eqnarray}
Therefore, the transverse penguin contractions vanish for $i=6,8$:
$P_{6,8}^{\pm,p}=0$. Note that we have factored out the $r_\chi^{M_2}$ term in
Eq. (\ref{eq:P6}) so that when the vertex correction $V_{6,8}$ is neglected, $a_6^0$ will
contribute to the decay amplitude in the product $r_\chi^{M_2}a_6^0\approx r_\chi^{M_2}P_6^0$.

For $i=8,10$ we find
\begin{equation}
   P_8^{h,p}(M_2) =  \frac{\alpha_{\rm em}}{9\pi N_c}\,(c_1+N_c c_2)\,
   \hat G^h_{M_2}(s_p) \,,
\end{equation}
\begin{equation}\label{PKEW}
   P_{10}^{h,p}(M_2) = \frac{\alpha_{\rm em}}{9\pi N_c} \Bigg\{
   (c_1+N_c c_2) \bigg[ G^h_{M_2}(s_p)+2g_{M_2}\bigg]
   - 3c^{\rm eff}_\gamma  G_g^h \bigg\}.
\end{equation}
For $i=7,9$,
 \be \label{radphoton}
  P_{7,9}^{-,p}(M_2)
 =  -{\alpha_{\rm em}\over 3\pi}C_{7\gamma}^{\rm eff}{m_Bm_b\over m_{M_2}^2}
  +{2\alpha_{\rm em}\over 27\pi}(c_1+N_c c_2)
   \left[\delta_{pc}\ln{m_c^2\over\mu^2}+\delta_{pu}\ln{\nu^2\over \mu^2}+1\right],
 \en
if $M_2=\rho^0, \omega, \phi$, otherwise $P_{7,9}^{-,p}(M_2)=0$. Here the first term is an electromagnetic penguin contribution to the transverse helicity amplitude enhanced by a factor of $m_Bm_b/m_{M_2}^2$, as first pointed out in  \cite{BenekeEWP}. Note that the quark loop contains an ultraviolet divergence for both transverse and longitudinal components
which must be subtracted in accordance with the scheme used to define the Wilson coefficients. The scale and scheme dependence after subtraction is required to cancel the scale and scheme dependence of the electroweak
penguin coefficients. Therefore, the scale $\mu$ in the above equation is the same as the one appearing in the expressions for the penguin corrections, e.g. Eq. (\ref{eq:GK}). On the other hand, the scale $\nu$ is referred  to the scale of the decay constant $f_{M_2} (\nu)$ as the operator $\bar q\gamma^\mu q$ has a non-vanishing anomalous dimension in the presence of electromagnetic interactions \cite{BN}. The $\nu$ dependence of Eq.~(\ref{radphoton}) is compensated by that of $f_{M_2} (\nu)$.

The relevant integrals for the dipole operators $O_{g,\gamma}$ are
  \begin{eqnarray}
&&  G^0_g = \int^1_0 du\,{\Phi^{M_2}_\|(u)\over \bar u}\,, \nonumber\\
&&  G^\pm_g =\int^1_0 {du\over \bar u}\,\Bigg[  \frac{1}{2}
 \bigg(\bar u \Phi_-^{M_2} (u) - u\Phi_+^{M_2}(u) \bigg) -\bar u \Phi_\pm^{M_2} (u) +
 \frac{1}{2}
 \bigg(\bar u \Phi_-^{M_2}(u) + u\Phi_+^{M_2}(u) \bigg) \Bigg].
 \label{eq:cg}
 \end{eqnarray}
Using Eq.~(\ref{eq:a-ww}), $G_g^\pm$ can be further reduced to
\begin{eqnarray}
G_g^+ &=& \int_0^1 du \Big[ \Phi_-^{M_2}(u) -\Phi_+^{M_2}(u) \Big] =0, \nonumber\\
G_g^- &=& 0.
\end{eqnarray}
Hence,  $G_g^{\pm}$ in Eq.~(\ref{eq:cg}) are actually equal to zero.
It was first pointed out by Kagan \cite{Kagan} that the dipole
operators $Q_{8g}$ and $Q_{7\gamma}$ do not contribute to the
transverse penguin amplitudes at ${\cal O}(\alpha_s)$ due to angular
momentum conservation.

\vskip 0.2in \noindent{\it \underline{Annihilation topologies}} \vskip 0.1in
 The weak annihilation contributions to the decay  $\overline B\to
M_{1}M_2$ can be described in terms of the building blocks $b_i^{p,h}$ and $b_{i,{\rm EW}}^{p,h}$
\begin{eqnarray}\label{eq:h1ksann}
\frac{G_F}{\sqrt2} \sum_{p=u,c} \! \lambda_p\, \!\langle M_{1}M_2
|{\cal T_B}^{h,p} |\overline B^0\rangle &=&
i\frac{G_F}{\sqrt{2}}\sum_{p=u,c} \lambda_p
 f_B f_{M_1} f_{M_{2}}\sum_i (d_ib_i^{p,h}+d'_ib_{i,{\rm EW}}^{p,h}).
\end{eqnarray}
The building blocks have the expressions
 \be \label{eq:bi}
 b_1 &=& {C_F\over N_c^2}c_1A_1^i, \qquad\quad b_3={C_F\over
 N_c^2}\left[c_3A_1^i+c_5(A_3^i+A_3^f)+N_cc_6A_3^f\right], \non \\
 b_2 &=& {C_F\over N_c^2}c_2A_1^i, \qquad\quad b_4={C_F\over
 N_c^2}\left[c_4A_1^i+c_6A_2^f\right], \non \\
 b_{\rm 3,EW} &=& {C_F\over
 N_c^2}\left[c_9A_1^{i}+c_7(A_3^{i}+A_3^{f})+N_cc_8A_3^{i}\right],
 \non \\
 b_{\rm 4,EW} &=& {C_F\over
 N_c^2}\left[c_{10}A_1^{i}+c_8A_2^{i}\right],
 \en
where for simplicity we have omitted the superscripts $p$ and $h$ in above
expressions.  The subscripts 1,2,3 of $A_n^{i,f}$ denote the annihilation
amplitudes induced from $(V-A)(V-A)$, $(V-A)(V+A)$ and $(S-P)(S+P)$ operators,
respectively, and the superscripts $i$ and $f$ refer to gluon emission from the
initial and final-state quarks, respectively. Following \cite{BN} we choose the
convention that $M_1$  contains an antiquark from the weak vertex and $M_2$
contains a quark from the weak vertex. The explicit expressions of weak
annihilation amplitudes are:
\begin{eqnarray}
A_1^{i,\,0}(M_{1} M_{2}) &=&
 \pi\alpha_s \int_0^1\! du \,dv\, \Bigg\{
 C^{M_1 M_2}\,\Phi_\parallel^{M_1}(v) \Phi_\parallel^{M_2}(v)\,
 \left[\frac{1}{u(1-\bar u v)}+\frac{1}{u\bar v^2}\right]
 \nonumber\\
 && \ \ \ \ \ \  \ \ \  \ \ \ \ \ \ \ \ \
 -
 D^{M_1 M_2}\, r_\chi^{M_1}r_\chi^{M_2} \,\Phi_{m_1}(u)\, \Phi_{m_2}(v)
\frac{2}{u\bar{v}} \Bigg\}\,, \label{eq:A1i0} \\
 A_1^{i,\,-}(M_1 M_2)  &=& -
 \pi\alpha_s {2m_{M_1}m_{M_2}\over m_B^2}\int_0^1\! du \,dv\,
 \Bigg\{
 \,\Phi^{M_1}_-(u)\,\Phi_-^{M_2} (v)\left[  \frac{\bar u+\bar v}{u^2 \bar{v}^2}
 +{1\over (1-\bar u v)^2}\right]
 \Bigg\}\,,  \\
  A_1^{i,\,+}(M_1 M_2)  &=& -
 \pi\alpha_s {2m_{M_1}m_{M_2}\over m_B^2}\int_0^1\! du \,dv\,
 \Bigg\{
 \,\Phi^{M_1}_+(u)\,\Phi_+^{M_2} (v)\left[ {2\over u\bar v^3}- \frac{v}{(1-\bar uv)^2}
 -{v\over \bar v^2(1-\bar u v)}\right]
 \Bigg\}\,,\hspace{1cm}\label{eq:A1ip}
 \end{eqnarray}

\begin{eqnarray}
A_2^{i,\,0}(M_{1} M_{2}) &=&
 \pi\alpha_s \int_0^1\! du \,dv\, \Bigg\{
 C^{M_1 M_2}\,\Phi_\parallel^{M_1}(v) \Phi_\parallel^{M_2}(v)\,
 \left[\frac{1}{\bar v(1-\bar u v)}+\frac{1}{u^2\bar v}\right]
 \nonumber\\
 && \ \ \ \ \ \  \ \ \  \ \ \ \ \ \ \ \ \
 -
 D^{M_1 M_2}\, r_\chi^{M_1}r_\chi^{M_2} \,\Phi_{m_1}(u)\, \Phi_{m_2}(v)
\frac{2}{u\bar{v}} \Bigg\}\,, \label{eq:A2i0} \\
 A_2^{i,\,-}(M_1 M_2)  &=& -
 \pi\alpha_s {2m_{M_1}m_{M_2}\over m_B^2}\int_0^1\! du \,dv\,
 \Bigg\{ C^{M_1 M_2}\,\Phi^{M_1}_+(u)\,\Phi_+^{M_2} (v)
 \non \\
 && \times \left[  \frac{u+v}{u^2 \bar{v}^2}
 +{1\over (1-\bar u v)^2}\right]
 \Bigg\}\,,  \\
   A_2^{i,\,+}(M_1 M_2)  &=& -
 \pi\alpha_s {2m_{M_1}m_{M_2}\over m_B^2}\int_0^1\! du \,dv\,
 \Bigg\{C^{M_1 M_2} \,\Phi^{M_1}_-(u)\,\Phi_-^{M_2} (v)
 \non \\
 && \times\left[ {2\over u^3\bar v}
   - \frac{\bar u}{(1-\bar uv)^2}
 -{\bar u\over u^2(1-\bar u v)}\right]
 \Bigg\}\,, \hspace{1cm}\label{eq:A2ip}
 \end{eqnarray}

\begin{eqnarray}
A_3^{i,\,0}(M_{1} M_{2}) &=&
 \pi\alpha_s \int_0^1\! du \,dv\, \Bigg\{
 C^{M_1 M_2}\,r_\chi^{M_1}\Phi_{m_1}(v) \Phi_\parallel^{M_2}(v)\,
 \frac{2\bar u}{u\bar v(1-\bar u v)}
 \nonumber\\
 && \ \ \ \ \ \  \ \ \  \ \ \ \ \ \ \ \ \
 +
 D^{M_1 M_2}\, r_\chi^{M_2} \,\Phi_\parallel^{M_1}(v)\, \Phi_{m_2}(v)
\frac{2v}{u\bar{v}(1-\bar uv)} \Bigg\}\,, \label{eq:A3i0} \\
 A_3^{i,\,-}(M_1 M_2)  &=& -
 \pi\alpha_s \int_0^1\! du \,dv\,
 \Bigg\{   -C^{M_1 M_2}{m_{M_2}\over m_{M_1}} r_\chi^{M_1}
 \,\Phi^{M_1}_\perp(u)\,\Phi_-^{M_2} (v)\frac{2}{u\bar v(1-\bar uv)}\non \\
 && \ \ \ \ \ \ \ \ \ \ \ \ \ \ \ \ \ \ \ \ \ \
 + D^{M_1 M_2}{m_{M_1}\over m_{M_2}} r_\chi^{M_2}
 \,\Phi^{M_1}_+(u)\,\Phi_\perp^{M_2} (v)\frac{2}{u\bar v(1-\bar
 uv)}
 \Bigg\}\,, \hspace{1cm}\label{eq:A3im}
 \end{eqnarray}

\begin{eqnarray}
A_3^{f,\,0}(M_{1} M_{2}) &=&
 \pi\alpha_s \int_0^1\! du \,dv\, \Bigg\{
 C^{M_1 M_2}\, r_\chi^{M_1} \,\Phi_{m_1}(u) \Phi_\parallel^{M_2}(v)\,
 \frac{2(1+\bar v)}{u\bar{v}^2}
 \nonumber\\
 && \ \ \ \ \ \  \ \ \  \ \ \ \ \ \ \ \ \
 -
 D^{M_1 M_2}\, r_\chi^{M_2} \,\Phi_\parallel^{M_1}(u)\, \Phi_{m_2}(v)
\frac{2(1+u)}{u^2\bar{v}} \Bigg\}\,, \label{eq:A3f0} \\
 A_3^{f,\,-}(M_1 M_2)  &=& -
 \pi\alpha_s \int_0^1\! du \,dv\, \Bigg\{
  C^{M_1 M_2}\,
 \frac{m_{M_2}}{m_{M_1}} r_\chi^{M_1} \,\Phi^{M_1}_{\perp}(u)\,\Phi_-^{M_2} (v)
 \frac{2}{u^2 \bar{v}}
 \nonumber\\
 && \ \ \ \ \ \  \ \ \  \ \ \ \ \ \ \ \ \ \ \
  +D^{M_1 M_2}\, \frac{m_{M_1}}{m_{M_2}} r_\chi^{M_2}\,\Phi_+^{M_1}(u)\, \Phi^{M_2}_{\perp}(v) \,
 \frac{2}{u\bar v^2 }
 \Bigg\}\,, \hspace{1cm}\label{eq:A3fm}
 \end{eqnarray}
and $A_1^{f,h}=A_2^{f,h}=A_3^{i,+}=A_3^{f,+}=0$, where
 \be \label{eq:C&D}
 && A_1^{i,0},A_3^{i,0}: \qquad\qquad D^{VA}=D^{AV}=-1, \non \\
 && A_2^{i,0},A_2^{i,\pm}: \qquad\qquad  C^{VA}=C^{AV}=-1,  \non \\
 && A_3^{i,-},A_3^{f,0},A_3^{f,-}: \quad ~C^{AV}=C^{AA}=-1, \quad D^{VA}=D^{AV}=-1,
 \en
and the parameters $C$ and $D$ are equal to $+1$ for all other cases. Note that
our results for $A_n^{i(f),\pm}$ have opposite signs to that in \cite{BenekeVV}
as Beneke {\it et al.} adopted the Jacob convention for the transverse
polarization vectors.  We employ the same   convention as in \cite{BN} that
$M_1$ contains an antiquark from the weak vertex with longitudinal fraction
$\bar y$, while $M_2$ contains a quark from the weak vertex with momentum
fraction $x$.

Since the annihilation contributions $A_{1,2}^{i,\pm}$ are
suppressed by a factor of $m_1m_2/m_B^2$ relative to other terms,
in numerical analysis we will consider only the annihilation
contributions due to $A_3^{f,0}$, $A_3^{f,-}$, $A_{1,2,3}^{i,0}$ and $A_3^{i,-}$.

Finally, two remarks are in order: (i) Although the parameters $a_i(i\neq 6,8)$
and $a_{6,8}r_\chi$ are formally renormalization scale and $\gamma_5$ scheme
independent, in practice there exists some residual scale dependence in
$a_i(\mu)$ to finite order. To be specific, we shall evaluate the vertex
corrections to the decay amplitude at the scale $\mu=m_b$.  (The issue with the
renormalization scale $\mu$ will be discussed in more detail in Sec. IV). In
contrast, as stressed in \cite{BBNS}, the hard spectator and annihilation
contributions should be evaluated at the hard-collinear scale
$\mu_h=\sqrt{\mu\Lambda_h}$ with $\Lambda_h\approx 500 $ MeV. (ii) Power
corrections in QCDF always involve troublesome endpoint divergences. For
example, the annihilation amplitude has endpoint divergences even at twist-2
level and the hard spectator scattering diagram at twist-3 order is power
suppressed and posses soft and collinear divergences arising from the soft
spectator quark. Since the treatment of endpoint divergences is model
dependent, subleading power corrections generally can be studied only in a
phenomenological way. We shall follow \cite{BBNS} to model the endpoint
divergence $X\equiv\int^1_0 dx/\bar x$ in the annihilation and hard spectator
scattering diagrams as
 \be \label{eq:XA}
 X_A=\ln\left({m_B\over \Lambda_h}\right)(1+\rho_A e^{i\phi_A}), \qquad
 X_H=\ln\left({m_B\over \Lambda_h}\right)(1+\rho_H e^{i\phi_H}),
 \en
with the unknown real parameters $\rho_{A,H}$ and $\phi_{A,H}$. For simplicity,
we shall assume that $X_A^h$ and $X_H^h$ are helicity independent; that is,
$X_A^-=X_A^+=X_A^0$ and $X_H^-=X_H^+=X_H^0$.

\section{Numerical results}

The decay amplitude of  $B\to M_1M_2$ with $M=V,A$ has the general expression
of $\varepsilon^{*\mu}_{M_1}(\lambda_{M_1})\varepsilon^{*\nu}_{M_2}
(\lambda_{M_2})M_{\mu\nu}$ with $\lambda_{M_1,M_2}$ being the corresponding
helicities. Hence, the decay amplitude can be decomposed into three components,
one for each helicity of the final state: $\A_0,\A_+,\A_-$. The transverse
amplitudes defined in  the transversity basis are related to the helicity ones
via \be
 \A_{\parallel}=\frac{\A_++\A_-}{\sqrt2}, \qquad
 \A_{\bot}&=&\frac{\A_+-\A_-}{\sqrt2}.
 \label{eq:Atrans}
 \en
The decay rate can be expressed in
terms of these amplitudes as
 \be
 \Gamma=\frac{p_c}{8\pi m_B^2}(|\A_0|^2+|\A_+|^2+|\A_-|^2)
       =\frac{p_c}{8\pi m_B^2}(|\A_L|^2+|\A_{\parallel}|^2+|\A_{\bot}|^2),
 \en
with $p_c$ being the c.m. momentum of the final-state meson.
Polarization fractions are defined as
 \be
 f_\alpha\equiv \frac{\Gamma_\alpha}{\Gamma}
                     =\frac{|\A_\alpha|^2}{|\A_0|^2+|\A_\parallel|^2+|\A_\bot|^2},
 \label{eq:f}
 \en
with $\alpha=L,\parallel,\bot$. The relative phases are
\be
 \phi_\bot={\rm arg}(\A_\bot/\A_0), \qquad \phi_\parallel={\rm arg}(\A_\parallel/\A_0).
 \en
Note that the experimental results of $\phi_\parallel$ and $\phi_\bot$ obtained
by BaBar and Belle are for $B\to \phi K^*$ decays
\cite{BaBar:KVphia,BaBar:KVphib,Belle:KVphia}. According to the convention
given by BaBar and Belle,  $|\A_+|>|\A_-|$ and $\phi_\parallel=\phi_\bot=\pi$
for $B\to \phi K^*$ in the absence of final-state interactions. Since our
calculations are for $\ov B\to \phi\ov K^{*}$ decays, in Eq. (\ref{eq:phi})
below we shall transform BaBar and Belle results from $\phi_\parallel$ to
$\pi-\phi_\parallel$ and $\phi_\bot$ to $-\phi_\bot$ so that
$|\bar\A_+|<|\bar\A_-|$ in $\ov B\to \phi \ov K^{*}$. When strong phases
vanish, $\phi_\parallel=0$, $\phi_\bot=-\pi$ for $\ov B\to \phi \ov K^*$.

\subsection{$B\to VV$ decays}
The branching ratios and polarization fractions of charmless $\ov B\to VV$
decays have been measured for $\rho\rho,~\rho\omega,~\rho K^*,~\phi K^*,\omega
K^*$ and $K^*\bar K^*$ final states. It is naively expected that the helicity
amplitudes $\bar \A_h$ (helicities $h=0,-,+$ ) for $\ov B \to VV$ respect the
hierarchy pattern \be \label{eq:hierarchy} \bar \A_0:\bar \A_-:\bar
\A_+=1:\left({\Lambda_{\rm QCD}\over m_b}\right):\left({\Lambda_{\rm QCD}\over
m_b}\right)^2. \en Hence,  they are dominated by the longitudinal polarization
states and satisfy the scaling law, namely \cite{Kagan}
 \be \label{eq:scaling}
1-f_L={\cal O}\left({m^2_V\over m^2_B}\right), \qquad {f_\bot\over f_\parallel}=1+{\cal
O}\left({m_V\over m_B}\right),
 \en
with $f_L,f_\bot$ and $f_\parallel$ being the longitudinal, perpendicular, and
parallel polarization fractions, respectively. In sharp contrast to the
$\rho\rho$ case, the large fraction of transverse polarization observed in
$B\to K^*\rho$ and $B\to K^*\phi$ decays at $B$ factories (see Table
\ref{tab:VVBr} below) is thus a surprise and poses an interesting challenge for
any theoretical interpretation.  Therefore, in order to obtain a large
transverse polarization in $B\to K^*\rho,K^*\phi$, this scaling law must be
circumvented in one way or another. Various mechanisms  such as sizable
penguin-induced annihilation contributions \cite{Kagan}, final-state
interactions \cite{Colangelo,CCSfsi}, form-factor tuning \cite{HNLi} and new
physics \cite{Yang&Das,YDYangnew,NP-tensor,newphysics} (where only the models
with large scalar or tensor coupling can explain the observation for $f_\perp
\simeq f_\parallel$  \cite{Yang&Das,NP-tensor}) have been proposed for solving
the $B\to VV$ polarization puzzle. It has been shown that  when the data for
$\phi K^*$ and $K \eta^{(\prime)}$ modes are simultaneously taken in into
account, the standard model predictions with weak annihilation corrections can
explain the observation, while the new physics effect due to scalar-type
operators is negligible \cite{H-Y}.

Before proceeding, we would like to make a few remarks on the polarization
anomaly. First, the hierarchy of helicity amplitudes given by Eq.
(\ref{eq:hierarchy}) is valid only for factorizable $W$-emission amplitudes. It
may be violated in the presence of nonfactorizable corrections (e.g. vertex,
penguin and hard spectator scattering contributions) and annihilation
contributions. Indeed, we shall show below that the polarization pattern
(\ref{eq:scaling}) will  get modified when nonfactorizable contributions are
included in QCD factorization. We shall see later that the polarization anomaly
is not so serious as originally believed. Second, it is known that the
predicted rates for the penguin dominated $B\to VP,VV$ decays in QCD
factorization are generally too small by a factor of $2\sim 3$ compared to the
data. It is obvious that in order to have a reliable calculation for
polarization fractions, it is of great importance to first reproduce the decay
rates correctly. Otherwise, the estimation of $f_{L,\parallel,\bot}$ will not
be trustworthy. Hence, our first priority is to have a mechanism resolving the
branching ratio puzzle for the penguin dominated charmless $B\to VV$ decays and
hopefully the same mechanism also unravels the polarization anomaly.

\subsubsection{Tree-dominated decays}
Branching ratios and polarization fractions for tree-dominated $B\to \rho\rho$
and $\rho\omega$ are shown in Table \ref{tab:VVBr}. The theoretical errors
correspond to the uncertainties due to variation of (i) the Gegenbauer moments,
the  decay constants, (ii) the heavy-to-light form factors and the strange
quark mass, and (iii) the wave function of the $B$ meson characterized by the
parameter $\lambda_B$, the power corrections due to weak annihilation and hard
spectator interactions described by the parameters $\rho_{A,H}$, $\phi_{A,H}$,
respectively. To obtain the errors shown in Tables
\ref{tab:BRa1}-\ref{tab:BRK1A}, we first scan randomly the points in the
allowed ranges of the above nine parameters and then add errors in quadrature.
More specifically, the second error in the table is referred  to the
uncertainties caused by the variation of  $\rho_{A,H}$ and $\phi_{A,H}$, where
all other uncertainties are lumped into the first error. Here we consider the
default results for tree-dominated decays by setting the annihilation
parameters to be zero, i.e. $\rho_A=\phi_A=0$, though the predictions are
insensitive to the choice of them.

It is obvious from Table \ref{tab:VVBr} that the longitudinal amplitude
dominates the tree-dominated decays  except for the $\rho^0\omega$ mode where
the transverse polarization could be equally important. The naive expectation
of $f_L\approx 1-4m_\rho^2/m_B^2\approx 0.92$ is experimentally confirmed. The
calculated rates are also in agreement with experiment except that the
predicted rate for $B^-\to \rho^-\omega$  is slightly high. Its decay amplitude
reads \be
 \sqrt2\,{\cal A}_{B^-\to \rho^- \omega}^h
   &\approx&
   \bigg[\delta_{pu}(\alpha_2^h + \beta_2^h)
     + 2\alpha_3^{p,h}+ \alpha_4^{p,h}   + \beta_3^{p,h}
    \bigg] X^{(\overline B \rho, \omega)}_h \nonumber\\
   &+&  \left[\delta_{pu}(\alpha_1^h + \beta_2^h)
    + \alpha_4^{p,h}  + \beta_3^{p,h}
     \right] X^{(\overline B \omega,\rho)}_h.
\en It is obvious that this decay is dominantly governed by $B\to\omega$
transition form factors. The data of $B^-\to \rho^-\omega$ and $B\to K^*\omega$
to be discussed below seem to suggest that $B\to \omega$ form factors are
slightly smaller than what are expected from the light-cone sum rules
\cite{Ball}.

\begin{table}[t]
\caption{$CP$-averaged branching ratios (in units of $10^{-6}$) and
polarization fractions for $\bar B\to
\rho\rho,\rho\omega,K^*\rho,K^*\phi,K^*\omega,K^*\bar K^*$ decays. The
annihilation parameters are specified to be $\rho_A=0.78$ and
$\phi_A=-43^\circ$ for $K^*\rho, K^*\bar K^*$ and $\rho_A=0.65$ and
$\phi_A=-53^\circ$ for $K^*\phi$ and $K^*\omega$ by default.
%The theoretical errors correspond to the uncertainties due to %variation of (i) Gegenbauer moments, decay constants, quark %masses, form factors, the $\lambda_B$ parameter for the $B$ meson %wave function, and (ii) $\rho_{A,H}$, $\phi_{A,H}$, respectively.
For longitudinal polarization fraction, the theoretical
uncertainty is dominated by $\rho_{A,H}$ and $\phi_{A,H}$, and hence only this error is listed in the
table for $f_L$. Experimental results are taken from
\cite{BaBar:KVV,Belle:KVphib,BaBar:KVrhonew,Belle:KVrho,BaBar:KVphia,BaBar:KVphib,Belle:KVphia,BaBar:rhorho,Belle:rhorho,BaBar:rho0rho0,Belle:rho0rho0,BaBar:rhoprhom,Belle:rhoprhom,BaBar:omegarho,BaBar:KVKV,CLEO:KVKV,Belle:KVomega}
and the world averages from \cite{HFAG}.} \label{tab:VVBr}
\begin{ruledtabular}
\begin{tabular}{l c c c c c c}
 &  \multicolumn{2}{c}{$\B$}
 &   \multicolumn{2}{c}{$f_L$}  &   \multicolumn{2}{c}{$f_\bot$}\\ \cline{2-3} \cline{4-5} \cline{6-7}
\raisebox{2.0ex}[0cm][0cm]{Decay} & Theory & Expt & Theory &
Expt & Theory & Expt \\ \hline
$B^-\to\rho^-\rho^0$ & $20.0^{+4.0+2.0}_{-1.9-0.9}$ & $18.2\pm3.0$ & $0.96^{+0.02}_{-0.02}$ & $0.912^{+0.044}_{-0.045}$ & $0.02\pm0.01$ & \\
$\ov B^0\to\rho^+\rho^-$ &  $25.5^{+1.5+2.4}_{-2.6-1.5}$ & $24.2^{+3.1}_{-3.2}$ & $0.92^{+0.01}_{-0.02}$ & $0.978^{+0.025}_{-0.022}$ & $0.04^{+0.01}_{-0.00}$ & \\
$\ov B^0\to\rho^0\rho^0$  & $0.9^{+1.5+1.1}_{-0.4-0.2}$ & $0.68\pm0.27$ & $0.92^{+0.06}_{-0.36}$ & $0.70\pm0.15$ & $0.04^{+0.14}_{-0.03}$ & \\
\hline
$B^-\to\rho^-\omega$ & $19.2^{+3.3+1.7}_{-1.6-1.0}$ & $10.6^{+2.6}_{-2.3}$ & $0.96^{+0.02}_{-0.02}$ & $0.82\pm0.11$ & $0.02\pm0.01$ & \\
$\ov B^0\to\rho^0\omega$ &  $0.1^{+0.1+0.4}_{-0.1-0.0}$ & $<1.5$ & $0.55^{+0.47}_{-0.29}$ & $$ & $0.22^{+0.16}_{-0.23}$ & \\
%$\ov B^0\to\omega\omega$  & $$ & $<4.0$ & $$ & $$ & $$ & \\
\hline
$B^-\to \bar K^{*0}\rho^-$ \footnotemark[1] & $9.2^{+1.2+3.6}_{-1.1-5.4}$ & $9.2\pm1.5$ & $0.48^{+0.52}_{-0.40}$ & $0.48\pm0.08$ & $0.26^{+0.20}_{-0.26}$ & \\
$B^-\to K^{*-}\rho^0$ & $5.5^{+0.6+1.3}_{-0.5-2.5}$ & $<6.1$ & $0.67^{+0.31}_{-0.48}$ & $0.96^{+0.06}_{-0.16}$  \footnotemark[2] & $0.16^{+0.24}_{-0.15}$ & \\
$\ov B^0\to K^{*-}\rho^+$ & $8.9^{+1.1+4.8}_{-1.0-5.5}$ & $<12$ & $0.53^{+0.45}_{-0.32}$ & $$ & $0.24^{+0.16}_{-0.22}$ & \\
$\ov B^0\to \bar K^{*0}\rho^0$ & $4.6^{+0.6+3.5}_{-0.5-3.5}$ & $5.6\pm1.6$ & $0.39^{+0.60}_{-0.31}$ & $0.57\pm0.12$ & $0.30^{+0.15}_{-0.30}$ & \\
\hline
$B^-\to K^{*-}\phi$ \footnotemark[3] & $10.0^{+1.4+12.3}_{-1.3-~6.1}$ & $10.0\pm1.1$ & $0.49^{+0.51}_{-0.42}$ & $0.50\pm0.05$
& $0.25^{+0.21}_{-0.25}$ & $0.20\pm0.05$ \\
$\ov B^0\to \bar K^{*0}\phi$ & $9.5^{+1.3+11.9}_{-1.2-~5.9}$ & $9.5\pm0.8$ & $0.50^{+0.50}_{-0.42}$ & $0.484\pm0.034$ & $0.25^{+0.21}_{-0.25}$ & $0.256\pm0.032$ \\
\hline
$B^-\to K^{*-}\omega$ & $3.5^{+0.4+3.0}_{-0.4-1.7}$ & $<3.4$ & $0.66^{+0.32}_{-0.38}$ & $$
& $0.17^{+0.20}_{-0.17}$ & $$ \\
$\ov B^0\to \bar K^{*0}\omega$ & $3.0^{+0.5+2.9}_{-0.4-1.8}$ & $<2.7$ & $0.57^{+0.44}_{-0.46}$ & $$ & $0.21^{+0.25}_{-0.22}$ & $$ \\
\hline
$B^-\to K^{*0}K^{*-}$ & $0.6^{+0.1+0.3}_{-0.1-0.3}$ & $<71$ & $0.45^{+0.55}_{-0.38}$ & & $0.27^{+0.19}_{-0.27}$ \\
$\ov B^0\to K^{*-}K^{*+}$ & $0.1^{+0.0+0.1}_{-0.0-0.1}$ & $<141$ & $1$ & & $0$ \\
$\ov B^0\to K^{*0}\bar K^{*0}$ & $0.6^{+0.1+0.2}_{-0.1-0.3}$ & $1.28^{+0.37}_{-0.32}$ &  $0.52^{+0.48}_{-0.48}$ & $0.80^{+0.12}_{-0.13}$ & $0.24^{+0.24}_{-0.24}$ \\
\end{tabular}
\footnotetext[1]{This mode is employed as an input for extracting the parameters $\rho_A$ and $\phi_A$ for $B\to K^*\rho$ decays.}
\footnotetext[2]{A recent BaBar measurement gives $f_L(K^{*-}\rho^0)=0.9\pm0.2$ \cite{BaBar:KVrhonew}, but it has only $2.5\sigma$ significance. }
\footnotetext[3]{This mode is employed as an input for extracting the parameters $\rho_A$ and $\phi_A$ for $B\to K^*\phi$ decays.}
\end{ruledtabular}
\end{table}

\subsubsection{Penguin-dominated decays}
The decays of interest in this category are $B\to K^*\rho, K^*\phi,K^*\omega$ and $K^*\bar K^*$.

\vskip 0.1in \noindent{\it \underline{$B\to K^*\rho$}} \vskip 0.1in We first
consider $\bar B\to K^*\rho$ decays. Retaining the leading contributions,
their decay amplitudes are approximated by \be \label{eq:KVrhoamp}
 \A_{B^-\to \ov K^{*0}\rho^-} &\approx& V_c(\alpha_4^{c,h}+\beta_3^h) X_{\rho\bar K^*}^h, \non \\
 \sqrt{2}\A_{B^-\to K^{*-}\rho^0} &\approx& \left[V_u\alpha_1^h+V_c(\alpha_4^{c,h}+\beta_3^h) \right] X_{\rho\bar K^*}^h+\left[V_u\alpha_2^h+V_c{3\over 2}\alpha_{3,{\rm EW}}^h\right]X_{\bar K^*\rho}^h,  \non \\
 \A_{\ov B^0\to K^{*-}\rho^+} &\approx& \left[V_u\alpha_1^h+V_c(\alpha_4^{c,h}+\beta_3^h) \right] X_{\rho\bar K^*}^h, \non \\
 -\sqrt{2}\A_{\ov B^0\to \ov K^{*0}\rho^0} &\approx& V_c(\alpha_4^{c,h}+\beta_3^h) X_{\rho\bar K^*}^h-\left[V_u\alpha_2^h+V_c{3\over 2}\alpha_{3,{\rm EW}}^h\right]X_{\bar K^*\rho}^h,
 \en
where $V_p\equiv V_{pb}V_{ps}^*$ with $|V_c|\gg |V_u|$, $\beta_3$ characterizes
the penguin-induced weak annihilation (see Eq. (\ref{eq:beta}) for definition)
and $X^h_{\rho\bar K^*}$ is a shorthand notation for $X^h_{\bar B\to \rho\bar
K^*}$ with its explicit expression shown in Eq. (\ref{eq:Xh}). The expressions
of the flavor parameters $\alpha_i^{h,p}$ in terms of the coefficients
$a_i^{h,p}$ can be found in Eq. (\ref{eq:alphai}). To proceed, we shall first
neglect annihilation completely by setting $\beta_3=0$. In the absence of NLO
nonfactorizable corrections, the parameters $\alpha_i^{h,p}$ are helicity
independent and hence the hierarchy relation (\ref{eq:hierarchy}) for helicity
amplitudes is respected as $|X_{\bar K^*\rho}^0|:|X_{\bar K^*\rho}^-|:|X_{\bar
K^*\rho}^+|=1:0.26:0.03$ and $|X_{\rho \bar K^*}^0|:|X_{\rho \bar
K^*}^-|:|X_{\rho \bar K^*}^+|=1:0.30:0.005$. When vertex, penguin and hard
spectator corrections are taken into account, we see from Table
\ref{tab:VVparameter} that $\alpha_2^h,\alpha_4^{p,h}$ and $\alpha_{3,{\rm
EM}}^h$ for negative helicity differ significantly from that the longitudinal
ones. For example, the real parts of $\alpha_2^h$ and $\alpha_{3,{\rm EW}}^h$
have opposite signs for $h=0$ and $h=-$ .  Let us consider two extreme cases
for the longitudinal polarization fraction. From Eq. (\ref{eq:KVrhoamp}) we
have \be \label{eq:ampratio} \left.{ \A^-\over \A^0}\right|_{\bar B^0\to \bar
K^{*0}\rho^0} &\approx& \left({\alpha_4^{c,-}-{3\over 2}\alpha_{3,{\rm EW}}^-
\over \alpha_4^{c,0}-{3\over 2}\alpha_{3,{\rm EW}}^0
}\right)\left({X^-_{\bar K^*\rho}\over X^0_{\bar K^*\rho}}\right), \non \\
\left.{ \A^-\over \A^0}\right|_{B^-\to  K^{*-}\rho^0} &\approx&
\left({\alpha_4^{c,-}+{3\over 2} \alpha_{3,{\rm EW}}^- \over
\alpha_4^{c,0}+{3\over 2}\alpha_{3,{\rm EW}}^0 }\right)\left({X^-_{\bar
K^*\rho}\over X^0_{\bar K^*\rho}}\right). \en From Table \ref{tab:VVparameter}
we see that the interference between $\alpha_4^{c,h}$ and $\alpha_{3,{\rm
EW}}^h$ is constructive for $h=-$ and destructive for $h=0$ for the decay $\bar
B^0\to \bar K^{*0}\rho^0$ and the other way around for $B^-\to K^{*-}\rho^0$.
As a consequence, $\A^-$ is comparable to $\A^0$ for the former but is highly
suppressed relative to $\A^0$ for the latter. The longitudinal polarization fraction for the penguin dominated processes can be approximated as
\begin{eqnarray}
f_L(\rho K^*) &\simeq& 1- \frac{|\alpha_4^{c,-}+ c_v\alpha_{3,\rm EW}^{c,-} +\beta_3^-|^2 \left|X^-_{\rho K^*}\right|^2 }
{\sum_{h=0,-} |\alpha_4^{c,h} + c_v\alpha_{3,\rm EW}^{c,-} + \beta_3^h |^2
 \left| X^h_{\rho K^*}\right|^2}, \nonumber\\
f_L(K^* \phi) &\simeq& 1- \frac{| \alpha_3^- + \alpha_4^{c,-}+ \frac{1}{2}\alpha_{3,\rm EW}^{c,-} +\beta_3^-|^2 \left|X^-_{K^*\phi} \right|^2}
{\sum_{h=0,-} |\alpha_3^h + \alpha_4^{c,h} + \frac{1}{2}\alpha_{3,\rm EW}^{c,-} + \beta_3^h |^2 \left| X^h_{K^*\phi}\right|^2} ,
\end{eqnarray}
where $|X^-_{\rho K^*}/X^0_{\rho K^*}|^2 \simeq  (m_{K^*}/m_B)^2 A_0^{B\to \rho}/F_-^{B\to \rho} \propto (m_{K^*}/m_B)^2$, $c_v=0$ for $\overline K^{*0} \rho^-$ and $K^{*-} \rho^+$ modes, $c_v=1$ for $K^{*-} \rho^0$ and $c_v=-1$ for $\overline K^{*0} \rho^0$ (see Ref.~\cite{BenekeVV} for the definitions of the $A_0$ and $F_-$ form factors). The calculated branching ratios
and the longitudinal polarization fractions $f_L$ in QCDF are shown in the case (i) of Table \ref{tab:KVrhoBr}. Indeed, we find $f_L(\bar K^{*0}\rho^0)=0.46$ and $f_L(K^{*-}\rho^0)=0.97$. If the coefficients $a_i^h$ are helicity independent, we will have $f_L(\bar K^{*0}\rho^0)=0.91$ rather than 0.46\,! However, the NLO corrections to $a_i^-$ will render the negative helicity amplitude $\A^-(\bar K^{*0}\rho^0)$ comparable to the longitudinal one $\A^0(\bar K^{*0}\rho^0)$ so that even at the short-distance level, $f_L$ for $\ov B^0\to \bar K^{*0}\rho^0$ can be as low as 50\%. Similar detailed discussions for $\phi K^*$ modes will be  given latter.

\begin{table}[t]
\caption{Longitudinal- and negative-helicity amplitude parameters.} \label{tab:VVparameter}
\begin{center}
\begin{ruledtabular}
\begin{tabular}{l r r | l r r }
Parameter & $h=0$~~~ & ~~~~$h=-$~~~~~~ & Parameter~~~ & $h=0$~~~ & $h=-$~~~
\\ \hline
 $\alpha_1(\rho K^*)$ & $0.96+~0.01i$ & $1.11+~0.03i$ &
 $\alpha_{3,{\rm EW}}(K^*\rho)$ & $-0.009-~0.000i$ & $0.010-~0.000i$  \\
$\alpha_2(K^*\rho)$ & $0.24-~0.08i$ & $-0.16-~0.16i$ & $\alpha_{4,{\rm EW}}(K^*\rho)$ &
$-0.002 +~0.001i$ & $0.001+~0.001i$ \\
$\alpha_4^u(\rho K^*)$ & $-0.022-~0.014i$ & $-0.048-~0.016i$ &
 $\beta_3(\rho K^*)$ & $0.008-~0.018i$ & $-0.031+0.060i$ \\
$\alpha_{4}^c(\rho K^*)$ & $-0.030-~0.010i$ & $-0.047-~0.002i$ &
\\ \hline
$\alpha_3(K^*\phi)$ & $0.005-~0.001i$ & $-0.004-~0.001i$
& $\alpha_{3,{\rm EW}}(K^*\phi)$ & $-0.009-~0.000i$ & $0.002-~0.000i$  \\
$\alpha_4^u(K^*\phi)$ & $-0.022-~0.014i$ & $-0.048-~0.016i$
& $\alpha_4^c(K^*\phi)$ & $-0.030-~0.010i$ & $-0.046-~0.002i$ \\
 $\beta_3(K^*\phi)$ & $0.008-~0.019i$ & $-0.028+0.053i$ \\
\end{tabular}
\end{ruledtabular}
\end{center}
\end{table}

\begin{table}[h]
\caption{$CP$-averaged branching ratios (in units of $10^{-6}$) and the
longitudinal polarization fraction $f_L$ for $B\to K^*\rho$ and $K^*\phi$ decays for three cases: (i) no annihilation contribution, (ii) adding annihilation contributions with $\rho_A=0.78$, $\phi_A=-43^\circ$ for $K^*\rho$ and $\rho_A=0.65$, $\phi_A=-53^\circ$ for $K^*\phi$. The predictions by Beneke,
Rohrer and Yang \cite{BenekeVV} are shown in the last two columns. For simplicity, only the central values are exhibited. The theoretical
uncertainties in case (ii) are shown in Table \ref{tab:VVBr}. Experimental results are taken from \cite{BaBar:KVrhonew,Belle:KVrho,BaBar:KVV}.}
\label{tab:KVrhoBr}
\begin{ruledtabular}
\begin{tabular}{l c c c c c c c c }
 &  \multicolumn{2}{c}{Expt} &  \multicolumn{2}{c}{$(i)$}
 &   \multicolumn{2}{c}{$(ii)$}
  &   \multicolumn{2}{c}{BRY} \\
  \cline{2-3} \cline{4-5} \cline{6-7} \cline{8-9}
\raisebox{2.0ex}[0cm][0cm]{Decay} & $\B$ & $f_L$ & $\B$ & $f_L$ & $\B$ & $f_L$ & $\B$ & $f_L$ \\ \hline
$B^-\to \bar K^{*0}\rho^-$ & $9.2\pm1.5$ & $0.48\pm0.08$ & 4.0 & $0.82$ & 9.2 & $0.48$ & 5.9 & 0.56\\
$B^-\to K^{*-}\rho^0$ & $<6.1$ & $0.96^{+0.06}_{-0.16}$ & 3.8 & $0.97$ & 5.5 & $0.67$ & 4.5 & 0.84 \\
$\ov B^0\to K^{*-}\rho^+$ & $<12$ & $-$ & 3.8 & $0.86$ & 8.9 & $0.53$ & 5.5 & 0.61 \\
$\ov B^0\to \bar K^{*0}\rho^0$ & $5.6\pm1.6$ & $0.57\pm0.12$ & 1.1 & $0.50$ & 4.6 & $0.39$ & 2.4 & 0.22 \\
\hline
$B^-\to K^{*-}\phi$ & $10.0\pm1.1$ & $0.50\pm0.05$ & 4.1 & 0.62 & 10.0 & 0.49& 10.1 & 0.45 \\
$\ov B^0\to \bar K^{*0}\phi$ & $9.5\pm0.8$ & $0.484\pm0.034$ & 3.8 & 0.62 & 9.5 & 0.50 & 9.3 & 0.44\\
\end{tabular}
\end{ruledtabular}
\end{table}

Comparing with the data,  it appears that even though the naive estimate of
$f_L$ is  too large for $\bar K^{*0}\rho^-$, the experimental observation of a
large $f_L$ for $K^{*-}\rho^0$ and a small $f_L$ for $\bar K^{*0}\rho^0$ are
well accommodated. However, as stressed before, in order to have a trustworthy
estimate of polarization fractions one has to reproduce the rates correctly as
the predicted branching fractions for $\bar K^{*0}\rho^-$ and $\bar
K^{*0}\rho^0$ are too small compared to experiment (see Table
\ref{tab:KVrhoBr}). In the present work, we shall follow \cite{Kagan} to
ascribe the necessary enhancement to a potentially large penguin annihilation
characterized by the parameter $\beta_3$. We fit the data of $B^-\to \bar
K^{*0}\rho^-$ by adjusting the parameters $\rho_A$ and $\phi_A$ that
characterize the nonperturbative effects of soft gluon exchanges in
annihilation diagrams. From Fig. \ref{fig:KVrho}(a) we see that $\rho_A$ is
preferred to be around 0.78 in order to fit the rate, while the corresponding
$f_L$ is around 0.48 (see Fig. \ref{fig:KVrho}(c)). Only the theoretical
uncertainty due to the variation of the phase $\phi_A$ is considered in Fig.
\ref{fig:KVrho}. It is clear that the total branching ratio and the
longitudinal one $\B_L$ increase with $\rho_A$, whereas $f_L$ decreases slowly
with $\rho_A$.
To fit the rate and $f_L$ simultaneously for $B^-\to \bar K^{*0}\rho^-$, we
find $\rho_A\approx 0.78$ and $\phi_A\approx -43^\circ$. Using this set of
parameters, we are able to predict branching ratios and polarization fractions
for other $\ov B\to K^*\rho$ decays as exhibited in Table \ref{tab:VVBr} and in
case (ii) of Table \ref{tab:KVrhoBr}.  In the presence of penguin annihilation,
the parameter $\alpha_4^{c,h}$ in Eqs. (\ref{eq:KVrhoamp}) and
(\ref{eq:ampratio}) should be replaced by $\alpha_4^{c,h}+\beta_3^h$. From
Table \ref{tab:VVparameter}, one can check that both $f_L(\bar K^{*0}\rho^0)$ and  $f_L(K^{*-}\rho^0)$ will
be decreased when penguin annihilation
is turned on.

Within the QCDF framework, Beneke, Rohrer and Yang (BRY) have employed the
choice $\rho_A=0.6$ and $\phi_A=-40^\circ$ obtained from a fit to the data of
$K^*\phi$ to study $\bar K^*\rho$ decays \cite{BenekeVV}. They have noticed
that the calculated $K^*\rho$ branching fractions are systematically below the
measurements. This is not a surprise as their $\rho_A$ is smaller than 0.78
[see also Fig. \ref{fig:KVrho}(a)], since as emphasized before, the estimation of polarization fractions
will not be reliable unless the calculated partial rate agrees with experiment and as shown below that $K^*\phi$ and $K^*\rho$ data cannot be fitted
simultaneously by two universal parameters $\rho_A$ and $\phi_A$. This may be a potential problem for QCDF.

\begin{figure}[t]
\vspace{0cm} \centerline{\epsfig{figure=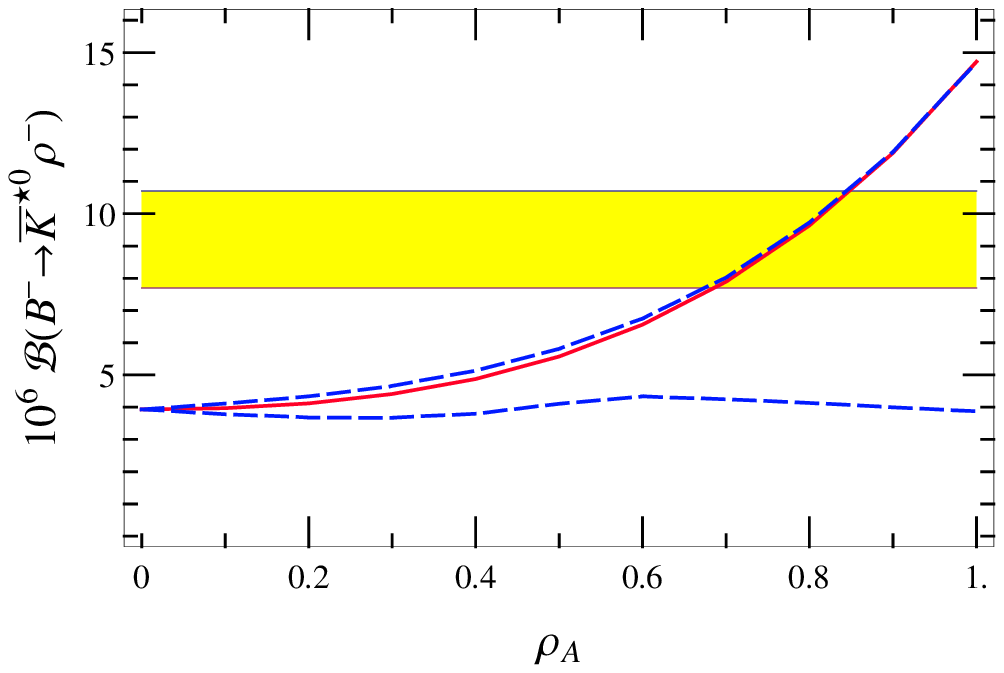,width=4.8cm}
              \hspace{0.7cm}
              \epsfig{figure=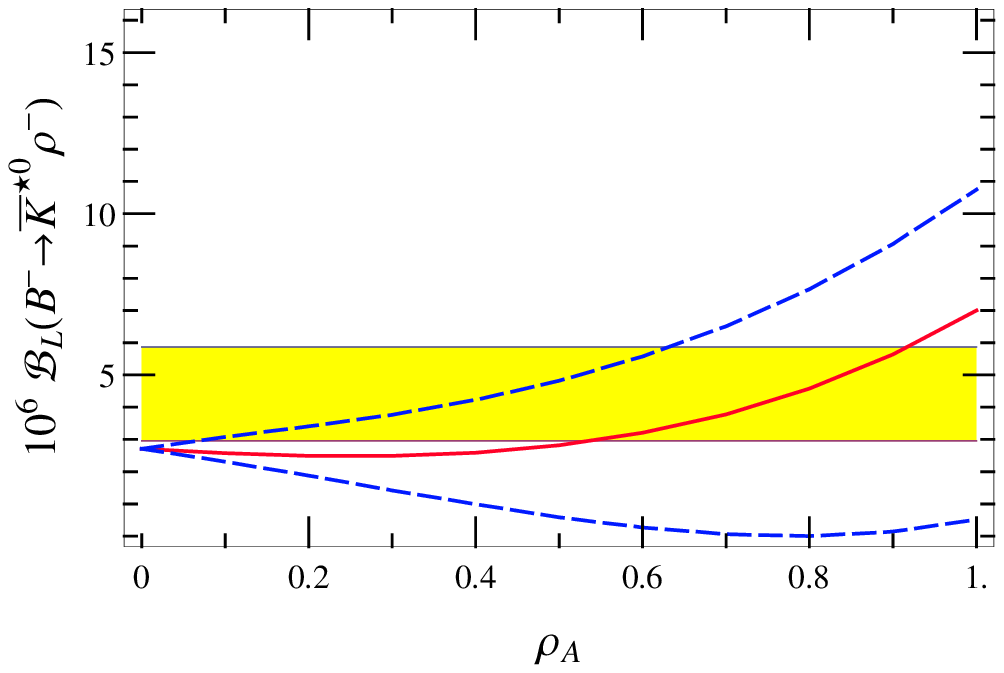,width=4.8cm}
              \hspace{0.7cm}
              \epsfig{figure=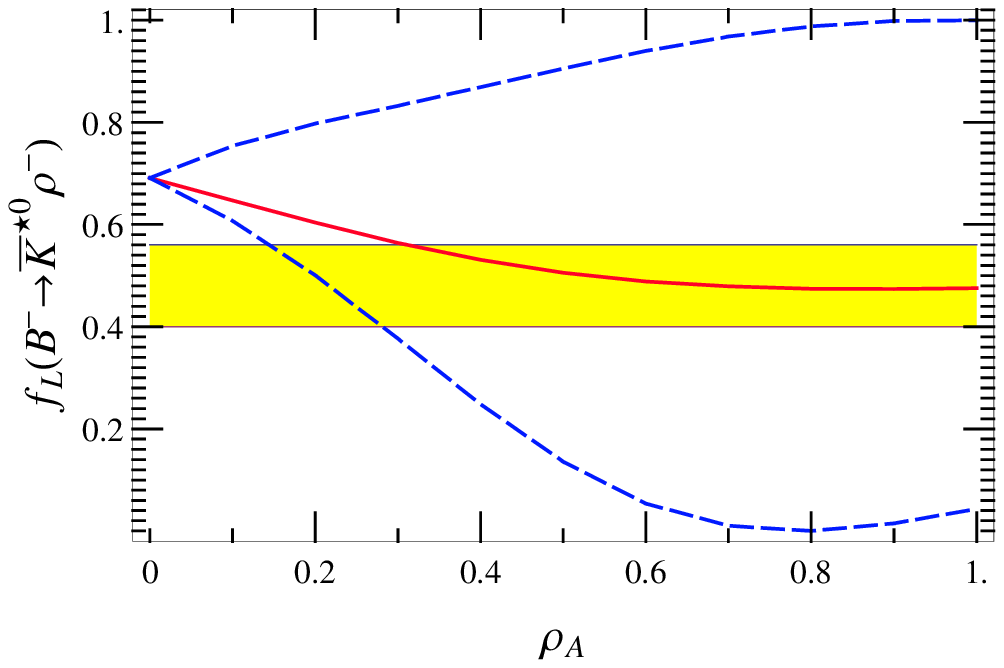,width=4.8cm}
%              \hspace{0.7cm}
%              \epsfig{figure=KVrho_4.eps,width=3.3cm}
              }
\centerline{\hspace{0.2cm}(a)\hspace{5.0cm} (b)\hspace{5.3cm} (c)\hspace{0.2cm}
} \caption[]{\small Predicted branching ratios [(a) and (b)  are for the total
and longitudinal branching ratios, respectively] and (c) the longitudinal
polarization fraction for $B^-\to \bar K^{*0}\rho^-$ as a function of the
parameter $\rho_A$. The solid and dashed curves correspond to the central value
and the allowed theoretical uncertainty due to the variation of $\phi_A$,
respectively. The horizontal band represents experimental values with one sigma
errors. }
    \label{fig:KVrho}
\end{figure}

The large longitudinal polarization fraction of $B^-\to K^{*-}\rho^0$,
$f_L=0.96^{+0.06}_{-0.16}$, measured by BaBar \cite{BaBar:KVV} seems to be
peculiar as a smaller $f_L$ of order 0.5 is observed in other $K^*\rho$ modes
such as $\bar K^{*0}\rho^-$ and $\bar K^{*0}\rho^0$. At first sight, it appears
that the BRY's prediction of $f_L(K^{*-}\rho^0)=0.84^{+0.02+0.16}_{-0.03-0.25}$
can account for the BaBar measurement. However, as we note in Appendix D, there
are sign errors in the expressions of the annihilation terms $A_3^{f,0}$ and
$A_3^{i,0}$ (see Eqs. (\ref{eq:vv-XAA3-1}) and (\ref{eq:A3i0VV}), respectively)
by BRY: The signs of the $r_\chi^{V_2}$ terms in these two equations are
erroneous in \cite{BenekeVV}. Because of the (incorrect) cancelation between
$r_\chi^{V_1}$ and $r_\chi^{V_2}$ terms in $A_3^{f,0}$, BRY claimed (wrongly)
that the longitudinal penguin annihilation amplitude $\beta_3^0$ is strongly
suppressed, while the $\beta_3^-$ term receives sizable penguin annihilation
contribution. If a wrong sign for $r_\chi^{V_2}$ terms is used, both rates and
longitudinal polarization fractions will be reduced, especially for the $\bar
K^{*0}\rho^0$ mode where $f_L$ is reduced by more than a factor of 2 (see the case (iii) of Table \ref{tab:KVrhoBr}). For comparison, BRY's predictions are shown in the last two columns of the same table. Using the correct expressions for $A_3^{f,0}$ and $A_3^{i,0}$, we find that $f_L(K^{*-}\rho^0)$ is reduced to the 70\% level  and $f_L(\bar K^{*0}\rho^0)$ is predicted to be
$0.39^{+0.60}_{-0.31}$. The latter agrees with the experimental value $f_L(\bar
K^{*0}\rho^0)=0.57\pm0.12$ \cite{HFAG}. As explained above, the corresponding
prediction  $0.22^{+0.03+0.53}_{-0.03-0.14}$ by BRY is too small owing to the
incorrect signs in their $A_3^{f,0}$ amplitudes.

In short, we have the pattern (see also \cite{BenekeVV}) \be \label{eq:KVrhofL}
f_L(K^{*-}\rho^0)> f_L(K^{*-}\rho^+)> f_L(\bar K^{*0}\rho^-)> f_L(\bar
K^{*0}\rho^0) \en for the longitudinal fractions in $B\to K^*\rho$ decays. Note
that the quoted experimental value $f_L(K^{*-}\rho^0)=0.96^{+0.06}_{-0.16}$ in
Tables \ref{tab:VVBr} and \ref{tab:KVrhoBr} was obtained by BaBar in a previous
measurement where $K^{*-}\rho^0$ and $K^{*-}f_0(980)$ were not separated
\cite{BaBar:KVV}. This has been overcome in a recent BaBar measurement, but the
resultant value $f_L(K^{*-}\rho^0)=0.9\pm0.2$ has only 2.5$\sigma$ significance
\cite{BaBar:KVrhonew}. At any rate,
 it would be important to have a refined
measurement of longitudinal polarization fraction for $K^{*-}\rho^0$ and $\bar
K^{*0}\rho^0$ and a new measurement of $f_L(K^{*-}\rho^+)$ to test the
hierarchy pattern (\ref{eq:KVrhofL}).

\vskip 0.2in \noindent{\it \underline{$B\to K^*\phi$}} \vskip 0.1in
Experimentally, $B\to K^*\phi$ decays have been studied with full angular
analysis from which information on final-state interactions can be extracted.
Historically, it was the observation of large transverse polarization in these
decays that had triggered the theoretical and experimental interest in the
study of charmless $B\to VV$ decays.

Theoretically, $B\to K^*\phi$ decays can be analyzed in the same manner as the
$K^*\rho$ modes. The decay amplitude of $B^-\to K^{*-}\phi$ can be
approximated as
\be
\A^h_{B^-\to K^{*-}\phi}\approx
V_c(\alpha_3^h+\alpha_4^{c,h}+\beta_3^h-{1\over 2}\alpha^h_{3,{\rm
EW}})X^h_{K^*\phi}. \en When the  penguin annihilation contribution
$\beta_3$ is turned off, we have \be \label{eq:ratioKphi} \left.{ \A^-\over
\A^0}\right|_{B^-\to K^{*-}\phi} &\approx&
\left({\alpha_3^-+\alpha_4^{c,-}-{1\over 2}\alpha_{3,{\rm EW}}^- \over
\alpha_3^0+\alpha_4^{c,0}-{1\over 2}\alpha_{3,{\rm EW}}^0
}\right)\left({X^-_{K^*\phi}\over X^0_{K^*\phi}}\right). \en From the
amplitude parameters given in Table \ref{tab:VVparameter}, it is clear that
there exists a constructive (destructive) interference in the $\A^-$ ($\A^0$)
amplitude. As a consequence, although the factorizable amplitudes respect the
hierarchy $|X^0_{K^*\phi}|:|X^-_{K^*\phi}|:|X^+_{
K^*\phi}|=1:0.35:0.007$ due to the $(V-A)$ structure of weak interactions and
helicity conservation in strong interactions, the negative- and
longitudinal-helicity amplitudes are comparable in magnitude. Numerically, we
indeed find $f_L=0.62$ and $f_\parallel\sim f_\bot= 0.19$ (see Table \ref{tab:KVrhoBr}). This is very
similar to the decay $\ov B^0\to \bar K^{*0}\rho^0$ where $f_L$ is also found
to be small, of order 0.50\,. Experimentally, the naive expectation of
$f_L\approx 1-4m_V/m_B^2\sim 0.90$ is strongly violated in charmless
penguin-dominated $VV$ modes. Nevertheless, a small $f_L$ for $K^*\phi$ is
quite natural in QCD factorization because the parameters $a_i^h$ are helicity
dependent. The fact that real parts of $\alpha_3$ and $\alpha_{\rm 3,EW}$ flip
signs from $h=0$ to $h=-$ and that $\alpha^h_4$ is smaller in magnitude for the
longitudinal amplitude (see Table \ref{tab:VVparameter}) will render the
negative helicity amplitude comparable to the longitudinal one.

Even though the longitudinal polarization fraction is reduced to 60\% level in
the absence of penguin annihilation, this does not mean that the polarization
anomaly is resolved. As stated before, irrespective of the predictions for
polarization fractions, the first task we need to focus on is to reproduce the
correct rate for $B\to K^*\phi$ because the calculated branching ratio of order
$4.1\times 10^{-6}$ is too small by a factor of $\sim 2.5$ compared to the
measured one, $\sim 10\times 10^{-6}$ (cf. Table \ref{tab:VVBr}). Assuming weak
annihilation to account for the discrepancy between theory and experiment, we
can fit the data of branching ratios and $f_L$ simultaneously by adjusting the
parameters $\rho_A$ and $\phi_A$. However, this also means that QCDF loses its
predictive power in this manner. We find that the rate and $f_L$ can be
accommodated by having $\rho_A\approx 0.65$ and $\phi_A\approx -53^\circ$. This
set of the annihilation parameters differs slightly from that extracted from
$B\to K^*\rho$ decays, namely, $\rho_A(K^*\rho)\approx 0.78$ and
$\phi_A(K^*\rho)\approx -43^\circ$. Therefore, within the framework of QCDF,
one cannot account for all charmless $B\to VV$ data by a universal set of
$\rho_A$ and $\phi_A$ parameters. This could be an indiction that large penguin
annihilation cannot be the ultimate story for understanding $B\to VV$ decays.

Since the complete angular analysis of $\ov B\to \bar K^*\phi$ has been
performed by both BaBar and Belle, information on the parallel and
perpendicular polarizations and their phases relative to the longitudinal one
is available. We see from Table \ref{tab:VVBr} that $f_\bot$ and $f_\parallel$
are very similar, of order 0.25. Experimentally, the phases $\phi_\parallel$
and $\phi_\bot$ deviate from either $\pi$ or zero by more than $4.6\sigma$ and
$3.3\sigma$ for $K^{*-}\phi$ and $5.5\sigma$ for $\bar K^{*0}\phi$ \cite{HFAG}.
This implies the presence of final-state interactions. The relative phases are
calculated to be \be \label{eq:phi}
 \phi_\parallel(K^{*-}\phi)&=&(80^{+43}_{-83})^\circ \qquad ({\rm expt}: ~(46\pm10)^\circ), \non \\
 \phi_\parallel(\bar K^{*0}\phi) &=&(78^{+43}_{-81})^\circ \qquad ({\rm expt}: ~(44^{+8}_{-7})^\circ),
 \en
and $\phi_\parallel(K^{*-}\phi)-\phi_\bot(K^{*-}\phi)-\pi= \phi_\parallel(\bar
K^{*0}\phi)-\phi_\bot(\bar K^{*0}\phi)-\pi\approx 0.7^\circ$. They are
consistent with the data.

Thus far we have chosen  the renormalization scale to be $\mu=m_b(m_b)$ in
calculations. We now address the issue with $\mu$. In principle, physics should
be independent of the choice of $\mu$, but in practice there exists some
residual $\mu$ dependence in the truncated calculations. We have checked
explicitly that the decay rates without annihilation are indeed essentially
stable against $\mu$. However, when penguin annihilation is turned on, it is
sensitive to the choice of the renormalization scale because the penguin
annihilation contribution characterized by the parameter $b_3$ is dominantly
proportional to $\alpha_s(\mu_h)c_6(\mu_h)$  at the hard-collinear scale
$\mu_h=\sqrt{\mu\Lambda_h}$.  For the hadronic scale $\Lambda_h\approx 500$
MeV, we have $\mu_h\approx 1.45$ GeV and 1 GeV for $\mu=4.2$ GeV and 2.1 GeV,
respectively. At the amplitude level, the enhancement of penguin annihilation
at $\mu=2.1$ GeV is of order $\alpha_s(1)c_6(1)/[\alpha_s(1.45)c_6(1.45)]\sim
1.8$. We find that if the renormalization scale is chosen to be
$\mu=m_b(m_b)/2=2.1$ GeV, we cannot fit the branching ratios and polarization
fractions simultaneously for both $B\to K^*\phi$ and $B\to K^*\rho$ decays. For
example, the rate of the former can be accommodated with $\rho_A\sim 0.25$, but
the corresponding $f_L\sim 0.28$ is too small. Likewise, although $\B(B^-\to
\bar K^{*0}\rho^-)$ can be fitted well with $\rho_A\sim 0.55$, the resultant
$f_L\lsim 0.12$ is highly suppressed. This is ascribed to the fact that at the
scale $\mu=2.1$ GeV, the negative-helicity amplitude receives much more
enhancement than the longitudinal one and hence the longitudinal polarization
is suppressed at the small $\mu$ scale. In order to ensure the validity of the
penguin-annihilation mechanism for describing $B\to VV$ decays, we will confine
ourselves to the renormalization scale $\mu=m_b(m_b)$ in the ensuing study.

\vskip 0.2in \noindent{\it \underline{$B\to K^*\omega$}} \vskip 0.1in
The decay amplitudes for $B\to K^*\omega$ read
\be
 \sqrt{2}\A_{B^-\to K^{*-}\omega} &\approx&
 \left[V_u\alpha_1^h+ V_c(\alpha_4^{c,h}+\beta_3^h) \right] X_{\omega\bar K^*}^h
 +\left[V_u\alpha_2^h+V_c(2\alpha_3^h+{1\over 2}\alpha_{3,{\rm EW}}^h)\right]
 X_{\bar K^*\omega}^h,  \non \\
 \sqrt{2}\A_{\bar B^0\to \bar K^{*0}\omega} &\approx&
 \left[V_c(\alpha_4^{c,h}+\beta_3^h) \right] X_{\omega\bar K^*}^h
 +\left[V_u\alpha_2^h+V_c(2\alpha_3^h+{1\over 2}\alpha_{3,{\rm EW}}^h)\right]
 X_{\bar K^*\omega}^h.
\en From the previous analysis of $K^*\rho$ and $K^*\phi$ decays, we found two
distinct sets of the penguin annihilation parameters $\rho_A$ and $\phi_A$. If
the  set of parameters inferred from $B\to K^*\rho$ decays, namely,
$\rho_A=0.78$ and $\phi_A=-43^\circ$, is employed, we obtain $\B(B^-\to
K^{*-}\omega)\approx 4.5\times 10^{-6}$ and $\B(\ov B^0\to \bar
K^{*0}\omega)\approx 3.9\times 10^{-6}$, which are slightly higher than the
respective  experimental upper limits, $3.4\times 10^{-6}$ and $2.7\times
10^{-6}$ \cite{BaBar:omegarho,Belle:KVomega}. By contrast, if the parameters
$\rho_A=0.65$ and $\phi_A=-53^\circ$ extracted from $K^*\phi$ modes are used,
the resultant predictions  $\B(B^-\to K^{*-}\omega)\approx 3.5\times 10^{-6}$
and $\B(\ov B^0\to \bar K^{*0}\omega)\approx 3.0\times 10^{-6}$ are consistent
with experiment (see Table \ref{tab:VVBr}). Of course, if the $B\to\omega$ form
factors are smaller than what we expected as implied by the measurement of
$B^-\to \rho^-\omega$, then $\B(B\to K^*\omega)$ will be safely below the
current bounds.

\vskip 0.2in \noindent{\it \underline{$B\to K^*\bar K^*$}} \vskip 0.1in
The expressions of $B\to K^*\bar K^*$ decay amplitudes read
\be
{\cal A}_{B^-\to K^{*0} K^{*-}}
   &=&  \frac{G_F}{\sqrt{2}}\sum_{p=u,c}\lambda_p^{(d)}
    \Big[ \delta_{pu}\beta_2+ \alpha_4^p - \frac{1}{2}\alpha_{4,{\rm
    EW}}^p +\beta_3^p +\beta^p_{3,{\rm EW}} \Big]   X^{(\overline B \bar K^{*},K^*)},
\non \\
    {\cal A}_{\ov B^0\to K^{*-} K^{*+}}
   &=&  \frac{G_F}{\sqrt{2}}\sum_{p=u,c}\lambda_p^{(d)}\Bigg\{
 \Big[ \delta_{pu}\beta_1+\beta_4^p +\beta^p_{4,{\rm EW}} \Big]
 X^{(\overline B \bar K^*,K^*)}
    +f_Bf_{K^*}^2\left[b_4^p-{1\over 2}b^p_{4,{\rm EW}}\right]_{K^*\bar K^*}\Bigg\},
\non \\
    {\cal A}_{\ov B^0\to \bar K^{*0} K^{*0}}
   &=&  \frac{G_F}{\sqrt{2}}\sum_{p=u,c}\lambda_p^{(d)}\Bigg\{
    \Big[ \alpha_4^p-{1\over 2}\alpha^p_{4,{\rm EW}}+\beta^p_3
    +\beta_4^p -{1\over 2}\beta^p_{3,{\rm EW}} -{1\over 2}\beta^p_{4,{\rm EW}}
    \Big]   X^{(\overline B \bar  K^*,K^*)}  \non \\
    &+&   f_Bf_{K^*}^2\left[b_4^p-{1\over 2}b^p_{4,{\rm EW}}\right]_{K^*\bar K^*}\Bigg\}.
\en Both $\ov B^0\to \bar K^{*0}K^{*0}$ and $B^-\to K^{*0}K^{*-}$ are $b\to d$
penguin-dominated decays, while $\ov B^0\to K^{*-}K^{*+}$ proceeds only through
weak annihilation. Hence, their branching ratios are expected to be small, of
order $\lsim 10^{-6}$. Recently, the $\bar K^{*0}K^{*0}$ mode was first
measured by BaBar with the branching ratio $(1.28^{+0.37}_{-0.32})\times
10^{-6}$ \cite{BaBar:KVKV}. Our prediction is slightly smaller, about 1$\sigma$
away from the BaBar measurement (Table \ref{tab:VVBr}). The absence of
transverse polarization in the $K^{*-}K^{*+}$ mode is due to the approximation
we have adapted; that is, we have neglected the transverse annihilation
contributions $A_{1,2}^\pm$ relative to other terms. Hence, transverse
polarization does not receive contributions from the annihilation terms
$b_1^\pm,b_2^\pm,b_4^\pm,b_{\rm 4,EW}^\pm$ within our approximation [see Eq.
(\ref{eq:bi})].

\vskip 0.2in \noindent{\it \underline{Comparison with other works}} \vskip
0.1in Within the framework of QCD factorization, we have studied charmless
$B\to VV$ decays closely to the works of Kagan \cite{Kagan} and Beneke, Rohrer
and Yang (BRY) \cite{BenekeVV}. Nevertheless, there are some differences
between our work and theirs as we are going to discuss below.

Without penguin annihilation, Kagan found $f_L\approx 0.90$ for the $\bar
K^{*0}\phi$ mode, while BRY got $f_L\approx 0.67$ and  we obtained $f_L\approx
0.62$. Kagan did not consider vertex corrections and hard spectator
interactions in his realistic calculations of $a_i^{p,h}$, though he has
discussed the latter briefly. It seems to us that $a_i^h$ are essentially
helicity independent in the Kagan's calculation and this accounts for the
difference in the estimation of $f_L$. Moreover, what is the initial value of
$f_L$ in the absence of penguin annihilation is immaterial because we will use
the unknown annihilation parameters to {\it accommodate}  the data of branching
ratios and $f_L$ rather than to {\it predict} them.

We differ from BRY mainly for using different $\rho_A$ and $\phi_A$ parameters for describing $B\to K^*\rho$ decays. If we follow BRY to use the set of $\rho_A$ and $\phi_A$ parameters extracted from $B\to K^*\phi$ decays to describe $K^*\rho$ modes, the rates for the latter will be systematically below the measurements. Since it is necessary to reproduce the measured rates first in order to have a reliable estimate of polarization fractions, we need to fit the
$K^*\rho$ data separately. The resultant $\rho_A$ and $\phi_A$ parameters
differ from the ones determined from $K^*\phi$ modes. This can be viewed as a
potential problem of QCDF.

In the pQCD approach, the calculated branching ratio of $\ov B\to \bar K^*\phi$
is too large, of order $15\times 10^{-6}$ and $f_L\sim 0.75$ \cite{Mishima}. It
has been proposed in \cite{HNLi} that a smaller form factor
$A_0^{BK^*}(0)\approx 0.30$ will bring down both the rates and $f_L$ and bring
up $f_\parallel$ and $f_\bot$. While this sounds plausible, the NLO corrections
to helicity-dependent coefficients $a_i^h$ should be taken into account in this
approach as we have demonstrated that NLO corrections to $a_i^-$ and $a_i^0$
will bring down $f_L$ significantly. It is also important to compute the rates
and polarization fractions for $\ov B\to \bar K^*\rho$ decays in this framework
and compare with experiment.

Another plausible solution is to consider the long-distance rescattering
contributions from some charm intermediate states such as $D^{(*)}D_s^{(*)}$
\cite{CCSfsi,Ladisa,Colangelo}. The idea is simple: First,  $B\to D^*D_s^*$
decays are CKM favored and hence final-state interactions via charm
intermediate states will bring up the $K^*\phi$ rates. This is welcome since
the short-distance predictions of the branching fractions for penguin-dominated
$B\to VV$ modes in most of the models under consideration are too small
compared to experiment. Second, large transverse polarization induced from
$B\to D^*D_s^*$ will be propagated to $\phi K^*$ via final-state rescattering.
The unknown parameter in the model for final-state rescattering is fixed by the
measured rate \cite{CCSfsi}. However, this approach has one drawback. That is,
while the longitudinal polarization fraction can be reduced significantly from
short-distance predictions due to final-state interaction effects, no sizable
perpendicular polarization $f_\bot$ is found owing mainly to the large
cancelations occurring in the processes $B\to D_s^* D\to\phi K^*$ and $B\to D_s
D^*\to\phi K^*$ and this can be understood as a consequence of CP and SU(3)
symmetry \cite{CCSfsi}. That is, final-state rescattering leads to the
suppression of $f_T(=f_\bot+f_\parallel)$ at the expense of $f_\parallel\gg
f_\bot$. As pointed out in \cite{CCSfsi}, one possibility to circumvent the
aforementioned cancelation is to consider the contributions from the
even-parity charmed meson intermediate states. In view of the fact that even at
the short-distance level, $f_L$ can be as small as 40\% to 70\% after NLO
corrections to effective Wilson coefficients are taken into account, it is
worth re-examining this type of solution again.

In soft-collinear effective theory (SCET) \cite{SCET}, large transverse
polarization in penguin-dominated $VV$ modes may arise from the long-distance
charming penguin contribution. Indeed, the aforementioned mechanism of
final-state rescattering of charm intermediate states mimics the charming
penguin in SCET, while both QCDF and pQCD approaches rely on penguin
annihilation to resolve the polarization anomaly.\footnote{Ways of
distinguishing penguin annihilation from rescattering have been recently
proposed in \cite{Datta}.}

\subsection{$B\to VA$ decays}
The calculated branching ratios and longitudinal polarization fractions for the
decays $B\to A\rho,~AK^*,~A\omega,~A\phi$ with
$A=a_1(1260),b_1(1235),K_1(1270),K_1(1400),f_1(1285),f_1(1420),h_1(1170),h_1(1380)$
are collected in Tables \ref{tab:BRa1}-\ref{tab:BRf1}.

\subsubsection{$\ov B\to a_1V,~b_1V$ decays}
The decays $\ov B^0\to (a_1^+,b_1^+)(\rho^-,\pi^-)$ are governed by the decay
constants of the $\rho$ and $\pi$, respectively. Since $f_\rho\gg f_\pi$, we
thus expect to have $\B(\ov B^0\to a_1^+\rho^-)\gg \B(\ov B^0\to a_1^+\pi^-)$
and $\B(\ov B^0\to b_1^+\rho^-)\gg \B(\ov B^0\to b_1^+\pi^-)\sim 10\times
10^{-6}$. These features  are borne out in our realistic calculations (see
Table \ref{tab:BRa1}). Calder\'on, Mu\~noz and Vera (CMV) \cite{Calderon} found
the other way around:  $(a_1^+,b_1^+)\rho^-$ modes have rates smaller than
$(a_1^+,b_1^+)\pi^-$ ones, which we strongly disagree. Since the modes
$(a_1^-,b_1^-)(\rho^+,\pi^+)$ are governed by $f_{a_1}$ and $f_{b_1}$,
respectively, and since $f_{a_1}\sim f_\rho$ and $f_{b_1}$ is very small
(vanishing for the neutral $b_1$), we anticipate that $a_1^-\rho^+$ and
$a_1^-\pi^+$ have comparable rates and the $b_1^-\rho^+$ mode is highly
suppressed relative to the $b_1^+\rho^-$ one. \footnote{As explained in
\cite{CY:AP}, within the QCD factorization approach, the suppression of
$b_1^-(\pi^+,\rho^+)$ modes is not directly related to the smallness of the
$b_1$ decay constant, but is ascribed to the tiny coefficient $a_1$. However,
the smallness of $a_1$ has to do with the decay constant suppression. For more
details, see \cite{CY:AP}.} The decays $(a_1^-,b_1^-)\rho^0$ receive both
color-allowed and color-suppressed contributions: \be
\A_{B^-\to a_1^-\rho^0} \propto (a_1^h+\cdots)X_h^{(\bar B\rho,a_1)}
 +(a_2^h+\cdots)X_h^{(\bar Ba_1,\rho)}, \non \\
\A_{B^-\to b_1^-\rho^0} \propto (a_1^h+\cdots)X_h^{(\bar B\rho,b_1)}
 +(a_2^h+\cdots)X_h^{(\bar Bb_1,\rho)},
\en where $X^{(\bar BM_1,M_2)}$ is the factorizable amplitude defined by Eq.
(\ref{eq:Xamplitude}). Since the color-allowed amplitude of the $b_1^-\rho^0$
mode is highly suppressed by the smallness of the $b_1$ decay constant and the
color-suppressed amplitude is suppressed by the small ratio of $a_2/a_1$,  it
is clear that $\B(B^-\to a_1^-\rho^0)\gg\B(B^-\to b_1^-\rho^0)$.
 The decays $(a_1^-,b_1^-)\omega$ should have rates similar to $(a_1^-,b_1^-)\rho^0$.
So far there is only one experimental measurement of $B\to VA$ decays, namely,
$B^0\to a_1^\pm \rho^\mp$ with the result \cite{BaBara1rho} \be \B(B^0\to
a_1^\pm \rho^\mp)\B(a_1^\pm\to (3\pi)^\pm)<61\times 10^{-6}. \en Assuming that
$a_1^\pm$ decays exclusively to $\rho^0\pi^\pm$, we then have the upper limit
of $61\times 10^{-6}$ for the branching ratio of $B^0\to a_1^\pm \rho^\mp$. Our
prediction $\B(B^0\to a_1^\pm \rho^\mp)\approx 59\times 10^{-6}$ is thus
consistent with experiment.

\begin{table}[t]
\caption{Branching ratios (in units of $10^{-6}$) and the longitudinal
polarization fraction (in parentheses) for the decays $B\to
(a_1,~b_1)(\rho,~\omega,~\phi,~K^*)$ with $a_1=a_1(1260)$ and $b_1=b_1(1235)$.
The theoretical errors correspond to the uncertainties due to variation of (i)
Gegenbauer moments, decay constants, quark masses, form factors, the
$\lambda_B$ parameter for the $B$ meson wave function, and (ii) $\rho_{A,H}$,
$\phi_{A,H}$, respectively. For longitudinal polarization fractions, we add all
errors in quadrature as the theoretical uncertainty is usually dominated by
(ii). Default results are  for $\rho_A=0.65$ and $\phi_A=-53^\circ$. We use the
light-cone sum rule results for the $B\to a_1$ and $B\to b_1$ transition form
factors (see Table \ref{tab:FFinLCSR}). The model predictions by Calder\'on,
Mu\~noz and Vera (CMV) \cite{Calderon} are also displayed here for comparison.}
 \label{tab:BRa1}
\begin{ruledtabular}
\begin{tabular}{l r r | l r r}
Mode & This work & CMV~~~ & Mode & This work & CMV \\
\hline
 $\overline B^0\to a_1^+ \rho^-$  & $ 23.9^{+10.5+3.2}_{-~9.2-0.4} \!\!$ ($0.82^{+0.05}_{-0.13}$)
 & 4.3~~~ & $\overline B^0\to b_1^+ \rho^-$ & $32.1^{+16.5+12.0}_{-14.7-~4.6} \!\!\! $
  ($0.96^{+0.01}_{-0.02}$) & 1.6 \\
 $\overline B^0\to a_1^-\rho^+$  & $36.0^{+3.5+3.5}_{-4.0-0.7}$ ($0.84^{+0.02}_{-0.06}$)
 & 4.7~~~ & $\overline B^0\to b_1^-\rho^+$ & $0.6^{+0.6+1.9}_{-0.3-0.2} ~$
  ($0.98^{+0.00}_{-0.33}$) & 0.55 \\
 $\overline B^0\to a_1^0 \rho^0$  & $1.2^{+2.0+5.1}_{-0.7-0.3}$ ($0.82^{+0.06}_{-0.68}$)
 & 0.01~~~ & $\overline B^0\to b_1^0 \rho^0$ & $3.2^{+5.2+1.7}_{-2.0-0.4} ~$
  $(0.99^{+0.00}_{-0.18})$  & 0.002 \\
 $B^-\to a_1^0 \rho^-$  & $17.8^{+10.1+3.1}_{-~6.4-0.2} \!\!$ ($0.91^{+0.03}_{-0.10}$) & 2.4~~~
 & $B^-\to b_1^0 \rho^-$ & $29.1^{+16.2+5.4}_{-10.6-5.9} \! $
  ($0.96^{+0.01}_{-0.06}$) & 0.86\\
 $B^-\to a_1^- \rho^0$   & $23.2^{+3.6+4.8}_{-2.9-0.1}$ ($0.89^{+0.11}_{-0.18}$) &3.0~~~
 & $B^-\to b_1^- \rho^0$ & $0.9^{+1.7+2.6}_{-0.6-0.5} \, $
  ($0.90^{+0.05}_{-0.38}$) & 0.36 \\
 $\ov B^0\to a_1^0\omega$ & $0.2^{+0.1+0.4}_{-0.1-0.0}$ ($0.75^{+0.11}_{-0.65}$) & 0.003~~~
 & $\ov B^0\to b_1^0\omega$ & $0.1^{+0.2+1.6}_{-0.0-0.0} \,$ ($0.04^{+0.96}_{-0.00}$) & 0.004\\
 $B^-\to a_1^-\omega$ &  $22.5^{+3.4+3.0}_{-2.7-0.7}$ ($0.88^{+0.10}_{-0.14}$) & 2.2~~~
 & $B^-\to b_1^-\omega$ & $0.8^{+1.4+3.1}_{-0.5-0.3} \,$
  ($0.91^{+0.07}_{-0.33}$) & 0.38\\
 $\ov B^0\to a_1^0\phi$ & $0.002^{+0.002+0.009}_{-0.001-0.000}$ ($0.94^{+0.00}_{-0.69}$)
 &0.0005~~~ & $\ov B^0\to b_1^0\phi$ & $0.01^{+0.01+0.01}_{-0.00-0.00}$ ($0.98^{+0.01}_{-0.33}$) & 0.0002\\
 $B^-\to a_1^-\phi$ & $0.01^{+0.01+0.04}_{-0.00-0.00}$ ($0.94^{+0.01}_{-0.69}$) &0.001~~~
 & $B^-\to b_1^-\phi$ &  $0.02^{+0.02+0.03}_{-0.01-0.00}$ ($0.98^{+0.01}_{-0.33}$) & 0.0004\\
 $\overline B^0\to a_1^+ K^{*-}$  &  $10.6^{+5.7+31.7}_{-4.0-~8.1} \!\!$ ($0.37^{+0.39}_{-0.29}$)
 & 0.92~~~ & $\overline B^0\to b_1^+ K^{*-}$ & $12.5^{+4.7+20.1}_{-3.7-~9.0} \!$  ($0.82^{+0.18}_{-0.41}$) & 0.32\\
 $\overline B^0\to a_1^0 \overline K^{*0}$ & $4.2^{+2.8+15.5}_{-1.9-4.2} \!\! $ ($0.23^{+0.45}_{-0.19}$)
 & 0.64~~~ & $\overline B^0\to b_1^0 \overline K^{*0}$ & $6.4^{+2.4+8.8}_{-1.7-4.8} \, $ ($0.79^{+0.21}_{-0.73}$) & 0.15 \\
 $B^-\to a_1^- \overline K^{*0}$ & $11.2^{+6.1+31.9}_{-4.4-~9.0} \!\! $ ($0.37^{+0.48}_{-0.37}$)
 & 0.51~~~ & $B^-\to b_1^- \overline K^{*0}$ & $12.8^{+5.0+20.1}_{-3.8-~9.6} \!$   ($0.79^{+0.21}_{-0.74}$)  & 0.18 \\
 $B^-\to a_1^0  K^{*-}$ & $7.8^{+3.2+16.3}_{-2.5-~4.3} \!\! $ ($0.52^{+0.41}_{-0.42}$) &0.86~~~
 &  $B^-\to b_1^0  K^{*-}$ & $7.0^{+2.6+12.0}_{-2.0-~4.8} \! $ ($0.82^{+0.16}_{-0.26}$) & 0.12\\
\end{tabular}
\end{ruledtabular}
\end{table}

To discuss the effect of the annihilation contribution, let us take the
penguin-dominated decays $\ov B^0\to (a_1^+,b_1^+)K^{*-}$ as an example. From
Eq. (A3) of \cite{CY:AP} we have
\be
\A_{\bar B^0\to a_1^+K^{*-}} &\propto& ( \alpha_4^c +
\alpha_{4,{\rm EW}}^c)X^{(\bar Ba_1,K^*)} +
if_Bf_{a_1}f_{K^*}(b_3^c -{1\over 2} b_{3,{\rm EW}}^c)_{a_1K^*}, \non \\
\A_{\bar B^0\to b_1^+K^{*-}} &\propto& ( \alpha_4^c +
\alpha_{4,{\rm EW}}^c)X^{(\bar Bb_1,K^*)} + if_Bf^\bot_{b_1}f_{K^*}(b_3^c
-{1\over 2} b_{3,{\rm EW}}^c)_{b_1K^*}, \en where we have replaced the decay
constant $f_{b_1}$ by $f_{b_1}^\bot$ as explained before (see the paragraph
after Eq. (\ref{eq:nor})).
From penguin-dominated $B\to VV$ decays we learn that the predicted rates in
default are typically too small by a factor of $2\sim 3$.  In the absence of
the experimental information for penguin-dominated $B\to VA$ decays, we shall
use the penguin-annihilation parameters $\rho_A=0.65$ and $\phi_A=-53^\circ$
inferred from $B\to K^*\phi$ decays as a guidance for annihilation enhancement.
Since the magnitude of $b_3$ is large for the $b_1K^*$ modes (specifically,
$b_3^0(b_1K^*)=-1.78+9.92i$ and $b_3^0(a_1K^*)=-0.19+4.11i$), $B\to b_1K^*$ decays receive
more enhancement from penguin annihilation than $B\to a_1K^*$ ones.  When
penguin annihilation is turned off, we have, for example,
\be
\B(\ov B^0\to a_1^+K^{*-})=(3.6^{+1.6+0.5}_{-1.3-0.1})\times 10^{-6}, \qquad
 & \B(\ov B^0\to b_1^+K^{*-}) =(4.1^{+2.3+0.3}_{-2.0-0.3})\times 10^{-6}, \non \\
\B(B^-\to a_1^-\bar K^{*0})=(4.1^{+2.0+1.7}_{-1.6-0.1})\times 10^{-6}, \qquad
 & \B( B^-\to b_1^-\bar K^{*0})=(4.0^{+2.0+0.7}_{-2.5-0.6})\times 10^{-6}.
\en We see from Table \ref{tab:BRa1} that $a_1K^*$ and $b_1K^*$ modes are
substantially enhanced by penguin annihilation. Experimentally, it is thus very
important to measure them to test the importance of the penguin annihilation
mechanism.

We have checked explicitly that, in the absence of penguin annihilation, the
longitudinal polarization fractions  are close to one half in $a_1K^*$ modes
and 90\% in $b_1K^*$ ones. This can be seen from the ratio of the negative- and
longitudinal-helicity amplitudes \be \left.{ \A^-\over \A^0}\right|_{\bar B\to
a_1^+K^{*-}} &\approx& \left({\alpha_4^{c,-}+\alpha_{4,{\rm EW}}^- \over
\alpha_4^{c,0}+\alpha_{4,{\rm EW}}^0 }\right)\left({X^-_{\bar Ba_1,K^*}\over
X^0_{\bar Ba_1,K^*}}\right). \en As discussed in the section of $B\to VV$
decays, the interference between $\alpha_4^{c,h}$ and $\alpha_{4,{\rm EW}}^h$
is generally constructive for $h=-$ and destructive for $h=0$. Since
$|X^0_{\bar Ba_1,K^*}|:|X^-_{\bar Ba_1,K^*}|:|X^+_{\bar
Ba_1,K^*}|=1:0.50:0.06$, and $|X^0_{\bar Bb_1,K^*}|:|X^-_{\bar
Bb_1,K^*}|:|X^+_{\bar Bb_1,K^*}|=1:0.21:0.03$, it is obvious that the $\A^-$
amplitude of $a_1K^*$ channels has more chance to be comparable to $\A^0$ than
the $b_1K^*$ ones. When penguin annihilation is turned on, it is evident from
Table \ref{tab:BRa1} that $a_1K^*$ modes are dominated by transverse
polarization amplitudes, whereas $b_1K^*$ are governed by longitudinal
polarization states.

The decays $B\to (a_1,b_1)\phi$ are highly suppressed relative to the
tree-dominated $(a_1,b_1)\rho$ modes as they proceed through $b\to d$ penguin
process and are thus suppressed by the small coefficients for penguin
operators. Moreover, they do not receive any annihilation contribution!

On the experimental ground, our calculations suggest that the tree-dominated
channels $a_1^+\rho^-$, $a_1^-\rho^-$, $a_1^0\rho^-$, $a_1^-\rho^0$,
$a_1^-\omega$, $b_1^+\rho^-$ and $b_1^0\rho^-$ should be readily accessible to
$B$ factories. Measurements of the penguin-dominated modes $a_1K^*$ and
$b_1K^*$ are crucial for testing the mechanism of penguin annihilation.

\subsubsection{$B\to K_1(1270)V,~K_1(1400)V$ decays}

\begin{table}[!]
\caption{Same as Table \ref{tab:BRa1} except for $b\to s$ penguin-dominated
decays (top) $B\to K_1(\rho,K^*,\omega)$ and $b\to d$ penguin-dominated ones
(bottom) $B\to K_1\bar K^*$ for two different mixing angles
$\theta_{K_1}=-37^\circ$ and $-58^\circ$.   Default results are for
$\rho_A=0.65$ and $\phi_A=-53^\circ$.}
 \label{tab:BRK1V}
\begin{ruledtabular}
\begin{tabular}{l l c l c  }
&  \multicolumn{2}{c}{$\theta_{K_1}=-37^\circ$}
 &   \multicolumn{2}{c}{$\theta_{K_1}=-58^\circ$} \\
\cline{2-3} \cline{4-5}
\raisebox{2.0ex}[0cm][0cm]{Decay} & $\B$ & $f_L$ & $\B$ &
$f_L$  \\ \hline
 $\overline B^0\to K_1^-(1270) \rho^+$
 & $\!\!\! 16.8^{+9.8+54.7}_{-6.8-13.8}$ & $0.57^{+0.39}_{-0.30}$ & $\!\!\! 19.4^{+12.1+47.6}_{-~8.5-14.8}$ & $0.49^{+0.48}_{-0.36}$ \\
 $\overline B^0\to \ov K_1^0(1270) \rho^0$
 & $9.1^{+5.3+34.2}_{-3.5-~8.6}$ & $0.50^{+0.39}_{-0.37}$ & $9.8^{+6.0+30.4}_{-4.1-~8.0}$ & $0.40^{+0.49}_{-0.30}$ \\
 $B^-\to \ov K_1^0(1270) \rho^-$ & $\!\!\! 17.0^{+10.7+53.0}_{-~7.3-15.3}$ & $0.52^{+0.47}_{-0.36}$ & $\!\!\! 20.1^{+12.5+48.5}_{-~8.9-15.2}$ & $0.47^{+0.51}_{-0.46}$  \\
 $B^-\to K_1^-(1270) \rho^0$ &  $8.2^{+5.0+20.4}_{-3.7-~6.1}$ & $0.56^{+0.39}_{-0.34}$ &  $\!\!\! 10.3^{+5.9+19.1}_{-4.5-~6.7}$ & $0.56^{+0.41}_{-0.37}$   \\
 $\ov B^0\to \ov K_1^0(1270) \omega$ & $7.3^{+4.7+24.0}_{-3.1-~7.5}$ & $0.59^{+0.39}_{-0.32}$ & $8.2^{+5.3+21.5}_{-3.7-~7.4}$ & $0.48^{+0.50}_{-0.47}$ \\
 $B^-\to K_1^-(1270) \omega$ & $7.4^{+4.7+22.6}_{-3.2-~6.9}$ & $0.56^{+0.41}_{-0.22}$ & $8.7^{+5.4+20.6}_{-3.7-~7.1}$ & $0.52^{+0.46}_{-0.34}$   \\
 $\ov B^0\to \ov K_1^0(1270) \phi$ & $3.6^{+1.7+4.8}_{-1.3-2.9}$ & $0.67^{+0.33}_{-0.64}$ & $3.2^{+2.1+5.2}_{-1.4-2.7}$ & $0.31^{+0.69}_{-0.31}$  \\
 $B^-\to K_1^-(1270) \phi$ & $3.8^{+1.9+5.1}_{-1.5-3.1}$ & $0.67^{+0.33}_{-0.64}$ & $3.4^{+2.2+5.5}_{-1.5-2.8}$ & $0.31^{+0.69}_{-0.37}$ \\
 $\overline B^0\to K_1^-(1400) \rho^+$
 & $8.6^{+2.8+13.0}_{-2.3-~4.3}$ & $0.64^{+0.30}_{-0.23}$ & $5.7^{+1.2+17.5}_{-1.0-~4.7}$ & $0.87^{+0.09}_{-0.43}$ \\
 $\overline B^0\to \ov K_1^0(1400) \rho^0$ & $\!\!\! 11.4^{+3.2+15.5}_{-2.8-~5.8}$ & $0.65^{+0.32}_{-0.21}$ & $9.8^{+2.1+22.4}_{-2.0-~8.0}$ & $0.90^{+0.07}_{-0.28}$   \\
 $B^-\to \ov K_1^0(1400) \rho^-$ & $\!\!\! 10.9^{+2.6+15.2}_{-3.1-~5.4}$ & $0.65^{+0.33}_{-0.26}$  & $7.5^{+2.06+19.4}_{-1.6-~6.0}$ & $0.85^{+0.11}_{-0.44}$   \\
 $B^-\to K_1^-(1400) \rho^0$ &$3.8^{+1.3+6.4}_{-1.1-1.9}$ & $0.61^{+0.32}_{-0.22}$ & $1.5^{+0.7+7.7}_{-0.3-1.0}$ & $0.70^{+0.24}_{-0.48}$   \\
 $\ov B^0\to \ov K_1^0(1400) \omega$ & $5.6^{+2.8+6.8}_{-2.2-2.5}$ & $0.72^{+0.31}_{-0.36}$ & $4.6^{+2.8+9.5}_{-1.7-3.5}$ & $0.90^{+0.07}_{-0.28}$ \\
 $B^-\to K_1^-(1400) \omega$ & $4.5^{+1.8+6.0}_{-1.4-2.1}$ & $0.68^{+0.32}_{-0.32}$ & $3.1^{+1.3+8.1}_{-0.8-2.4}$ & $0.87^{+0.09}_{-0.40}$  \\
 $\ov B^0\to \ov K_1^0(1400) \phi$ & $\!\!\! 10.4^{+7.9+38.3}_{-5.1-10.4}$ & $0.46^{+0.26}_{-0.02}$ & $\!\!\! 10.7^{+7.1+37.69}_{-4.6-10.4}$ & $0.57^{+0.31}_{-0.22}$ \\
 $B^-\to K_1^-(1400) \phi$ & $\!\!\! 11.1^{+8.5+41.1}_{-5.4-11.4}$ & $0.45^{+0.13}_{-0.09}$ & $\!\!\! 11.3^{+7.5+40.2}_{-4.9-11.1}$ & $0.57^{+0.32}_{-0.22}$  \\
 \hline
 $\overline B^0\to K_1^-(1270) K^{*+}$  &  $0.01^{+0.01+0.03}_{-0.01-0.00}$ & 1.0 & $0.00^{+0.00+0.01}_{-0.00-0.00}$ & 1.0 \\
 $\overline B^0\to K_1^+(1270) K^{*-}$  &  $0.06^{+0.03+1.43}_{-0.02-0.02}$ & 1.0 & $0.06^{+0.02+0.91}_{-0.02-0.00}$ & 1.0 \\
 $\overline B^0\to K_1^-(1400) K^{*+}$  & $0.08^{+0.04+0.28}_{-0.03-0.00}$ & 1.0 & $0.09^{+0.05+0.30}_{-0.03-0.00}$ & 1.0 \\
 $\overline B^0\to K_1^+(1400) K^{*-}$  & $0.00^{+0.00+0.20}_{-0.00-0.00}$ & 1.0 & $0.00^{+0.01+0.69}_{-0.00-0.00}$ & 1.0 \\
 $\overline B^0\to \ov K_1^0(1270) K^{*0}$ & $0.40^{+0.25+0.65}_{-0.26-0.35}$ & $0.86^{+0.08}_{-0.25}$ & $0.29^{+0.19+0.41}_{-0.20-0.07}$ & $0.84^{+0.16}_{-0.59}$   \\
 $\overline B^0\to K_1^0(1270) \ov K^{*0}$ & $0.09^{+0.04+3.52}_{-0.03-0.00}$ & $0.34^{+0.62}_{-0.24}$ & $0.25^{+0.09+3.53}_{-0.08-0.00}$ & $0.52^{+0.44}_{-0.52}$  \\
 $B^-\to K_1^0(1270) K^{*-}$ & $0.05^{+0.04+2.21}_{-0.02-0.00}$ & $0.33^{+0.58}_{-0.06}$ & $0.13^{+0.04+2.00}_{-0.05-0.00}$ & $0.54^{+0.38}_{-0.49}$   \\
 $B^-\to K_1^-(1270) K^{*0}$ & $0.19^{+0.13+0.28}_{-0.13-0.15}$ & $0.84^{+0.07}_{-0.30}$ & $0.15^{+0.10+0.20}_{-0.10-0.03}$ & $0.80^{+0.08}_{-0.62}$  \\
 $\overline B^0\to \ov K_1^0(1400) K^{*0}$ & $0.01^{+0.01+3.41}_{-0.01-0.00}$ & $0.97^{+0.00}_{-0.97}$ & $0.08^{+0.05+3.01}_{-0.06-0.01}$ & $0.89^{+0.06}_{-0.79}$  \\
 $\overline B^0\to K_1^0(1400) \ov K^{*0}$ & $0.51^{+0.08+1.30}_{-0.11-0.29}$ & $0.71^{+0.28}_{-0.29}$ & $0.36^{+0.08+1.60}_{-0.01-0.30}$ & $0.85^{+0.11}_{-0.54}$   \\
 $B^-\to K_1^0(1400) K^{*-}$ & $0.28^{+0.04+0.53}_{-0.06-0.14}$ & $0.77^{+0.20}_{-0.34}$ & $0.19^{+0.04+0.73}_{-0.04-0.10}$ & $0.85^{+0.12}_{-0.18}$   \\
 $B^-\to K_1^-(1400) K^{*0}$ & $0.01^{+0.00+2.00}_{-0.01-0.00}$ & $0.93^{+0.00}_{-0.72}$ & $0.05^{+0.03+1.77}_{-0.04-0.01}$ & $0.92^{+0.05}_{-0.67}$   \\
\end{tabular}
\end{ruledtabular}
\end{table}

To obtain the branching ratios and $f_L$ listed in Table \ref{tab:BRK1V} for
$B\to K_1V$ decays, we have used the light-cone sum rule results for the $B\to
K_{1A}$ and $B\to K_{1B}$ form factors given in Table \ref{tab:FFinLCSR}. The
decays $B\to K_1\phi$ have been considered in \cite{ChenK1} based on the
generalized factorization framework where nonfactorizable effects are lumped
into $N_c^{\rm eff}$, the effective number of colors. It is interesting to note
that the results of \cite{ChenK1} for $B\to K_1(1270)\phi$ are similar to ours
when $N_c^{\rm eff}$ is close to 5, but the predicted rates for $K_1(1400)\phi$ are smaller than ours irrespective of the value of $N_c^{\rm eff}$. From Eqs. (\ref{eq:K1cphi}) and (\ref{eq:K1nphi}) we have the decay amplitudes given by
\be \label{eq:K1phi}
 \A_{B^-\to K_1^-(1270)\phi}^h &\propto & [\alpha_3^c+\alpha_4^c+\beta^c_3]X_h^{(\bar BK_1(1270), \phi)}  \non \\ &\propto&
 [\alpha_3^c+\alpha_4^c+\beta^c_3]_{_{ K_{1A}\phi}}F^{BK_{1A}}\sin\theta_{K_1}+
 [\alpha_3^c+\alpha_4^c+\beta^c_3]_{_{K_{1B}\phi}}F^{BK_{1B}} \cos\theta_{K_1}, \non \\
 \A_{B^-\to K_1^-(1400)\phi}^h &\propto & [\alpha_3^c+\alpha_4^c+\beta^c_3]X_h^{(\bar BK_1(1400), \phi)}   \\ &\propto&
 [\alpha_3^c+\alpha_4^c+\beta^c_3]_{_{ K_{1A}\phi}}F^{BK_{1A}}\cos\theta_{K_1}-
 [\alpha_3^c+\alpha_4^c+\beta^c_3]_{_{K_{1B}\phi}}F^{BK_{1B}} \sin\theta_{K_1}, \non
\en
where $F^{BK_{1A}}$ denotes generic form factors for the $B\to K_{1A}$ transition and likewise for $F^{BK_{1B}}$. In our convention, form factors $F^{BK_{1A}}$ and $F^{BK_{1B}}$ have opposite signs (see Table
\ref{tab:FFinLCSR}). Since the mixing angle $\theta_{K_1}$ is negative, it follows that the two amplitudes in Eq. (\ref{eq:K1phi}) contribute constructively to $B^-\to K_1(1270)^-\phi$ and destructively to $B^-\to K_1(1400)^-\phi$. Hence, it is naively expected that the former has a rate larger than the latter. Indeed, when the penguin annihilation ($\beta_3$) is turned off, we find $\B(B^-\to K_1(1270)^-\phi)\approx 3.2\times 10^{-6}\gg \B(B^-\to K_1(1400)^-\phi)\approx 3.1\times 10^{-7}$. However, this feature is dramatically changed in the presence of weak annihilation with $\rho_A=0.65$ and $\phi_A=-53^\circ$. Since $\beta_3(K_{1A}\phi)$ and $\beta_3(K_{1B}\phi)$ are opposite in sign, the interference  becomes destructive in $B^-\to K_1(1270)^-\phi$  and constructive in   $B^-\to K_1(1400)^-\phi$. This explains why we have $\B(B^-\to K_1(1270)^-\phi)<\B(B^-\to K_1(1400)^-\phi)$ in Table \ref{tab:BRK1V}. If this relation is not borne out by experiment, this will indicate that the parameter $\rho_A$ and hence weak annihilation are small.

The decays $B\to K_1(1270)\rho$ have rates larger than  that of $B\to
K_1(1400)\rho$ and this can be understood as follows. Their decay amplitudes
have the expressions, for example,
\be
 \A_{\ov B^0\to K_1^-(1270)\rho^+}^h &\propto & [\alpha_4^c+\beta^c_3]X_h^{(\bar B\rho, K_1(1270))}  \non \\ &\propto&
 [\alpha_4^c+\beta^c_3]_{_{\rho K_{1A}}}f_{K_{1A}}\sin\theta_{K_1}+
 [\alpha_4^c+\beta^c_3]_{_{\rho K_{1B}}}f^\bot_{K_{1B}} \cos\theta_{K_1}, \non \\
 \A_{\ov B^0\to K_1^-(1400)\rho^+}^h &\propto & [\alpha_4^c+\beta^c_3]X_h^{(\bar B\rho,K_1(1400))} \non \\  &\propto&
 [\alpha_4^c+\beta^c_3]_{_{\rho K_{1A}}}f_{K_{1A}}\cos\theta_{K_1}-[\alpha_4^c+\beta^c_3]_{_{\rho K_{1B}}}f_{K^\bot_{1B}}\sin\theta_{K_1}.
\en
Just as the case for $B\to K_1\phi$ decays, the interference is
constructive (destructive) in  $K_1^-(1270)\rho^+$ and destructive (constructive) in
$K_1^-(1400)\rho^+$ in the presence (absence) of weak annihilation with $\rho_A=0.65$ and $\phi_A=-53^\circ$. This explains why the rates of the former are larger than the latter,
especially for $\theta_{K_1}=-58^\circ$, in Table \ref{tab:BRK1V}. Hence, measurements of the relative rates of $K_1(1270)\rho$ and $K_1(1400)\rho$ will enable us to see the role played by the weak annihilation effect. If $\B(B\to K_1(1270)\rho)<\B(B\to K_1(1400)\rho)$ is observed, this will hint at the smallness of weak annihilation.
The reader may notice that the decay modes
involving $K_1(1270)$ and $K_1(1400)$ always have opposite dependence on the
mixing angle $\theta_{K_1}$. For example, $K_1^-(1270)\rho^+$ gets enhanced
whereas  $K_1^-(1400)\rho^+$ is suppressed when $\theta_{K_1}$ is changed from
$-37^\circ$ to $-58^\circ$.

Decay rates of $B\to K_1(1270)K^*$ and $K_1(1400)K^*$ are generally small
because they proceed through $b\to d$ penguin diagrams and are suppressed by
the smallness of the penguin Wilson coefficients. Their branching ratios  are
of order $10^{-7}-10^{-8}$. The decay modes $K_1^-K^{*+}$ and $K_1^+K^{*-}$ are
of particular interest as they are the only $AV$ modes which receive
contributions solely from weak annihilation. Just as the decay $\ov B^0\to
K^{*+}K^{*-}$ discussed before, the absence of transverse polarization in the
$K_1^-K^{*+}$ and $K_1^+K^{*-}$ modes is due to the fact that the annihilation
terms  $b_1^\pm,b_2^\pm,b_4^\pm,b_{\rm 4,EW}^\pm$ vanish under our
approximation.

From Table \ref{tab:BRK1V}, it is clear that the channels $K_1^-\rho^+$, $\bar
K_1^0\rho^-$ for $K_1=K_1(1270)$ and $K_1(1400)$ have sizable rates and the
experimental search of them would be encouraging.

\subsubsection{$B\to f_1V,~h_1V$ decays}

\begin{table}[t]
\caption{Same as Table \ref{tab:BRa1} except for the decays $B\to
(f_1,h_1)(\rho,\omega,K^*,\phi)$ with $f_1=f_1(1285),f_1(1420)$
and $h_1=h_1(1170),h_1(1380)$. We use two different sets of mixing
angles: (i) $\theta_{^3P_1}= 27.9^\circ$ and $\theta_{^1P_1}=
25.2^\circ$ (in first entry), corresponding to $\theta_{K_1}=-37^\circ$, and (ii)
$\theta_{^3P_1}= 53.2^\circ$, $\theta_{^1P_1}= 0^\circ$
(in second entry), corresponding to $\theta_{K_1}=-58^\circ$.}
\label{tab:BRf1}
\begin{ruledtabular}
\begin{tabular}{l c r |l  c r }
Mode & $\B$ & $f_L$ ~~~~ & Mode & $\B$ & $f_L$   \\
\hline
 $B^-\to f_1(1285)\rho^-$ & $\, 10.2^{+5.5+0.5}_{-3.5-0.4}$ & $0.91^{+0.02}_{-0.02}$ &
 $B^-\to f_1(1420)\rho^-$ & $0.1^{+0.2+0.1}_{-0.1-0.0}$ & $0.70^{+0.19}_{-0.07}$ \\
 &~~$8.9^{+5.1+0.4}_{-3.2-0.3}$ & $0.90^{+0.04}_{-0.03}$ &
 & $1.3^{+0.6+0.2}_{-0.3-0.0}$ & $0.93^{+0.04}_{-0.03}$  \\
 $\ov B^0\to f_1(1285)\rho^0$ & ~~$0.2^{+0.2+0.3}_{-0.1-0.0}$ & $0.77^{+0.09}_{-0.40}$ &
 $\ov B^0\to f_1(1420)\rho^0$ & $0.01^{+0.03+0.02}_{-0.00-0.00}$ & $0.38^{+0.57}_{-0.22}$ \\
 & ~~$0.2^{+0.1+0.3}_{-0.1-0.0}$ & $0.71^{+0.09}_{-0.36}$ &
 &  $0.04^{+0.12+0.08}_{-0.03-0.00}$ & $0.87^{+0.08}_{-0.40}$ \\
 $\ov B^0\to f_1(1285)\omega$ & ~~$1.0^{+1.1+2.5}_{-0.4-0.2}$ & $0.87^{+0.07}_{-0.62}$  &
 $\ov B^0\to f_1(1420)\omega$ \, & $0.02^{+0.03+0.05}_{-0.01-0.00}$ & $0.53^{+0.31}_{-0.39}$ \\
 &~ $0.9^{+1.0+2.2}_{-0.4-0.1}$ & $0.86^{+0.07}_{-0.62}$  &
 & $0.1^{+0.2+0.3}_{-0.1-0.0}$ & $0.86^{+0.04}_{-0.76}$   \\
 $B^-\to f_1(1285)K^{*-}$ & ~~~$5.8^{+8.3+10.8}_{-2.8-~2.3}$ & $0.90^{+0.11}_{-0.74}$ &
 $B^-\to f_1(1420)K^{*-}$ & $15.9^{+8.4+18.0}_{-5.3-~7.0}$ & $0.50^{+0.48}_{-0.52}$ \\
 & ~~~$5.7^{+3.8+21.4}_{-2.2-~4.8}$ & $0.47^{+0.49}_{-0.45}$ &
 & ~$15.6^{+10.9+10.4}_{-~5.2-~4.7}$ & $0.64^{+0.37}_{-0.61}$  \\
 $\ov B^0\to f_1(1285)\ov K^{*0}$ & ~~~$5.5^{+7.9+10.1}_{-2.7-~1.7}$ & $0.89^{+0.14}_{-0.79}$ &
 $\ov B^0\to f_1(1420)\ov K^{*0}$ & $14.8^{+8.0+17.4}_{-5.0-~6.7}$ & $0.49^{+0.49}_{-0.50}$  \\
 & ~~\, $5.1^{+3.6+20.0}_{-2.1-~4.7}$ & $0.45^{+0.55}_{-0.50}$ &
 & \, $14.9^{+10.2+10.1}_{-~5.0-~4.6}$ & $0.64^{+0.38}_{-0.61}$   \\
 $\ov B^0\to f_1(1285)\phi$ &~ $0.002^{+0.002+0.010}_{-0.001-0.000}$ & $0.93^{+0.02}_{-0.68}$ &
 $\ov B^0\to f_1(1420)\phi$ & \,$0.0001^{+0.0001+0.0006}_{-0.0000-0.0000}$ & $0.97^{+0.03}_{-0.75}$\\
 & ~~$0.002^{+0.002+0.009}_{-0.001-0.000}$ & $0.90^{+0.03}_{-0.71}$ &
 & \,$0.0008^{+0.0009+0.0009}_{-0.0001-0.0001}$ & $0.98^{+0.02}_{-0.44}$   \\
 $B^-\to h_1(1170)\rho^-$ & ~~$17.4^{+10.1+2.8}_{-~6.6-3.3}$ & $0.96^{+0.01}_{-0.06}$ &
 $B^-\to h_1(1380)\rho^-$ & \,$0.9^{+0.5+0.2}_{-0.3-0.2}$ & $0.95^{+0.00}_{-0.08}$ \\
 &  ~$10.9^{+6.5+1.8}_{-4.2-2.1}$ & $0.96^{+0.01}_{-0.06}$ &
 &  $5.9^{+3.2+1.0}_{-2.1-1.1}$ & $0.95^{+0.01}_{-0.07}$ \\
 $\ov B^0\to h_1(1170)\rho^0$ & ~ $0.05^{+0.11+1.16}_{-0.03-0.00}$ & $0.22^{+0.75}_{-0.11}$ &
 $\ov B^0\to h_1(1380)\rho^0$ & \,$0.01^{+0.01+0.04}_{-0.00-0.00}$ & $0.30^{+0.59}_{-0.13}$ \\
 & ~ $0.04^{+0.09+0.78}_{-0.02-0.00}$ & $0.30^{+0.61}_{-0.16}$ &
 & \,$0.02^{+0.03+0.40}_{-0.01-0.00}$ & $0.01^{+0.99}_{-0.00}$    \\
 $\ov B^0\to h_1(1170)\omega$ & ~~$1.5^{+2.4+1.2}_{-0.9-0.2}$ & $0.99^{+0.01}_{-0.10}$  &
 $\ov B^0\to h_1(1380)\omega$ & \,$0.1^{+0.1+0.1}_{-0.1-0.0}$ & $0.97^{+0.02}_{-0.14}$\\
 & ~~$0.9^{+1.4+0.8}_{-0.6-0.1}$ & $0.99^{+0.01}_{-0.10}$ &
 &  \,$0.5^{+0.7+0.4}_{-0.3-0.1}$ & $0.98^{+0.01}_{-0.12}$ \\
 $B^-\to h_1(1170)K^{*-}$ & ~~~$5.3^{+2.5+12.8}_{-1.6-~4.3}$ & $0.84^{+0.13}_{-0.16}$ &
 $B^-\to h_1(1380)K^{*-}$ &  \, $8.1^{+4.0+21.3}_{-2.8-~6.6}$ & $0.87^{+0.13}_{-0.75}$ \\
 & ~~ $7.7^{+5.1+31.6}_{-3.0-~7.1}$ & $0.81^{+0.17}_{-0.21}$ &
 & \,$3.7^{+2.0+7.8}_{-1.3-2.2}$ & $0.88^{+0.12}_{-0.53}$  \\
 $\ov B^0\to h_1(1170)\ov K^{*0}$ & ~~ $4.5^{+2.2+11.5}_{-1.4-~4.2}$ & $0.82^{+0.18}_{-0.40}$ &
 $\ov B^0\to h_1(1380)\ov K^{*0}$ & \, $8.3^{+4.4+21.8}_{-2.9-~6.9}$ & $0.88^{+0.12}_{-0.80}$  \\
 & ~~~$7.1^{+5.1+30.1}_{-2.9-~6.9}$ & $0.81^{+0.19}_{-0.42}$ &
 &  $3.9^{+1.9+8.3}_{-1.3-2.6}$ & $0.88^{+0.12}_{-0.64}$  \\
 $\ov B^0\to h_1(1170)\phi$ &~~ $0.006^{+0.007+0.010}_{-0.002-0.005}$ & $0.97^{+0.02}_{-0.90}$ &
 $\ov B^0\to h_1(1380)\phi$ & ~~ $0.004^{+0.003+0.230}_{-0.002-0.000}$ & $1.00^{+0.00}_{-0.04}$ \\
 &~~ $0.001^{+0.005+0.074}_{-0.003-0.000}$ & $0.93^{+0.07}_{-0.55}$ &
 & ~~ $0.007^{+0.005+0.170}_{-0.003-0.001}$ & $0.99^{+0.01}_{-0.14}$  \\
\end{tabular}
\end{ruledtabular}
\end{table}

Results for the decays $B\to (f_1,h_1)(\rho,\omega,K^*,\phi)$ with
$f_1=f_1(1285),f_1(1420)$ and $h_1=h_1(1170),h_1(1380)$ for two distinct sets
of the  mixing angles $\theta_{^3P_1}$ and $\theta_{^1P_1}$ are summarized in
Table \ref{tab:BRf1}. Among tree-dominated decays, the channels
$h_1(1380)\rho^-$ for $\theta_{^1P_1}=0^\circ$,  $f_1(1285)\rho^-$ and
$h_1(1170)\rho^-$ have branching ratios of order $10^{-5}$ as they receive
color-allowed tree contributions. Many of the penguin-dominated modes e.g.
$f_1(1420)K^*$ have branching ratios in the range of $(5\sim 15)\times
10^{-6}$. It is of interest to notice that the decays involving $h_1(1380)$ in
the final state have a sharp dependence of the rates on the mixing angle
$\theta_{^1P_1}$.

\subsection{$B\to AA$ decays}

For the axial vector mesons $a_1(1260),b_1(1235), f_1(1285),f_1(1420),
h_1(1170),h_1(1380)$ and $K_1(1270),K_1(1400)$, there exist many possible $B\to
AA$ two-body decay channels. We will classify them into tree- and
penguin-dominated decays. The latter involves the strange axial-vector meson
$K_1$.

\subsubsection{Tree-dominated decays}

\begin{table}[!]
\caption{Branching ratios (in units of $10^{-6}$) and the longitudinal
polarization fraction  for tree-dominated $B\to AA$ decays. For decays
involving $f_1$ and $h_1$ states, we use two different sets of mixing angles:
(i) $\theta_{^3P_1}= 27.9^\circ$ and $\theta_{^1P_1}= 25.2^\circ$ (in first
entry) and (ii) $\theta_{^3P_1}= 53.2^\circ$, $\theta_{^1P_1}= 0^\circ$ (in
second entry).} \label{tab:a1a1}
\begin{ruledtabular}
\begin{tabular}{l r r  |l  r r }
Mode & $\B$~~~ & $f_L$~~~~~  & Mode &$\B$~~~ & $f_L$~~~  \\
\hline
 $B^-\to a_1^-a_1^0$ & $22.4^{+10.7+6.6}_{-~8.2-1.5}$ & $0.74^{+0.24}_{-0.32}$~~ &
 $\ov B^0\to a_1^+a_1^-$ &  $37.4^{+16.1+9.7}_{-13.7-1.4}$ & $0.64^{+0.07}_{-0.17}$ \\
 $B^-\to a_1^-b_1^0$ &  $37.8^{+23.9+11.4}_{-15.3-~5.3}$ & $0.92^{+0.02}_{-0.24}$~~ &
 $\ov B^0\to a_1^0a_1^0$ & $0.5^{+0.8+9.3}_{-0.2-0.0}$ & $0.60^{+0.00}_{-0.70}$ \\
 $B^-\to a_1^0b_1^-$  & $1.0^{+1.6+6.2}_{-0.5-0.1}$ & $0.73^{+0.12}_{-0.82}$~~  &
 $\ov B^0\to a_1^-b_1^+$ & $41.3^{+20.7+16.6}_{-18.2-~3.4}$ & $0.90^{+0.02}_{-0.05}$
  \\
 $B^-\to b_1^-b_1^0$ & $1.4^{+2.5+2.8}_{-1.0-0.0}$ & $0.95^{+0.00}_{-0.82}$~~ &
 $\ov B^0\to a_1^+b_1^-$ & $0.8^{+1.09+3.6}_{-0.4-0.1}$ & $0.98^{+0.00}_{-0.80}$ \\
 $\ov B^0\to b_1^0b_1^0$ & $3.2^{+5.6+11.0}_{-2.3-~0.8}$ & $0.95^{+0.02}_{-0.80}$~~ &
 $\ov B^0\to a_1^0b_1^0$ & $3.8^{+6.2+2.6}_{-2.3-0.5}$ & $0.98^{+0.01}_{-0.31}$ \\
 $\ov B^0\to b_1^+b_1^-$ & $1.0^{+1.6+15.7}_{-0.7-~0.3}$ & $0.96^{+0.03}_{-0.65}$~~ & \\
 \hline
 $B^-\to a_1^-f_1(1285)$ & $12.4^{+5.6+6.9}_{-4.3-~0.7}$ & $0.73^{+0.22}_{-0.32}$~~ &
 $\ov B^0\to a_1^0f_1(1285)$ & $0.1^{+0.1+3.1}_{-0.1-0.0}$ & $0.53^{+0.13}_{-0.59}$ \\
 & $11.0^{+5.4+6.0}_{-4.1-0.8}$ & $0.71^{+0.23}_{-0.31}$~~ & & $0.1^{+0.1+2.7}_{-0.0-0.0}$ &
 $0.49^{+0.06}_{-0.52}$ \\
 $B^-\to a_1^-f_1(1420)$ & $0.2^{+0.2+0.3}_{-0.1-0.0}$ & $0.42^{+0.42}_{-0.19}$~~ &
 $\ov B^0\to  a_1^0f_1(1420)$ & $0.02^{+0.02+0.12}_{-0.01-0.01}$ & $0.14^{+0.75}_{-0.10}$ \\
 & $1.5^{+0.4+0.9}_{-0.3-0.0}$ & $0.77^{+0.16}_{-0.33}$~~ & & $0.02^{+0.02+0.12}_{-0.01-0.01}$ &
 $0.18^{+0.80}_{-0.15}$ \\
 $B^-\to a_1^-h_1(1170)$ & $22.4^{+14.5+5.3}_{-~9.3-3.1}$ & $0.91^{+0.02}_{-0.22}$~~ &
 $\ov B^0\to  a_1^0h_1(1170)$ & $0.1^{+0.3+2.1}_{-0.1-0.0}$ & $0.24^{+0.76}_{-0.26}$ \\
 & $14.1^{+9.5+3.4}_{-6.0-1.9}$~ & $0.91^{+0.02}_{-0.22}$~~ & & $0.08^{+0.17+1.36}_{-0.06-0.02}$
 & $0.30^{+0.70}_{-0.34}$ \\
 $B^-\to a_1^-h_1(1380)$ & $1.2^{+0.7+0.3}_{-0.5-0.1}$ & $0.90^{+0.02}_{-0.24}$~~ &
 $\ov B^0\to  a_1^0h_1(1380)$ & $0.01^{+0.01+0.12}_{-0.01-0.00}$ & $0.32^{+0.67}_{-0.21}$ \\
 & $7.6^{+4.7+1.6}_{-3.0-0.7}$ & $0.89^{+0.03}_{-0.24}$~~ & & $0.05^{+0.07+0.79}_{-0.03-0.00}$ &
 $0.08^{+0.92}_{-0.03}$ \\
 $B^-\to b_1^-f_1(1285)$ & $0.8^{+1.3+4.3}_{-0.5-0.3}$ & $0.82^{+0.16}_{-0.56}$~~ &
 $\ov B^0\to b_1^0f_1(1285)$ & $0.2^{+0.4+2.7}_{-0.1-0.1}$ & $0.48^{+0.53}_{-0.38}$ \\
 & $0.7^{+1.0+3.5}_{-0.4-0.1}$ & $0.79^{+0.19}_{-0.57}$~~ & & $0.2^{+0.3+2.5}_{-0.1-0.0}$ &
 $0.36^{+0.63}_{-0.22}$ \\
 $B^-\to b_1^-f_1(1420)$ & $0.03^{+0.17+0.16}_{-0.02-0.01}$ & $0.73^{+0.23}_{-0.33}$~~ &
 $\ov B^0\to b_1^0f_1(1420)$ & $0.01^{+0.08+0.10}_{-0.01-0.01}$ & $0.66^{+0.29}_{-0.49}$ \\
 & $0.2^{+0.6+0.8}_{-0.1-0.1}$ & $0.89^{+0.08}_{-0.51}$~~ & & $0.1^{+0.3+0.5}_{-0.1-0.0}$ &
 $0.81^{+0.14}_{-0.49}$ \\
 $B^-\to b_1^-h_1(1170)$ & $1.2^{+2.0+9.2}_{-0.9-~0.5}$ & $0.95^{+0.03}_{-0.76}$~~ &
 $\ov B^0\to b_1^0h_1(1170)$ & $0.2^{+0.2+5.1}_{-0.1-0.0}$ & $0.86^{+0.12}_{-0.79}$ \\
 & $0.8^{+1.3+3.0}_{-0.3-0.1}$~ & $0.94^{+0.03}_{-0.79}$~~ & & $0.1^{+0.2+3.4}_{-0.1-0.0}$ &
 $0.86^{+0.12}_{-0.79}$ \\
 $B^-\to b_1^-h_1(1380)$ & $0.05^{+0.10+0.34}_{-0.04-0.00}$ & $0.85^{+0.06}_{-0.80}$~~ &
 $\ov B^0\to b_1^0h_1(1380)$ & $0.01^{+0.01+0.20}_{-0.01-0.00}$ & $0.47^{+0.45}_{-0.46}$ \\
 & $0.3^{+0.5+3.0}_{-0.2-0.2}$ & $0.94^{+0.03}_{-0.77}$~~ & & $0.04^{+0.04+1.73}_{-0.02-0.00}$ &
 $0.81^{+0.17}_{-0.70}$ \\
 $\ov B^0\to f_1(1285)f_1(1285)$ & $0.3^{+0.3+3.1}_{-0.1-0.0}$ & $0.67^{+0.06}_{-0.84}$~~ &
 $\ov B^0\to h_1(1170)h_1(1170)$ & $0.8^{+1.6+1.6}_{-0.6-0.3}$ & $0.97^{+0.01}_{-0.55}$ \\
 & $0.2^{+0.2+2.5}_{-0.1-0.0}$ & $0.66^{+0.07}_{-0.84}$~~ & & $0.4^{+0.7+1.2}_{-0.3-0.1}$ & $0.97^{+0.02}_{-0.60}$ \\
 $\ov B^0\to f_1(1285)f_1(1420)$ & $0.01^{+0.01+0.10}_{-0.01-0.00}$ & $0.26^{+0.31}_{-0.28}$~~  &
 $\ov B^0\to h_1(1170)h_1(1380)$ & $0.1^{+0.1+0.1}_{-0.0-0.0}$ & $0.97^{+0.01}_{-0.74}$ \\
 & $0.05^{+0.05+0.63}_{-0.02-0.00}$ & $0.57^{+0.10}_{-0.66}$~~ & & $0.3^{+0.6+0.7}_{-0.3-0.1}$ &
 $0.96^{+0.01}_{-0.70}$ \\
 $\ov B^0\to f_1(1420)f_1(1420)$ & $0.01^{+0.00+0.03}_{-0.01-0.00}$ & $0.94^{+0.05}_{-0.34}$~~  &
 $\ov B^0\to h_1(1380)h_1(1380)$ & $0.01^{+0.01+0.42}_{-0.01-0.00}$ & $0.97^{+0.03}_{-0.39}$ \\
 & $0.01^{+0.01+0.06}_{-0.00-0.00}$ & $0.68^{+0.23}_{-0.58}$~~ & & $0.08^{+0.14+0.79}_{-0.05-0.02}$ &
 $0.96^{+0.03}_{-0.62}$ \\
 $\ov B^0\to f_1(1285)h_1(1170)$ & $1.1^{+1.9+1.1}_{-0.7-0.2}$ & $0.98^{+0.02}_{-0.07}$~~  &
 $\ov B^0\to f_1(1285)h_1(1380)$ & $0.05^{+0.09+0.07}_{-0.03-0.01}$ & $0.97^{+0.02}_{-0.22}$  \\
 & $0.6^{+1.1+0.6}_{-0.4-0.1}$ & $0.97^{+0.02}_{-0.09}$~~ & & $0.3^{+0.6+0.3}_{-0.2-0.0}$ &
 $0.97^{+0.03}_{-0.11}$\\
 $\ov B^0\to f_1(1420)h_1(1170)$ & $0.02^{+0.06+0.03}_{-0.01-0.00}$ & $0.87^{+0.11}_{-0.23}$~~ &
 $\ov B^0\to f_1(1420)h_1(1380)$ & $0.001^{+0.001+0.003}_{-0.000-0.000}$ & $0.74^{+0.22}_{-0.25}$  \\
 & $0.08^{+0.17+0.10}_{-0.04-0.01}$ & $0.96^{+0.03}_{-0.12}$~~ & & $0.04^{+0.10+0.08}_{-0.03-0.01}$ &
 $0.95^{+0.04}_{-0.12}$ \\
\end{tabular}
\end{ruledtabular}
\end{table}

The decay amplitudes for some of tree-dominated $B\to AA$ decays are shown in
Appendix B. Since the decay constant of the $b_1$ is either vanishing or very
small, it is anticipated that $b_1b_1$ channels are highly suppressed relative
to $a_1a_1$. Some of $a_1b_1$ decays are comparable to $a_1a_1$. We find that
$a_1^+a_1^-$ and $a_1^-a_1^0$ modes have rates larger than the corresponding
$\rho^+\rho^-$ and $\rho^-\rho^0$ ones, but $a_1^0a_1^0$ is very similar to
$\rho^0\rho^0$. While $a_1^+a_1^-$, $a_1^-a_1^0$, $a_1^-b_1^+$ and $a_1^-b_1^0$
modes have branching ratios of order $(20\sim 40)\times 10^{-6}$, the other
channels are suppressed by the smallness of either $f_{b_1}$ or the coefficient
$a_2$.

Among various $B\to (a_1,b_1)(f_1,h_1)$ decays, we see from Table
\ref{tab:a1a1} that only  $a_1^-f_1(1285)$ and $a_1^-h_1(1170)$ modes and
$a_1^-h_1(1380)$  with $\theta_{^1P_1}=0^\circ$ can have sizable rates and all
other charged and neutral channels are suppressed.

\begin{table}[!]
\caption{Branching ratios (in units of $10^{-6}$) and the longitudinal
polarization fraction  for penguin-dominated $B\to K_1A$ decays for the mixing
angles: (i) $\theta_{K_1}=-37^\circ$, $\theta_{^3P_1}= 27.9^\circ$ and
$\theta_{^1P_1}= 25.2^\circ$, and (ii) $\theta_{K_1}=-58^\circ$,
$\theta_{^3P_1}= 53.2^\circ$ and $\theta_{^1P_1}= 0^\circ$.
  Default results are for $\rho_A=0.65$ and $\phi_A=-53^\circ$.}
\label{tab:BRK1A}
\begin{ruledtabular}
\begin{tabular}{l l c l c }
 &  \multicolumn{2}{c}{$\theta_{K_1}=-37^\circ$}
 &   \multicolumn{2}{c}{$\theta_{K_1}=-58^\circ$} \\
\cline{2-3} \cline{4-5}
\raisebox{2.0ex}[0cm][0cm]{Decay} & $\B$ & $f_L$ & $\B$ &
$f_L$  \\ \hline
 $\ov B^0\to K_1^-(1270)a_1^+$ & $42.3^{+58.2+165.6}_{-27.3-~41.1}$ &
  $0.24^{+0.16}_{-0.07}$ & $46.1^{+60.9+176.8}_{-29.8-~43.7}$ & $0.16^{+0.28}_{-0.06}$ \\
 $\ov B^0\to \bar K_1^0(1270)a_1^0$ & $21.6^{+29.7+81.2}_{-13.9-20.9}$
  & $0.27^{+0.69}_{-0.21}$ & $22.4^{+30.8+88.7}_{-15.0-22.2}$ & $0.15^{+0.28}_{-0.03}$ \\
 $B^-\to \bar K_1^0(1270)a_1^-$ & $44.3^{+60.9+165.9}_{-28.6-~43.5}$
  & $0.23^{+0.09}_{-0.19}$ & $48.3^{+63.4+175.7}_{-31.1-~47.0}$ & $0.15^{+0.25}_{-0.14}$ \\
 $B^-\to K_1^-(1270)a_1^0$ & $23.8^{+30.1+84.7}_{-14.3-21.4}$ &
  $0.26^{+0.35}_{-0.15}$ & $26.3^{+31.6+87.6}_{-15.6-23.2}$ & $0.20^{+0.41}_{-0.11}$ \\
 $\ov B^0\to K_1^-(1400)a_1^+$ & $12.0^{+10.6+12.0}_{-~6.1-~7.9}$ &
  $0.36^{+0.38}_{-0.32}$ & \, $7.5^{+7.3+21.7}_{-3.3-~6.1}$ & $0.96^{+0.04}_{-0.43}$ \\
 $\ov B^0\to \bar K_1^0(1400)a_1^0$ & \, $6.7^{+5.5+7.4}_{-3.3-4.5}$ &
  $0.45^{+0.48}_{-0.41}$ & \, $5.5^{+4.2+12.4}_{-2.3-~4.1}$ & $0.98^{+0.13}_{-0.36}$ \\
 $B^-\to \bar K_1^0(1400)a_1^-$  & $13.8^{+11.4+17.1}_{-~6.8-10.1}$ &
  $0.42^{+0.58}_{-0.40}$ & \, $9.0^{+8.4+23.8}_{-3.9-~7.4}$ & $0.98^{+0.15}_{-0.43}$\\
 $B^-\to K_1^-(1400)a_1^0$ & \, $6.0^{+5.4+7.0}_{-3.0-4.3}$ &
  $0.33^{+0.35}_{-0.30}$ & \, $3.1^{+3.6+10.2}_{-1.5-~2.6}$ & $0.94^{+0.01}_{-0.59}$ \\
 $\ov B^0\to K_1^-(1270)b_1^+$ & $14.8^{+12.9+65.8}_{-~7.2-13.7}$ &
  $0.28^{+0.52}_{-0.12}$ & $14.1^{+15.7+63.1}_{-~6.7-11.1}$ & $0.13^{+0.71}_{-0.14}$\\
 $\ov B^0\to \bar K_1^0(1270)b_1^0$ & \, $7.3^{+6.3+33.0}_{-3.5-~6.7}$ &
  $0.29^{+0.48}_{-0.18}$ & \, $6.9^{+7.4+34.1}_{-3.1-~5.2}$ & $0.12^{+0.69}_{-0.16}$ \\
 $B^-\to \bar K_1^0(1270)b_1^-$ & $15.3^{+13.8+72.6}_{-~7.5-14.9}$ & $0.31^{+0.31}_{-0.13}$ & $13.0^{+15.1+69.5}_{-~6.0-10.8}$ & $0.06^{+0.67}_{-0.09}$ \\
 $B^-\to K_1^-(1270)b_1^0$ & \, $9.0^{+8.3+36.3}_{-4.5-~7.7}$ &
  $0.39^{+0.49}_{-0.20}$ & \, $8.1^{+9.2+32.2}_{-3.9-~5.9}$ & $0.22^{+0.65}_{-0.22}$ \\
 $\ov B^0\to K_1^-(1400)b_1^+$ & $25.0^{+27.3+205.5}_{-11.1-~22.9}$ &
  $0.91^{+0.03}_{-0.34}$ & $26.2^{+24.2+209.2}_{-12.0-~24.7}$ & $0.99^{+0.01}_{-0.56}$ \\
 $\ov B^0\to \bar K_1^0(1400)b_1^0$  & $13.0^{+13.4+110.1}_{-~5.6-~12.3}$ &
  $0.91^{+0.05}_{-0.66}$ & $13.7^{+12.3+109.4}_{-~6.2-~13.1}$ & $0.98^{+0.02}_{-0.92}$  \\
 $B^-\to \bar K_1^0(1400)b_1^-$ & $27.7^{+28.6+231.4}_{-12.0-~26.1}$ &
  $0.91^{+0.05}_{-0.66}$ & $30.4^{+27.2+235.5}_{-13.9-~29.3}$ & $0.98^{+0.02}_{-0.97}$ \\
 $B^-\to K_1^-(1400)b_1^0$ & $13.3^{+14.4+107.9}_{-~5.9-~12.2}$ &
  $0.92^{+0.01}_{-0.35}$ & $14.4^{+13.4+112.7}_{-~6.7-~13.8}$ & $0.99^{+0.01}_{-0.74}$ \\
 \hline
 $\ov B^0\to \bar K_1^0(1270)f_1(1285)$ & $14.5^{+20.1+68.4}_{-~8.2-11.1}$ &
  $0.56^{+0.52}_{-0.70}$ & \, $5.7^{+8.8+31.0}_{-3.7-~5.2}$ & $0.22^{+0.72}_{-0.29}$ \\
 $\ov B^0\to \bar K_1^0(1270)f_1(1420)$ & \, $10.4^{+9.3+51.7}_{-3.5-~6.3}$ &
  $0.93^{+0.08}_{-0.55}$ & $18.9^{+20.9+82.1}_{-~8.0-10.7}$ & $0.64^{+0.48}_{-0.67}$ \\
 $\ov B^0\to \bar K_1^0(1270)h_1(1170)$ & \, $5.3^{+8.7+25.7}_{-3.4-~4.4}$ &
  $0.52^{+0.46}_{-0.60}$ & \, $4.1^{+11.3+7.8}_{-~3.4-2.3}$ & $0.83^{+0.17}_{-0.83}$ \\
 $\ov B^0\to  \bar K_1^0(1270)h_1(1380)$ & \, $8.5^{+13.3+38.4}_{-~5.6-~5.2}$ &
  $0.93^{+0.05}_{-0.76}$ & \, $8.5^{+11.5+33.5}_{-~4.9-~4.3}$ & $0.50^{+0.45}_{-0.45}$ \\
 $B^-\to K_1^-(1270)f_1(1285)$ & $15.7^{+21.5+73.4}_{-~8.7-11.7}$ &
  $0.60^{+0.46}_{-0.75}$ & \, $6.2^{+9.0+34.2}_{-3.8-~5.2}$ & $0.29^{+0.60}_{-0.35}$ \\
 $B^-\to K_1^-(1270)f_1(1420)$ & $10.9^{+9.7+53.7}_{-3.7-~6.4}$ &
  $0.93^{+0.09}_{-0.55}$ & $19.7^{+21.8+85.6}_{-~8.2-11.0}$ & $0.65^{+0.46}_{-0.69}$ \\
 $B^-\to K_1^-(1270)h_1(1170)$ & \, $6.5^{+9.3+27.4}_{-3.9-~4.9}$ & $0.56^{+0.38}_{-0.62}$ & \, $5.6^{+12.2+9.2}_{-~4.0-0.8}$ & $0.85^{+0.13}_{-0.76}$\\
 $B^-\to  K_1^-(1270)h_1(1380)$ & \, $9.2^{+14.7+40.4}_{-~6.1-~5.8}$ &
  $0.93^{+0.05}_{-0.73}$ & \, $8.9^{+12.3+33.5}_{-~5.2-~4.4}$ & $0.50^{+0.25}_{-0.73}$  \\
 $\ov B^0\to \bar K_1^0(1400)f_1(1285)$ & \, $4.0^{+5.3+12.2}_{-2.1-~4.5}$ &
  $0.12^{+0.48}_{-0.13}$ &  \, $2.2^{+2.2+10.7}_{-0.5-~1.7}$ & $0.52^{+0.39}_{-0.56}$ \\
 $\ov B^0\to \bar K_1^0(1400)f_1(1420)$ & $18.9^{+25.4+87.6}_{-16.7-19.0}$ &
  $0.04^{+0.95}_{-0.03}$  &  $21.5^{+26.2+95.4}_{-13.5-17.6}$ & $0.26^{+0.74}_{-0.16}$ \\
 $\ov B^0\to \bar K_1^0(1400)h_1(1170)$ & \, $8.9^{+10.4+75.4}_{-~4.5-~8.3}$ &
  $0.96^{+0.02}_{-0.76}$ & $21.0^{+21.6+148.3}_{-11.4-~19.7}$ & $0.90^{+0.09}_{-0.81}$\\
 $\ov B^0\to  \bar K_1^0(1400)h_1(1380)$ & $16.6^{+22.0+81.9}_{-10.1-16.9}$ &
  $0.55^{+0.10}_{-0.11}$ & \, $6.5^{+8.4+33.2}_{-3.9-~6.5}$ & $0.39^{+0.16}_{-0.08}$ \\
 $B^-\to K_1^-(1400)f_1(1285)$ & \,  $4.3^{+5.9+13.2}_{-2.4-~4.4}$ &
  $0.15^{+0.57}_{-0.10}$ & \, $2.2^{+2.5+10.4}_{-0.5-~1.7}$ & $0.50^{+0.25}_{-0.73}$  \\
 $B^-\to K_1^-(1400)f_1(1420)$ & $19.4^{+26.2+89.7}_{-13.0-19.8}$ &
  $0.03^{+0.95}_{-0.03}$ & $22.5^{+27.3+99.9}_{-14.1-18.1}$ & $0.27^{+0.73}_{-0.15}$   \\
 $B^-\to K_1^-(1400)h_1(1170)$ & \, $9.1^{+11.1+74.5}_{-~4.6-~8.3}$ &
  $0.96^{+0.02}_{-0.44}$ & $22.0^{+23.2+154.5}_{-12.1-~20.8}$ & $0.91^{+0.08}_{-0.83}$ \\
 $B^-\to  K_1^-(1400)h_1(1380)$ & $17.7^{+23.4+87.8}_{-10.8-18.0}$ &
  $0.56^{+0.12}_{-0.09}$ & \,  $7.0^{+9.1+35.9}_{-4.2-~6.9}$ & $0.42^{+0.45}_{-0.18}$ \\
\end{tabular}
\end{ruledtabular}
\end{table}

\begin{table}[h]
{Table XI. (Continued)}
\begin{ruledtabular}
\begin{tabular}{l l c l c }
 &  \multicolumn{2}{c}{$\theta_{K_1}=-37^\circ$}
 &   \multicolumn{2}{c}{$\theta_{K_1}=-58^\circ$} \\
\cline{2-3} \cline{4-5}
\raisebox{2.0ex}[0cm][0cm]{Decay} & $\B$ & $f_L$ & $\B$ &
$f_L$  \\ \hline
 $\ov B^0\to K_1^-(1270)K_1^+(1270)$ & $0.07^{+0.12+1.07}_{-0.05-0.00}$ & 1 & $0.10^{+0.10+1.23}_{-0.06-0.00}$ & 1 \\
 $\ov B^0\to K_1^-(1270)K_1^+(1400)$ & $0.01^{+0.02+0.35}_{-0.01-0.00}$ & 1 & $0.01^{+0.02+0.34}_{-0.01-0.00}$ & 1 \\
 $\ov B^0\to K_1^-(1400)K_1^+(1270)$ & $0.20^{+0.20+3.72}_{-0.11-0.00}$ & 1 & $0.23^{+0.24+3.03}_{-0.17-0.00}$ & 1\\
 $\ov B^0\to K_1^-(1400)K_1^+(1400)$ & $0.04^{+0.05+0.42}_{-0.02-0.00}$ & 1 & $0.04^{+0.08+0.63}_{-0.03-0.00}$ & 1 \\
 $\ov B^0\to \bar K_1^0(1270)K_1^0(1270)$ & $0.16^{+0.17+1.03}_{-0.12-0.01}$ & $0.67^{+0.31}_{-0.62}$ &
  $0.44^{+0.33+8.35}_{-0.31-0.00}$ & $0.75^{+0.21}_{-0.73}$ \\
 $\ov B^0\to \bar K_1^0(1400)K_1^0(1400)$ & $0.29^{+0.23+4.99}_{-0.21-0.01}$ & $0.64^{+0.32}_{-0.72}$ &
  $0.06^{+0.06+0.13}_{-0.01-0.0}$ & $0.46^{+0.57}_{-0.29}$ \\
 $\ov B^0\to \bar K_1^0(1270)K_1^0(1400)$ & $0.85^{+0.44+10.95}_{-0.41-~0.38}$ & $0.77^{+0.25}_{-0.59}$ &
  $0.54^{+0.37+4.17}_{-0.27-0.23}$ & $0.83^{+0.18}_{-0.88}$ \\
 $\ov B^0\to K_1^0(1270)\bar K_1^0(1400)$ & $0.00^{+0.01+19.0}_{-0.00-~0.0}$ & $0.66^{+0.26}_{-0.70}$ &
  $0.02^{+0.02+14.12}_{-0.01-~0.0}$ & $0.40^{+0.39}_{-0.38}$ \\
 $B^-\to K_1^0(1270)K_1^-(1400)$ &$0.03^{+0.02+17.1}_{-0.01-~0.0}$ & $0.94^{+0.03}_{-0.86}$ &
  $0.10^{+0.05+13.07}_{-0.07-~0.00}$ & $0.91^{+0.04}_{-0.91}$ \\
 $B^-\to K_1^-(1270)K_1^0(1270)$ & $0.13^{+0.11+0.43}_{-0.07-0.02}$ & $0.51^{+0.23}_{-0.40}$ &
  $0.23^{+0.17+4.28}_{-0.13-0.02}$ & $0.42^{+0.26}_{-0.51}$ \\
 $B^-\to K_1^-(1270)K_1^0(1400)$ & $0.79^{+0.42+8.01}_{-0.39-0.29}$ & $0.74^{+0.28}_{-0.53}$ &
  $0.46^{+0.30+2.29}_{-0.21-0.13}$ & $0.79^{+0.20}_{-0.84}$ \\
 $B^-\to K_1^-(1400)K_1^0(1400)$ & $0.21^{+0.16+3.16}_{-0.12-0.02}$ & $0.41^{+0.28}_{-0.47}$ &
  $0.11^{+0.10+0.36}_{-0.05-0.04}$ & $0.61^{+0.34}_{-0.49}$\\
\end{tabular}
\end{ruledtabular}
\end{table}

\subsubsection{Penguin-dominated decays}
The penguin-dominated $B\to AA$ decays involve at least one $K_1$ meson.
Results for the decay modes $K_1(a_1,b_1,f_1,h_1,K_1)$ are summarized in Table
\ref{tab:BRK1A}. Some salient features are (i) $\Gamma(B\to
K_1(1270)a_1)>\Gamma(B\to K_1(1400)b_1)>\Gamma(B\to K_1(1270)b_1)>\Gamma(B\to
K_1(1400)a_1)$, (ii) $B\to K_1(a_1,b_1)$ decays are dominated by transverse
polarization amplitudes except for $K_1(1400)b_1$ and $K_1(1400)a_1$ with
$\theta_{K_1}=-58^\circ$, and (iii)  the charged and neutral $B$ decays have
similar rates and longitudinal polarization fractions. For example, $\B(B^-\to
K_1^-(f_1,h_1))\approx \B(\ov B^0\to \bar K_1(f_1,h_1))$, $\B(B^-\to
K_1^-(a_1^0,b_1^0))\approx \B(\ov B^0\to \bar K_1^0(a_1^0,b_1^0))$ and
$\B(B^-\to \bar K_1^0(a_1^-,b_1^-))\approx \B(\ov B^0\to  K_1^-(a_1^+,b_1^+))$.

The first feature can be understood as follows. Consider the decays  $\ov
B^0\to K_1^-(a_1^+,b_1^+)$ as an illustration. Their decay amplitudes are given
by
\be
 {\cal A}_{\ov B^0\to K^-_1(1270) a_1^+}^h &\propto&
  \Big[  \alpha_4^{c,h} + \alpha_{4,{\rm
    EW}}^{c,h} +\beta_3^{c,h} -{1\over 2}\beta^{c,h}_{3,{\rm EW}} \Big]_{a_1K_1}
 X^{(\overline Ba_1, \ov K_1)}_h  \non \\
  &\propto& [\cdots]_{_{a_1K_{1A}}}f_{K_{1A}}\sin\theta_{K_1}+
   [\cdots]_{_{a_1K_{1B}}}f_{K_{1B}}^\bot  \cos\theta_{K_1}, \non \\
 {\cal A}_{\ov B^0\to K^-_1(1400) a_1^+}^h
  &\propto& [\cdots]_{_{a_1K_{1A}}}f_{K_{1A}}\cos\theta_{K_1}-
   [\cdots]_{_{a_1K_{1B}}}f_{K_{1B}}^\bot \sin\theta_{K_1}, \non \\
 {\cal A}_{\ov B^0\to K^-_1(1270) b_1^+}^h
  &\propto& [\cdots]_{_{b_1K_{1A}}}f_{K_{1A}}\sin\theta_{K_1}+
   [\cdots]_{_{b_1K_{1B}}}f_{K_{1B}}^\bot \cos\theta_{K_1}, \non \\
 {\cal A}_{\ov B^0\to K^-_1(1400) b_1^+}^h
  &\propto& [\cdots]_{_{b_1K_{1A}}}f_{K_{1A}}\cos\theta_{K_1}-
   [\cdots]_{_{b_1K_{1B}}}f_{K_{1B}}^\bot \sin\theta_{K_1}.
\en Since $K_1a_1$ modes are dominated by transverse amplitudes and since the
negative-helicity parameters such as $\alpha_4^-(a_1K_{1A})$ and
$\alpha_4^-(a_1K_{1B})$ have opposite signs, it is clear that the interference
is constructive in $B\to K_1(1270)a_1$ and destructive in $B\to K_1(1400)a_1$
for a negative mixing angle $\theta_{K_1}$. This also explains why the former
increases and the latter decreases when the mixing angle is changed from
$-37^\circ$ to $-58^\circ$. For the $K_1(1400)b_1$ modes dominated by the
longitudinal amplitudes, they have large rates as $\alpha_4^0(b_1K_{1A})$ and
$\alpha_4^0(b_1K_{1B})$ are of the same sign. In general, $K_1a_1$ and $K_1b_1$
rates are insensitive to the value of $\theta_{K_1}$, $-37^\circ$ or
$-58^\circ$,  except for $K_1(1400)a_1$ modes. It is interesting to notice that
$K_1(1400)a_1$ channels are dominated by transverse amplitudes for
$\theta_{K_1}=-37^\circ$ and by longitudinal ones for $\theta_{K_1}=-58^\circ$.
Therefore, measurement of polarization fractions in $B\to K_1(1400)a_1$ will
yield a clear discrimination between the two different $K_{1A}\!-K_{1B}$ mixing
angles. Branching ratios of $B\to K_1(f_1,h_1)$ fall into the range of
$10^{-6}\sim 10^{-5}$. At first sight, it appears that they depend on the
mixing angles $\theta_{K_1}$ and $\theta_{^1P_1}$ or $\theta_{^3P_1}$. However,
the latter two angles are correlated to the first one [see Eq.
(\ref{eq:mixingangle})]. Consequently, the decays $B\to K_1(f_1,h_1)$ depend on
only one mixing angle $\theta_{K_1}$. From Table \ref{tab:BRK1A} it is clear
that the mode $\bar K_1(1400)h_1(1380)$ has a strong dependence on
$\theta_{K_1}$.

Just as $B\to K_1\bar K^*$ decays, branching ratios of the $b\to d$
penguin-dominated $K_1\bar K_1$ modes are small, of order $10^{-7}-10^{-8}$
owing to the smallness of the penguin coefficients.

In short, there are many penguin-dominated $B\to AA$ decays within the reach of
$B$ factories: $K_1(1270)a_1$, $K_1(1400)b_1$, $K_1(1270)b_1^\pm$,
$K_1(1400)a_1^\pm$, $K_1(1270)(f_1(1285),f_1(1420))$ and
$K_1(1400)(f_1(1420),h_1(1170))$. In most cases, transverse polarization is
large except for $K_1(1400)(b_1,h_1(1170))$, $K_1(1270)(f_1(1420),h_1(1380))$
with $\theta_{K_1}=-37^\circ$  and $K_1(1400)a_1$ with $\theta_{K_1}=-58^\circ$
where longitudinal polarization dominates.

\section{Conclusions}
In this work we have presented a detailed study of charmless two-body $B$
decays into final states involving two vector mesons ($VV$) or two axial-vector
mesons ($AA$) or
 one vector and one axial-vector meson ($VA$),  within the framework of QCD factorization,
where $A$ is either a $^3P_1$ or $^1P_1$ axial-vector meson.
Owing to the $G$-parity,
the chiral-even two-parton light-cone distribution amplitudes of
the $^3P_1$ ($^1P_1$) mesons are symmetric (antisymmetric) under
the exchange of quark and anti-quark momentum fractions in the
SU(3) limit. For chiral-odd light-cone distribution amplitudes, it
is other way around.
The main results are as follows.

\begin{itemize}

\item  We have worked out the hard spectator scattering and annihilation contributions
to $B\to VA$ and $B\to AA$ decays.

\item
NLO nonfactorizable corrections to longitudinal- and negative-helicity
effective Wilson coefficients $a_i^h$ generally differ in magnitude and even in
sign. For some $VV$ modes, the constructive (destructive) interference in the
negative-helicity (longitudinal-helicity) amplitude of the $\ov B\to VV$ decay
will render the former comparable to the latter and bring up the transverse
polarization. Any serious solution to the polarization puzzle should take into
account NLO effects on $a_i^h$.

\item
The measured rates and $f_L$ of penguin-dominated charmless $VV$ modes can be
{\it accommodated} (but cannot be {\it predicted} at first place) in QCD
factorization by allowing for sizable penguin annihilation contributions.
However, the parameters $\rho_A$ and $\phi_A$ fit to the data of $K^*\phi$ and
$K^*\rho$ are not the same.
%and this might be a potential problem for QCD factorization.
Hence, we do not have a good hint at the values of
$\rho_A$ and $\phi_A$ for $B\to AV$ and $B\to AA$ decays.

\item
While NLO contributions due to vertex, penguin and hard scattering corrections
are insensitive to the choice of the renormalization scale $\mu$, the penguin
annihilation contribution at the hard-collinear scale is sensitive to $\mu$. In
the present work, we choose $\mu=m_b(m_b)$ for the reason that if
$\mu=m_b(m_b)/2$ is selected, the decay rates and polarization fractions of
$B\to K^*\phi$ or $B\to K^*\rho$ cannot be simultaneously fitted by the
annihilation parameters $\rho_A$ and $\phi_A$.

\item
The predicted rates and longitudinal polarization fractions by QCD
factorization for tree-dominated $\rho\rho$ modes are in good agreement with
experiment, but the calculated $\B(B^-\to \rho^-\omega)$ is slightly high. The
latter may imply that the $B\to\omega$ transition form factors are slightly
smaller than what are expected from the light-cone sum rules. Only in the decay
$B^0\to \rho^0\omega$ where a large deviation from the naive expectation of
$f_L\sim 1$ is possible. We found $f_L(\rho^0\omega)\sim 0.55$.

\item
Using the measured $\bar K^{*0}\rho^-$ channel as an input, we predict the
branching ratios and polarization fractions for other $B\to K^*\rho$ decays and
find the relation $f_L(K^{*-}\rho^0)> f_L(K^{*-}\rho^+)> f_L(\bar
K^{*0}\rho^-)> f_L(\bar K^{*0}\rho^0)$. Experimentally, it is quite important
to measure them to test theory. Our result of $f_L(\bar K^{*0}\rho^0)\sim 0.39$
is consistent with experiment and is higher than the prediction  $\sim 0.22$
made by Beneke, Rohrer and Yang.

\item
The calculations suggest that the tree-dominated channels $a_1^+\rho^-$,
$a_1^-\rho^-$, $a_1^0\rho^-$, $a_1^-\rho^0$, $a_1^-\omega$, $b_1^+\rho^-$ and
$b_1^0\rho^-$ should be readily accessible to $B$ factories. One of the salient
features of the $^1P_1$ axial vector meson is that its axial-vector decay
constant is small, vanishing in the SU(3) limit. This can be tested by
measuring various $b_1\rho$ modes to see if $\Gamma(\ov B^0\to b_1^-\rho^+)\ll
\Gamma(\ov B^0\to b_1^+\rho^-)$ and $\Gamma(B^-\to b_1^-\rho^0)\ll
\Gamma(B^-\to b_1^0\rho^-)$.

\item
In the absence of the experimental guideline, we employed the penguin
annihilation parameters $\rho_A=0.65$ and $\phi_A=-53^\circ$ inferred from the
channel $B\to K^*\phi$  to describe penguin-dominated $B\to VA,AA$ decays. It
is very crucial to measure the penguin-dominated modes $a_1K^*$ and $b_1K^*$ to
test the importance of penguin annihilation. We found that $a_1K^*$ modes are
dominated by transverse polarization amplitudes, whereas $b_1K^*$ are governed
by longitudinal polarization states.

\item
For $B\to K_1V$ decays involving the $K_1(1270)$ or $K_1(1400)$ meson, the
channels $K_1^-\rho^+$, $\bar K_1^0\rho^-$  have sizable rates and the
experimental search of them would be encouraging.
Measurements of the relative strengths of $K_1(1270)\phi(\rho)$ and $K_1(1400)\phi(\rho)$ will enable us to test the importance of weak annihilation. The rates of $B\to
K_1(1270)K^*$ and $K_1(1400)K^*$ are generally very small. The decay modes
$K_1^-K^{*+}$ and $K_1^+K^{*-}$ are of particular interest as they are the only
$VA$ modes which receive contributions solely from weak annihilation.

\item
Among the decays $B\to (f_1,h_1)(\rho,\omega,K^*,\phi)$ with
$f_1=f_1(1285),f_1(1420)$ and $h_1=h_1(1170),h_1(1380)$,  the tree-dominated
modes $h_1(1380)\rho^-$,  $f_1(1285)\rho^-$, $h_1(1170)\rho^-$ and several of
the penguin-dominated channels e.g. $f_1(1420)K^*$ have appreciable rates.

\item
For tree-dominated $B\to AA$ decays, the $a_1^+a_1^-$, $a_1^-a_1^0$,
$a_1^-b_1^+$ and $a_1^-b_1^0$ modes have sizable branching ratios, of order
$(20\sim 40)\times 10^{-6}$. Among various $B\to (a_1,b_1)(f_1,h_1)$ decays,
only  $a_1^-f_1(1285)$ and $a_1^-h_1(1170)$ modes and $a_1^-h_1(1380)$  with
$\theta_{^1P_1}=0^\circ$ can have large rates and all other charged and neutral
channels are suppressed.

\item
There are two salient features for penguin-dominated $B\to AA$ decays: (i)
$\Gamma(B\to K_1(1270)a_1)>\Gamma(B\to K_1(1400)b_1)>\Gamma(B\to
K_1(1270)b_1)>\Gamma(B\to K_1(1400)a_1)$ and (ii) most of them are dominated by
transverse polarization amplitudes except for $K_1(1400)b_1$ and $K_1(1400)a_1$
with $\theta_{K_1}=-58^\circ$. Since the $K_1(1400)a_1$ channels are dominated
by transverse amplitudes for $\theta_{K_1}=-37^\circ$ and by longitudinal ones
for $\theta_{K_1}=-58^\circ$, measurement of polarization fractions in $B\to
K_1(1400)a_1$ will yield a clear discrimination between the two different
$K_{1A}\!-K_{1B}$ mixing angles. Many penguin-dominated $B\to AA$ decays are
are readily detectable at $B$ factories: $K_1(1270)a_1$, $K_1(1400)b_1$,
$K_1(1270)b_1^\pm$, $K_1(1400)a_1^\pm$, $K_1(1270)(f_1(1285),f_1(1420))$ and
$K_1(1400)(f_1(1420),h_1(1170))$.

\end{itemize}

\vskip 2.5cm \acknowledgments We are grateful to Hsiang-nan Li, James Smith and
Deshan Yang for valuable discussions and to Wei Wang for providing us Table
III.
 This research was supported in part by the
National Science Council of R.O.C. under Grant Nos.
NSC96-2112-M-001-003 and NSC96-2112-M-033-004-MY3.

\newpage
\appendix

\section{Flavor Operators}

The  coefficients of the flavor operators $\alpha_i^{h,p}$ can be
expressed in terms of $a_i^{h,p}$~\cite{BBNS,BN} as
follows:\footnote{The numerical values of the coefficients
$\alpha_i(M_1 M_2)$ also depend on the nature of the initial-state
$B$ meson. This dependence is not indicated explicitly in our
notation. The same remark applies to the annihilation coefficients
$b_i^p$ defined below.}
\begin{eqnarray}\label{eq:alphai}
   \alpha_1^{h}(M_1 M_2) &=& a_1^{h}(M_1 M_2) \,, \nonumber\\
   \alpha_2^{h}(M_1 M_2) &=& a_2^{h}(M_1 M_2) \,, \nonumber\\
   \alpha_3^{h,p}(M_1 M_2) &=& \left\{
    \begin{array}{cl}
     a_3^{h,p}(M_1 M_2) - a_5^{h,p}(M_1 M_2)
      & \quad \mbox{for~} M_1 M_2=VA, \, AA , \\
     a_3^{h,p}(M_1 M_2) + a_5^{h,p}(M_1 M_2)
      & \quad \mbox{for~} M_1 M_2=VV,\, AV ,
    \end{array}\right. \nonumber\\
   \alpha_4^{h,p}(M_1 M_2) &=& \left\{
    \begin{array}{cl}
     a_4^{h,p}(M_1 M_2) + r_{\chi}^{M_2}\,a_6^{h,p}(M_1 M_2)
      & \quad \mbox{for~} M_1 M_2=AV, \, VA , \\
     a_4^{h,p}(M_1 M_2) - r_{\chi}^{M_2}\,a_6^{h,p}(M_1 M_2)
      & \quad \mbox{for~} M_1 M_2=AA\,, VV,
    \end{array}\right.\\
   \alpha_{3,\rm EW}^{h,p}(M_1 M_2) &=& \left\{
    \begin{array}{cl}
     a_9^{h,p}(M_1 M_2) - a_7^{h,p}(M_1 M_2)
      & \quad \mbox{for~} M_1 M_2=VA, \, AA , \\
     a_9^{h,p}(M_1 M_2) + a_7^{h,p}(M_1 M_2)
      & \quad \mbox{for~} M_1 M_2=VV,\, AV ,
    \end{array}\right. \nonumber\\
   \alpha_{4,\rm EW}^{h,p}(M_1 M_2) &=& \left\{
    \begin{array}{cl}
     a_{10}^{h,p}(M_1 M_2) + r_{\chi}^{M_2}\,a_8^{h,p}(M_1 M_2)
      & \quad \mbox{for~} M_1 M_2=AV, \, VA , \\
     a_{10}^{h,p}(M_1 M_2) - r_{\chi}^{M_2}\,a_8^{h,p}(M_1 M_2)
      & \quad \mbox{for~} M_1 M_2=AA\,, VV.
     \end{array}\right.\nonumber
\end{eqnarray}
Note that the order of the arguments of $\alpha_i^p(M_1 M_2)$ and
$a_i^p(M_1 M_2)$ is relevant. For vector mesons we have
\begin{equation}
\label{eq:rchiV}
   r_\chi^V(\mu) = \frac{2m_V}{m_b(\mu)}\,\frac{f_V^\perp(\mu)}{f_V} \,,
\end{equation}
while for axial-vector mesons we have
\begin{equation}
\label{eq:rchiA}
   r_\chi^A(\mu) = \frac{2m_A}{m_b(\mu)}\,\frac{f_A^\perp(\mu)}{f_A} \,.
\end{equation}

\section{Decay amplitudes}

For simplicity, here we do not explicitly show the arguments, $M_1$
and $M_2$, of  $\alpha_i^{p,h}$ and $\beta_i^{p,h}$ coefficients. The order
of the arguments of $\alpha_i^p (M_1 M_2)$ and $\beta_i^p(M_1
M_2)$ is consistent with the order of the arguments of
$X^{(\overline B M_1, M_2)}$, where
\be \label{eq:beta}
 \beta_i^p (M_1 M_2) =\frac{i f_B f_{M_1}
f_{M_2}}{X^{(\overline B M_1,M_2)}}b_i^p.
\en
The decay amplitudes for
$(a_1,b_1)\rho,(a_1,b_1)K^*,(f_1,h_1)\rho,(f_1,h_1)K^*,K_1\rho$
and $K_1K^*$ can be obtained from Appendix A of \cite{CY:AP}
by the replacement of $P$ by $V$ with the same quark content.

For $VA$ modes, we only list those channels involving $\omega$ and $\phi$
vector mesons. For other $B\to VA$ decay amplitudes, the reader is referred to
Appendix A of \cite{CY:AP} with a simple replacement of the pseudoscalar meson
by the vector meson. \be
 \sqrt2\,{\cal A}_{B^-\to a_1^- \omega}^h
   &=&  \frac{G_F}{\sqrt{2}}\sum_{p=u,c}\lambda_p^{(d)}
   \Bigg\{ \bigg[ \delta_{pu}\,(\alpha_2^h + \beta_2^h)
     + 2\alpha_3^{p,h}+ \alpha_4^{p,h} + \frac{1}{2}\alpha_{3,{\rm EW}}^{p,h} \non \\
   && - \frac{1}{2}\alpha_{4,{\rm EW}}^{p,h}  + \beta_3^{p,h} + \beta_{3,{\rm EW}}^{p,h}
    \bigg] X^{(\overline B a_1, \omega)}_h \nonumber\\
   &+&  \left[ \delta_{pu}\,(\alpha_1^h + \beta_2^h)
    + \alpha_4^{p,h} + \alpha_{4,{\rm EW}}^{p,h} + \beta_3^{p,h}
    + \beta_{3,{\rm EW}}^{p,h} \right] X^{(\overline B \omega, a_1)}_h\Bigg\}  , \\
   -2\,{\cal A}_{\bar B^0\to a_1^0\omega}^h
   &=& \frac{G_F}{\sqrt{2}}\sum_{p=u,c}\lambda_p^{(d)}
   \Bigg\{ \Big[ \delta_{pu}\,(\alpha_2^h - \beta_1^h)
    + 2\alpha_3^{p,h}+ \alpha_4^{p,h} + \frac{1}{2}\alpha_{3,{\rm EW}}^{p,h}
    - \frac{1}{2}\alpha_{4,{\rm EW}}^{p,h}
    \nonumber\\[-0.1cm]
    &&\hspace*{1cm}
 + \beta_3^{p,h}-\, \frac{1}{2}\beta_{3,{\rm EW}}^{p,h} -
\frac{3}{2}\beta_{4,{\rm EW}}^{p,h} \Big]
    X^{(\overline Ba_1, \omega)}_h
    \nonumber\\
   && \hspace*{1cm} +\Big[ \delta_{pu}\,(-\alpha_2^h - \beta_1^h)
    + \alpha_4^{p,h}- \frac{3}{2}\alpha_{3,{\rm EW}}^{p,h}
    - \frac{1}{2}\alpha_{4,{\rm EW}}^{p,h}
    \nonumber\\[-0.1cm]
    &&\hspace*{1cm}
 + \beta_3^{p,h}- \,\frac{1}{2}\beta_{3,{\rm EW}}^{p,h} -
\frac{3}{2}\beta_{4,{\rm EW}}^{p,h} \Big]
     X^{(\overline B\omega, a_1)}_h \Bigg\}\,,   \\
 {\cal A}_{B^-\to a_1^- \phi}^h
   &=&  \frac{G_F}{\sqrt{2}}\sum_{p=u,c}\lambda_p^{(d)}
   \Bigg\{ \left[
    \alpha_3^{p,h} - \frac{1}{2}\alpha_{3,{\rm EW}}^{p,h}
    \right] X^{(\overline B a_1, \phi)}_h \Bigg\}\,,  \\
   -\sqrt{2}\,{\cal A}_{\bar B^0\to a_1^0\phi}^h
   &=& \frac{G_F}{\sqrt{2}}\sum_{p=u,c}\lambda_p^{(d)}
   \Bigg\{ \left[
    \alpha_3^{p,h} - \frac{1}{2}\alpha_{3,{\rm EW}}^{p,h}
    \right] X^{(\overline B a_1, \phi)}_h \Bigg\}\,,
\end{eqnarray}
for $\ov B\to a_1(\omega,\phi)$,
\begin{eqnarray}
 2{\cal A}_{\ov B^0\to f_1\omega}^h
   &=&  \frac{G_F}{\sqrt{2}}\sum_{p=u,c}\lambda_p^{(d)} \Bigg\{
    \Big[ \delta_{pu}(\alpha_2^h
    + \beta_1^h)+ 2\alpha_3^{p,h} +\alpha_4^{p,h} + \frac{1}{2}\alpha_{3,{\rm
    EW}}^{p,h}  \non \\ &-&  \frac{1}{2}\alpha_{4,{\rm EW}}^{p,h}
  + \beta_3^{p,h} + 2\beta_4^{p,h} -{1\over 2}\beta_{3,{\rm EW}}^{p,h} +{1\over
2}\beta_{4,{\rm EW}}^{p,h} \Big]   X^{(\overline B f_1^q,\omega)}_h
    \nonumber\\
    &+& \Big[  \delta_{pu}(\alpha_2^h
 + \beta_1^h) +2\alpha_3^{p,h} +\alpha_4^{p,h} + {1\over 2}\alpha_{3,{\rm
EW}}^{p,h}- {1\over 2}\alpha_{4,{\rm EW}}^{p,h}
    + \beta_3^{p,h} + 2\beta_4^{p,h} \non \\
    &-& {1\over 2}\beta_{3,{\rm EW}}^{p,h} +{1\over 2}\beta_{4,{\rm EW}}^{p,h}\Big]
 X^{(\overline B\omega,f_1^q)}_h  + \sqrt{2}\Big[\alpha_3^{p,h}-{1\over 2}\alpha^{p,h}_{3,{\rm EW}}\Big]
 X^{(\overline B \omega, f_1^s)}_h \Bigg\} \\
 2{\cal A}_{\ov B^0\to f_1\phi}^h
   &=&  \frac{G_F}{\sqrt{2}}\sum_{p=u,c}\lambda_p^{(d)} \Bigg\{
 \sqrt{2}\Big[ \alpha_3^{p,h}-{1\over 2}\alpha_{3,{\rm EW}}^{p,h}\Big]
X^{(\overline B f_1^q,\phi)}_h
    \nonumber\\
 &+& 2f_Bf_{f_1^s}f_\phi\Big[ \beta_4^{p,h} -{1\over 2}\beta_{4,{\rm
EW}}^{p,h}\Big]_{f_1^s\phi} +2f_Bf_{f_1^s}f_\phi\Big[ \beta_4^{p,h} -{1\over
2}\beta_{4,{\rm EW}}^{p,h}\Big]_{\phi f_1^s} \Bigg\}, \en for $\ov B\to
f_1(\omega,\phi)$, \be
  \sqrt{2}{\cal A}_{B^-\to K^-_1\omega}^h
   &=&  \frac{G_F}{\sqrt{2}}\sum_{p=u,c}\lambda_p^{(s)} \Bigg\{
 \Big[ \delta_{pu}(\alpha_1^h+\beta_2^h)+ \alpha_4^{p,h} +\alpha_{4,{\rm EW}}^{p,h}
 +\beta_3^{p,h} +\beta^{p,h}_{3,{\rm EW}} \Big]   X^{(\overline B \omega,\ov K_1)}_h  \non \\
   &+& \Big[\delta_{pu}\alpha_2^h+2\alpha_3^{p,h}+{1\over 2}\alpha^{p,h}_{3,{\rm EW}}\Big]
   X^{(\overline B\ov K_1,\omega)}_h\Bigg\},    \\
  \sqrt{2}{\cal A}_{\ov B^0\to \ov K^0_1\omega}^h
   &=&  \frac{G_F}{\sqrt{2}}\sum_{p=u,c}\lambda_p^{(s)} \Bigg\{
    \Big[ \alpha_4^{p,h} - {1\over 2}\alpha_{4,{\rm
    EW}}^{p,h} +\beta_3^{p,h} - {1\over 2}\beta^{p,h}_{3,{\rm EW}} \Big]
   X^{(\overline B \omega,\ov K_1)}_h  \non \\
   &+& \Big[\delta_{pu}\alpha_2^h+ 2\alpha_3^{p,h}+ {1\over 2}\alpha^{p,h}_{3,{\rm EW}}\Big]
   X^{(\overline B \ov
   K_1,\omega)}_h \Bigg\},    \\
  {\cal A}_{B^-\to K^-_1\phi}^h
   &=&  \frac{G_F}{\sqrt{2}}\sum_{p=u,c}\lambda_p^{(s)} \Bigg\{
    \Big[ \delta_{pu}\beta_2^h+ \alpha_3^{p,h}+ \alpha_4^{p,h} -{1\over 2}\alpha_{3,{\rm EW}}^{p,h} \non \\
&& -{1\over 2}\alpha_{4,{\rm
    EW}}^{p,h}
    +\beta_3^{p,h} +\beta^{p,h}_{3,{\rm EW}} \Big]
 X^{(\overline B \ov K_1,\phi)}_h  \Bigg\}, \label{eq:K1cphi} \\
    {\cal A}_{\ov B^0\to \ov K^0_1\phi}^h
   &=&  \frac{G_F}{\sqrt{2}}\sum_{p=u,c}\lambda_p^{(s)} \Bigg\{
    \Big[\alpha_3^{p,h}+ \alpha_4^{p,h} -{1\over 2}\alpha_{3,{\rm EW}}^{p,h} -{1\over 2}\alpha_{4,{\rm
    EW}}^{p,h} \non \\
    && +\beta_3^{p,h} -{1\over 2}\beta^{p,h}_{3,{\rm EW}} \Big]
 X^{(\overline B \ov K_1,\phi)}_h  \Bigg\}, \label{eq:K1nphi}
 \en
for $\ov B\to \ov K_1(\omega,\phi)$.

The relevant decay amplitudes for $B\to AA$ decays are
\be
 \sqrt2\,{\cal A}_{B^-\to a_1^- a_1^0}^h
   &=&  \frac{G_F}{\sqrt{2}}\sum_{p=u,c}\lambda_p^{(d)}
   \left[ \delta_{pu}\,(\alpha_1^h+\alpha_2^h )
    +\frac{3}{2}\alpha_{3,{\rm EW}}^{p,h}
    + \frac{3}{2}\alpha_{4,{\rm EW}}^{p,h}
    \right] X^{(\overline B a_1, a_1)}_h, \\
 {\cal A}_{\bar B^0\to a_1^- a_1^+}^h
   &=& \frac{G_F}{\sqrt{2}}\sum_{p=u,c}\lambda_p^{(d)}
   \Bigg[ \delta_{pu}\,(\alpha_1^h +\beta_1^h)+ \alpha_4^{p,h}
    + \alpha_{4,{\rm EW}}^{p,h} + \beta_3^{p,h} \non \\
    && + 2\beta_4^{p,h}
    - \frac{1}{2}\beta_{3,{\rm EW}}^{p,h} + \frac{1}{2}\beta_{4,{\rm EW}}^{p,h} \Bigg]
    X^{(\overline B a_1, a_1)}_h,  \\
 -\,{\cal A}_{\bar B^0\to a_1^0 a_1^0}^h
   &=& \frac{G_F}{\sqrt{2}}\sum_{p=u,c}\lambda_p^{(d)}
    \Bigg[ \delta_{pu}\,(\alpha_2^h - \beta_1^h)
    - \alpha_4^{p,h} + \frac{3}{2}\alpha_{3,{\rm EW}}^{p,h}
    + \frac{1}{2}\alpha_{4,{\rm EW}}^{p,h} - \beta_3^{p,h} - 2\beta_4^{p,h}
    \nonumber\\[-0.1cm]
    &&\hspace*{1cm}
    +\, \frac{1}{2}\beta_{3,{\rm EW}}^{p,h} - \frac{1}{2}\beta_{4,{\rm EW}}^{p,h} \Bigg]
    X^{(\overline B a_1, a_1)}_h,  \\
 \sqrt2\,{\cal A}_{B^-\to a_1^- f_1}^h
   &=&  \frac{G_F}{\sqrt{2}}\sum_{p=u,c}\lambda_p^{(d)}
   \Bigg\{ \bigg[ \delta_{pu}\,(\alpha_2^h + \beta_2^h)
     + 2\alpha_3^{p,h}+ \alpha_4^{p,h} + \frac{1}{2}\alpha_{3,{\rm EW}}^{p,h} \non \\
   && - \frac{1}{2}\alpha_{4,{\rm EW}}^{p,h}  + \beta_3^{p,h} + \beta_{3,{\rm EW}}^{p,h}
    \bigg] X^{(\overline B a_1, f_1^q)}_h
   +\sqrt{2}\left[ \alpha_3^{p,h}-{1\over 2}\alpha_{4,{\rm EW}}^{p,h}\right]X^{(\overline B a_1, f_1^s)}_h \non \\
   && + \left[ \delta_{pu}\,(\alpha_1^h + \beta_2^h)
    + \alpha_4^{p,h} + \alpha_{4,{\rm EW}}^{p,h} + \beta_3^{p,h}
    + \beta_{3,{\rm EW}}^{p,h} \right] X^{(\overline B f_1^q, a_1)}_h\Bigg\}  ,  \\
   -2\,{\cal A}_{\bar B^0\to a_1^0 f_1}^h
   &=& \frac{G_F}{\sqrt{2}}\sum_{p=u,c}\lambda_p^{(d)}
   \Bigg\{ \Big[ \delta_{pu}\,(\alpha_2^h - \beta_1^h)
    + 2\alpha_3^{p,h}+ \alpha_4^{p,h} + \frac{1}{2}\alpha_{3,{\rm EW}}^{p,h}
    - \frac{1}{2}\alpha_{4,{\rm EW}}^{p,h}
    \nonumber\\[-0.1cm]
    &&\hspace*{0cm}
    + \beta_3^{p,h}-\, \frac{1}{2}\beta_{3,{\rm EW}}^{p,h} - \frac{3}{2}\beta_{4,{\rm EW}}^{p,h} \Big]
    X^{(\overline B f_1^q, a_1)}_h +\sqrt{2}
    \left[ \alpha_3^{p,h}-{1\over 2}\alpha_{4,{\rm EW}}^{p,h}\right]
    X^{(\overline B a_1, f_1^s)}_h
    \nonumber\\
   && +\Big[ \delta_{pu}\,(-\alpha_2^h - \beta_1^h)
    + \alpha_4^{p,h}- \frac{3}{2}\alpha_{3,{\rm EW}}^{p,h}
    - \frac{1}{2}\alpha_{4,{\rm EW}}^{p,h}
    \nonumber\\[-0.1cm]
    &&\hspace*{0cm}
    + \beta_3^{p,h}- \,\frac{1}{2}\beta_{3,{\rm EW}}^{p,h} - \frac{3}{2}\beta_{4,{\rm EW}}^{p,h} \Big]
     X^{(\overline B a_1, f_1^q)}_h \Bigg\}\,,
\end{eqnarray}
for $\ov B\to a_1(a_1,f_1)$, and
\be
  \sqrt{2}{\cal A}_{B^-\to K^-_1 a_1^0}^h
   &=&  \frac{G_F}{\sqrt{2}}\sum_{p=u,c}\lambda_p^{(s)} \Bigg\{
    \Big[ \delta_{pu}(\alpha_1^h+\beta_2^h)+ \alpha_4^{p,h} +\alpha_{4,{\rm
    EW}}^{p,h} +\beta_3^{p,h} +\beta^{p,h}_{3,{\rm EW}} \Big]   X^{(\overline B a_1,\ov K_1)}_h  \non \\
   && +\Big[\delta_{pu}\alpha_2^h+{3\over 2}\alpha^{p,h}_{3,{\rm EW}}\Big]  X^{(\overline B\ov K_1,a_1)}_h\Bigg\},
       \\
 {\cal A}_{B^-\to \ov K^0_1 a_1^-}^h
   &=&  \frac{G_F}{\sqrt{2}}\sum_{p=u,c}\lambda_p^{(s)}
    \Big[ \delta_{pu}\beta_2^h+ \alpha_4^{p,h}  - \frac{1}{2}\alpha_{4,{\rm
    EW}}^{p,h}  +\beta_3^{p,h} +\beta^{p,h}_{3,{\rm EW}} \Big]   X^{(\overline Ba_1, \ov K_1)}_h,
       \\
 {\cal A}_{\ov B^0\to K^-_1 a_1^+}^h
   &=&  \frac{G_F}{\sqrt{2}}\sum_{p=u,c}\lambda_p^{(s)}
    \Big[ \delta_{pu}\alpha_1^h+ \alpha_4^{p,h} + \alpha_{4,{\rm
    EW}}^{p,h} +\beta_3^{p,h} -{1\over 2}\beta^{p,h}_{3,{\rm EW}} \Big]
       X^{(\overline Ba_1, \ov K_1)}_h,
       \\
  \sqrt{2}{\cal A}_{\ov B^0\to \ov K^0_1 a_1^0}^h
   &=&  \frac{G_F}{\sqrt{2}}\sum_{p=u,c}\lambda_p^{(s)} \Bigg\{
    \Big[ -\alpha_4^{p,h} + {1\over 2}\alpha_{4,{\rm
    EW}}^{p,h} -\beta_3^{p,h} + {1\over 2}\beta^{p,h}_{3,{\rm EW}} \Big]
       X^{(\overline B a_1,\ov K_1)}_h  \non \\
   &+& \Big[\delta_{pu}\alpha_2^h+ {3\over 2}\alpha^{p,h}_{3,{\rm EW}}\Big]
     X^{(\overline B \ov K_1,a_1)}_h\Bigg\},    \\
 \sqrt{2}{\cal A}_{B^-\to K^-_1 f_1}^h
   &=&  \frac{G_F}{\sqrt{2}}\sum_{p=u,c}\lambda_p^{(s)} \Bigg\{
    \Big[ \delta_{pu}(\alpha_1^h+\beta_2^h)+ \alpha_4^{p,h} +\alpha_{4,{\rm
    EW}}^{p,h} +\beta_3^{p,h} +\beta^{p,h}_{3,{\rm EW}} \Big]
      X^{(\overline B f_1^q,\ov K_1)}_h  \non \\
   && +\Big[\delta_{pu}\alpha_2^h+2\alpha_3^{p,h}+{1\over 2}\alpha^{p,h}_{3,{\rm EW}}\Big]
     X^{(\overline B\ov K_1,f_1^q)}_h+ \sqrt{2}\Big[ \delta_{pu}\beta_2^h+\alpha_3^{p,h}+
    \alpha_4^{p,h}-{1\over 2}\alpha^{p,h}_{3,{\rm EW}}   \non \\
   && -{1\over 2}\alpha^{p,h}_{4,{\rm EW}}+\beta_3^{p,h}+
   \beta^{p,h}_{3,{\rm EW}} \Big]X^{(\overline B\ov K_1,f_1^s)}_h \Bigg\},     \\
 \sqrt{2}{\cal A}_{\ov B^0\to \ov K^0_1 f_1}^h
   &=&  \frac{G_F}{\sqrt{2}}\sum_{p=u,c}\lambda_p^{(s)} \Bigg\{
    \Big[ \alpha_4^{p,h} -{1\over 2}\alpha_{4,{\rm
    EW}}^{p,h} +\beta_3^{p,h} -{1\over 2}\beta^{p,h}_{3,{\rm EW}} \Big]
       X^{(\overline B f_1^q,\ov K_1)}_h  \non \\
   && +\Big[\delta_{pu}\alpha_2^h+2\alpha_3^{p,h}+{1\over 2}\alpha^{p,h}_{3,{\rm EW}}\Big]
     X^{(\overline B\ov K_1,f_1^q)}_h+ \sqrt{2}\Big[ \alpha_3^{p,h}+ \alpha_4^{p,h}-{1\over 2}\alpha^{p,h}_{3,{\rm EW}}   \non \\
   && -{1\over 2}\alpha^{p,h}_{4,{\rm EW}}+\beta_3^{p,h}-
    {1\over 2} \beta^{p,h}_{3,{\rm EW}} \Big]X^{(\overline B\ov K_1,f_1^s)}_h   \Bigg\},
 \en
for $\ov B\to \ov K_1(a_1,f_1)$, where $\lambda_p^{(d)}\equiv V_{pb}V_{pd}^*$,
$\lambda_p^{(s)}\equiv V_{pb}V_{ps}^*$, and the helicity-dependent factorizable
amplitudes $X^{(\bar BM_1,M_2)}_h$ are defined in Eq. (\ref{eq:Xamplitude}).
The decay amplitudes for $\ov B\to b_1(\omega,\phi)$  are obtained from $\ov
B\to a_1(\omega,\phi)$  by replacing $a_1\to b_1$. Likewise, the expressions
for $\ov B\to h_1(\omega,\phi)$ decay amplitudes are obtained by setting
$(f_1\omega\to h_1\omega)$ and $(f_1 \phi\to h_1 \phi)$.

\section{An example of the Annihilation amplitudes in  $B\to AV$ decays}

\begin{figure}[h]
\vspace{0cm} \centerline{\epsfig{file=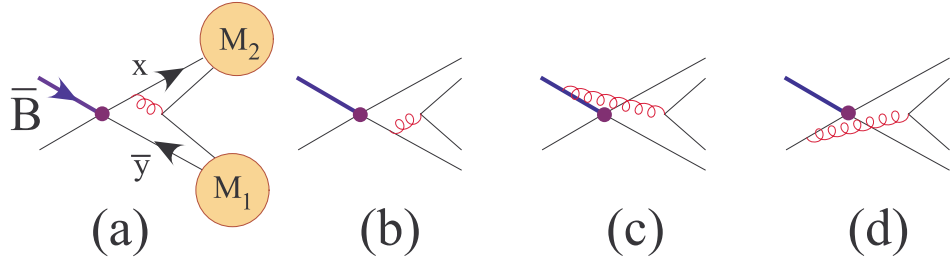,clip=46mm,
width=13cm} \vspace{0cm}} \caption{\small Annihilation contributions
to the decay amplitude of $\bar{B}\to A \, V$, where (a) and (b)
correspond to $A^f_n$,  (c) and (d) give rise to $A^i_n$.}
\label{fig:ann}
\end{figure}

In this appendix we show an explicit evaluation of the annihilation diagrams in
Fig. \ref{fig:ann} with $(M_1, M_2 )= (A,V)$ and the conventions $p_V^\mu
\simeq E n_-^\mu$ and $p_A^\mu \simeq E n_+^\mu$. The longitudinal annihilation
amplitudes of Figs. \ref{fig:ann}(a) and \ref{fig:ann}(b) read

%Fig.~\ref{fig:ann}(a):
\begin{eqnarray}
  && A_{\rm Fig. 2(a)}=-2 \langle 0| \bar d (1-\gamma_5)b|\bar B\rangle (ig_s)^2
\frac{{\rm Tr} (t^a t^a)}{N_c^2}
  \nonumber\\
  && \times \int_0^1 \int_0^1 dx dy (-1){\rm Tr}
  \Big[ M^A_\parallel(y)\gamma_\delta M_\parallel^V(x)
  \gamma^\delta \gamma_\alpha (1+\gamma_5)
  \Big]
 \frac{i(-i) (k + p_V)^\alpha}{(k + p_V)^2 (\bar{x} p_V + k )^2}\Bigg|_{k=y p_A},
\end{eqnarray}
and
\begin{eqnarray}
  && A_{\rm Fig. 2(b)}=-2 \langle 0| \bar d (1-\gamma_5)b|\bar B\rangle (ig_s)^2
\frac{{\rm Tr} (t^a t^a)}{N_c^2}
\nonumber\\
  && \times  \int_0^1 \int_0^1 dx dy (-1){\rm Tr}
  \Big[- \gamma_\alpha \gamma_\delta
 M^A_\parallel(y) \gamma^\delta M_\parallel^V(x) (1+\gamma_5)
  \Big]  \frac{i(-i) (p_{A} + \bar{k})^\alpha }
  {(\bar k + y p_A)^2 (\bar k + p_{A})^2} \Bigg|_{\bar k=\bar x p_V}.
\end{eqnarray}
The longitudinal projectors $M^A_\parallel$ and $M^V_\parallel$ are given in
Eqs. (\ref{eq:VproL}) and (\ref{eq:AproL}), respectively.

Case 1: Taking
\begin{eqnarray}
M^V_\parallel(x) & \rightarrow & -i\frac{f_V}{4} \,
\frac{m_V(\epsilon_V^*\cdot n_+)}{2}
 \not\! n_- \,\Phi_\parallel^V(x) , \non \\
 M^A_\parallel(y) & \rightarrow & -i
\frac{f_A^\perp m_A}{4}  \,{m_A(\epsilon_A^*\cdot n_-)\over 2E}
\Bigg\{\frac{i}{2}\,\sigma_{\mu\nu}\gamma_5 \, n_+^\mu  n_-^\nu \,
h_\parallel^{(t)}(y)
  - \gamma_5 {h'^{(p)}_\parallel(y)\over 2}\Bigg\},
\end{eqnarray}
and using
 \be
 \langle 0| \bar d (1-\gamma_5)b|\bar B\ra=i{f_Bm_B^2\over m_b+m_d},
 \en
we have
\begin{eqnarray}
A_{\rm Figs. 2(a)+2(b)}^{(1)}
 &=&   -i \pi \alpha_s
\frac{C_F}{N_c} \,f_Bf_Vf_A\,r_\chi^A
\int_0^1 \int_0^1 dx dy
 \,\Phi_\parallel^V(x)\left({h_\parallel^{(t)}(y)+{1\over 2}h'^{(p)}_\parallel(y)
 \over \bar xy}-{h'^{(p)}_\parallel(y)\over \bar x^2y}\right)\nonumber\\
&=&
 -2 i \pi \alpha_s
\frac{C_F}{N_c} \,f_Bf_Vf_A\,r_\chi^A
\int_0^1 \int_0^1 dx dy
 \,\Phi_\parallel^V(x)\Phi_a(y)\left({1\over \bar x^2y}\right), \label{app:case1}
\end{eqnarray}
where use of $\int^1_0 dy\,\Phi_a(y)=0$ has been made.

Case 2: Taking
\begin{eqnarray}
M^V_\parallel(x)  & \rightarrow & -i\frac{f_V^\perp m_V}{4} \,
\frac{m_V(\epsilon^*\cdot n_+)}{2 E}
\Bigg\{-\frac{i}{2}\,\sigma_{\mu\nu}\,  n_-^\mu  n_+^\nu \,
h_\parallel^{(t)} (x)
  + \frac{h_\parallel'^{(s)} (x)}{2}\Bigg\},
\end{eqnarray}
and
 \begin{eqnarray}
M^A_\parallel(y) & \rightarrow & -i\frac{f_A}{4} \,
\frac{m_V(\epsilon^*\cdot n_-)}{2}
 \not\! n_+ \gamma_5\,\Phi_\parallel^A(y) ,
\end{eqnarray}
we obtain
\begin{eqnarray}
 A_{\rm Figs. 2(a)+2(b)}^{(2)}
 &=&  -i \pi \alpha_s
\frac{C_F}{N_c} \,f_Bf_Vf_A\,r_\chi^V
\int_0^1 \int_0^1 dx dy
 \,\Phi_\parallel^A(y)\left({h_\parallel^{(t)}(x)+{1\over 2}h'^{(s)}_\parallel(x)
 \over \bar x^2y}+{h'^{(s)}_\parallel(x)\over \bar xy^2}\right)\nonumber\\
&=&
 2 i \pi \alpha_s
\frac{C_F}{N_c} \,f_Bf_Vf_A\,r_\chi^V
\int_0^1 \int_0^1 dx dy
 \,\Phi_\parallel^A(y)\Phi_v(x)\left({1\over \bar xy^2}\right). \label{app:case2}
\end{eqnarray}

Case 3: If
 \begin{eqnarray}
M^V_\parallel(x) & \rightarrow & -i\frac{f_V}{4} \,
\frac{m_V(\epsilon^*\cdot n_+)}{2}
 \not\! n_- \,\Phi_\parallel^V(x) ,
\end{eqnarray}
and
\begin{eqnarray}
 M^A_\parallel(y) & \rightarrow & -i
\frac{f_A^\perp m_A}{4}  \,{m_A(\epsilon^*\cdot n_-)\over 2E}
\Bigg\{  i E \,y\bar y\Phi_a(y)
     \sigma_{\mu\nu} \gamma_5 n_+^\mu
     \, \frac{\partial}{\partial k_\perp{}_\nu} \Bigg\},
\end{eqnarray}
then we have
\begin{eqnarray}
 A_{\rm Figs. 2(a)+2(b)}^{(3)}
&=&
 -2 i \pi \alpha_s
\frac{C_F}{N_c} \,f_Bf_Vf_A\,r_\chi^A
\int_0^1 \int_0^1 dx dy
 \,\Phi_\parallel^V(x)\Phi_a(y)\left({1\over \bar xy}\right). \label{app:case3}
\end{eqnarray}

Case 4: For
\begin{eqnarray}
M^V_\parallel(x)  & \rightarrow & i\frac{f_V^\perp m_V}{4} \,
\frac{m_V(\epsilon^*\cdot n_+)}{2 E} \Bigg\{i E\, x\bar x\Phi_v(x) \
 \sigma_{\mu\nu} n_-^\mu \, \frac{\partial}{\partial k_\perp{}_\nu}\Bigg\},
\end{eqnarray}
and
 \begin{eqnarray}
M^A_\parallel(x) & \rightarrow & -i\frac{f_A}{4} \,
\frac{m_A(\epsilon^*\cdot n_+)}{2}
 \not\! n_- \gamma_5\,\Phi_\parallel^A(x) ,
\end{eqnarray}
we have
\begin{eqnarray}
 A_{\rm Fig. 2(b)}^{(4)}
&=&
 2 i \pi \alpha_s
\frac{C_F}{N_c} \,f_Bf_Vf_A\,r_\chi^V
\int_0^1 \int_0^1 dx dy
 \,\Phi_\parallel^A(y)\Phi_v(x)\left({1\over \bar xy}\right).  \label{app:case4}
\end{eqnarray}
Finally, we obtain
\begin{eqnarray}
 A_{\rm Eqs.~2(a)+2(b)}^{(1)+(3)}
&=&
 -2 i \pi \alpha_s
\frac{C_F}{N_c} \,f_Bf_Vf_A\,r_\chi^A
\int_0^1 \int_0^1 dx dy
 \,\Phi_a(y)\Phi_\parallel^V(x)\left({1+\bar x\over \bar x^2y}\right),
\end{eqnarray}
and
\begin{eqnarray}
 A_{\rm Eqs.~2(a)+2(b)}^{(2)+(4)}
&=&
 2 i \pi \alpha_s
\frac{C_F}{N_c} \,f_Bf_Vf_A\,r_\chi^V
\int_0^1 \int_0^1 dx dy
 \,\Phi_\parallel^A(y)\Phi_v(x)\left({1+y\over \bar xy^2}\right).
\end{eqnarray}

We are led to
\begin{eqnarray}
A_3^{f,\,0}(A \ V) &=&
 \pi\alpha_s \int_0^1\! dx \,dy\, \Bigg\{
 -r_\chi^{A} \,\Phi_{a}(x) \Phi_\parallel^{V}(y)\,
 \frac{2(1+\bar y)}{x\bar{y}^2}
 \nonumber\\
 && \ \ \ \ \ \  \ \ \  \ \ \ \ \ \ \ \ \
 +r_\chi^{V} \,\Phi_\parallel^{A}(x)\, \Phi_{v}(y)
\frac{2(1+x)}{x^2\bar{y}} \Bigg\}\,. \label{A3f0}
 \en
Likewise,
\begin{eqnarray}
A_3^{f,\,0}(V_1V_2) &=&
 \pi\alpha_s \int_0^1\! dx \,dy\, \Bigg\{
 r_\chi^{V_1} \,\Phi_{v1}(x) \Phi_\parallel^{V_2}(y)\,
 \frac{2(1+\bar y)}{x\bar{y}^2}
 \nonumber\\
 && \ \ \ \ \ \  \ \ \  \ \ \ \ \ \ \ \ \
 -r_\chi^{V_2} \,\Phi_\parallel^{V_1}(x)\, \Phi_{v2}(y)
\frac{2(1+x)}{x^2\bar{y}} \Bigg\}, \non \\
A_3^{f,\,0}(A_1A_2) &=&
 \pi\alpha_s \int_0^1\! dx \,dy\, \Bigg\{
 -r_\chi^{A_1} \,\Phi_{a1}(x) \Phi_\parallel^{A_2}(y)\,
 \frac{2(1+\bar y)}{x\bar{y}^2}
 \\
 && \ \ \ \ \ \  \ \ \  \ \ \ \ \ \ \ \ \
 -r_\chi^{A_2} \,\Phi_\parallel^{A_1}(x)\, \Phi_{a2}(y)
\frac{2(1+x)}{x^2\bar{y}} \Bigg\},  \\
A_3^{f,\,0}(V\,A) &=&
 \pi\alpha_s \int_0^1\! dx \,dy\, \Bigg\{
 r_\chi^{V} \,\Phi_{v}(x) \Phi_\parallel^{A}(y)\,
 \frac{2(1+\bar y)}{x\bar{y}^2}
 \non \\
 && \ \ \ \ \ \  \ \ \  \ \ \ \ \ \ \ \ \
 +r_\chi^{A} \,\Phi_\parallel^{V}(x)\, \Phi_{a}(y)
\frac{2(1+x)}{x^2\bar{y}} \Bigg\}.
 \en
This implies
 \be
 && C^{VV}=C^{VA}=-C^{AV}=-C^{AA}=1, \non \\
 && D^{VV}=D^{AA}=-D^{AV}=-D^{VA}=1,
 \en
which lead to the third line of Eq. (\ref{eq:C&D}).

\section{Explicit expressions of annihilation amplitudes}
The general expressions of the helicity-dependent annihilation amplitudes are
given in Eqs. (\ref{eq:A1i0})-(\ref{eq:A3fm}). They can be further
simplified by considering the asymptotic distribution amplitudes
for $\Phi_V,\Phi_v,\Phi_\parallel^{^3P_1},\Phi_a^{^1P_1}$ and the
leading contributions to $\Phi_\parallel^{^1P_1},\Phi_a^{^3P_1}$:
 \be
&& \Phi^V_\parallel(u)=6u\bar u, \qquad
\Phi_\parallel^{^3P_1}(u)=6u\bar u,
 \qquad \Phi_\parallel^{^1P_1}(u)=18\,a_1^{\parallel,^1P_1}u\bar
 u(2u-1), \non \\
&& \Phi_v(u)=3(2u-1), \quad
\Phi_a^{^3P_1}(u)=3\,a_1^{\perp,^3P_1}(6u^2-6u+1),
 \qquad \Phi_a^{^1P_1}(u)=3(2u-1), \non \\
&& \Phi_\perp^V(u)=6u\bar u, \qquad
\Phi_\perp^{^3P_1}(u)=18\,a_1^{\perp,^3P_1}u\bar u(2u-1),
 \qquad \Phi_\perp^{^1P_1}(u)=6u\bar u, \non \\
&& \Phi^M_+=\int^1_u dv{\Phi^M_\parallel(v)\over v}, \qquad
\Phi^M_-=\int^u_0 dv{\Phi^M_\parallel(v)\over \bar v}.
 \en
We find
\begin{eqnarray}
  && A_3^{f,\,0} (V_1\,V_2) \approx  -18 \pi\alpha_s
 \bigg( r_\chi^{V_1} + r_\chi^{V_2}\bigg)
 (X_A^0-2)(2 X_A^0-1)  ,\label{eq:vv-XAA3-1} \\
 && A_3^{f,\,-} (V_1\, V_2) \approx -18 \pi\alpha_s
 \bigg(
  \frac{m_{V_2}}{m_{V_1}}r_\chi^{V_1} +
  \frac{m_{V_1}}{m_{V_2}}r_\chi^{V_2}
 \bigg)
 (2 X^-_A -3)(X^-_A -1) , \label{eq:vv-XAA3-2}
 \end{eqnarray}

\begin{eqnarray}
  A_3^{f,\,0} (V\, ^3P_1) & \approx & 18 \pi\alpha_s
(2 X_A^0-1) \bigg[
 a_1^{\perp,\, ^3P_1} r_\chi^{^3P_1} (X_A^0-3)
 - r_\chi^{V} (X_A^0-2)  \bigg] ,\label{eq:v3p1-XAA3-1}  \\
 A_3^{f,\,-} (V\, ^3P_1) & \approx & -18 \pi\alpha_s ( 2X^-_A - 3 )
 \non \\
  && \times \Bigg[\frac{m_{^3P_1}}{m_{V}}  r_\chi^{V} (X^-_A -1)
 - 3a_1^{\perp, ^3P_1} \frac{m_{V}}{m_{^3P_1}} \
 r_\chi^{^3P_1} ( X^-_A - 2 )
  \Bigg],
 \label{eq:v3p1-XAA3-2}
 \end{eqnarray}

\begin{eqnarray}
 A_3^{f,\,0}(V\, ^1P_1) & \approx&  18 \pi\alpha_s
 (X_A^0-2) \bigg[ r_\chi^{^1P_1} (2X_A^0-1)
 - a_1^{\parallel,\, ^1P_1}\, r_\chi^{V}
(6X_A^0-11)
 \bigg] ,\label{eq:v1p1-XAA3-1} \\
A_3^{f,\,-}(V\, ^1P_1) & \approx& -18 \pi\alpha_s (X^-_A-1)
\nonumber\\
 &&\times
 \Bigg[ -\frac{m_{V}}{m_{^1P_1}} r_\chi^{^1P_1}
 ( 2 X^-_A -3)
 + a_1^{\parallel,\, ^1P_1}
 \frac{m_{^1P_1}}{m_{V}}  r_\chi^{V}
 \biggl( 2 X^-_A -\frac{17}{3} \biggl) \Bigg],
 \label{eq:v1p1-XAA3-2}
 \end{eqnarray}

\begin{eqnarray}
 A_3^{f,\,0}( ^3P_1\, V) & =& -A_3^{f,\,0}(V\, ^3P_1), \qquad
  A_3^{f,\,-}( ^3P_1\, V) = -A_3^{f,\,-}(V\, ^3P_1),
 \end{eqnarray}

\begin{eqnarray}
 A_3^{f,\,0}( ^1P_1\, V) & =& A_3^{f,\,0}(V\, ^1P_1), \qquad
  A_3^{f,\,-}( ^1P_1\, V) = A_3^{f,\,-}(V\, ^1P_1),
 \end{eqnarray}

\begin{eqnarray}
  A_3^{f,\,0} ([^3P_1]_1\,\, [^3P_1]_2) & \approx & -18 \pi\alpha_s
(2 X_A^0-1)(X_A^0-3) \nonumber\\
 && \times \bigg[
 a_1^{\perp,\, [^3P_1]_1} r_\chi^{[^3P_1]_1}
 + a_1^{\perp,\, [^3P_1]_2} r_\chi^{[^3P_1]_2}  \bigg] ,
 \label{eq:3p13p1-XAA3-1}  \\
 A_3^{f,\,-} ([^3P_1]_1\,\, [^3P_1]_2) & \approx &  -54  \pi\alpha_s
 (2X^-_A- 3)( X^-_A - 2 ) \nonumber\\
 && \times \Bigg[
  a_1^{\perp, [^3P_1]_1} \frac{m_{[^3P_1]_2}}{m_{[^3P_1]_1}} \
 r_\chi^{[^3P_1]_1}
 + a_1^{\perp, [^3P_1]_2} \frac{m_{[^3P_1]_1}}{m_{[^3P_1]_2}} \
 r_\chi^{[^3P_1]_2} \Bigg] ,
 \label{eq:3p13p1-XAA3-2}
 \end{eqnarray}

\begin{eqnarray}
  A_3^{f,\,0} ([^1P_1]_1\,\, [^1P_1]_2) & \approx & 18 \pi\alpha_s
(X_A^0-2)(6 X_A^0-11) \nonumber\\
 && \times \bigg[
 a_1^{\parallel,\, [^1P_1]_2} r_\chi^{[^1P_1]_1}
 + a_1^{\parallel,\, [^1P_1]_1} r_\chi^{[^1P_1]_2}  \bigg] ,
 \label{eq:1p11p1-XAA3-1}  \\
 A_3^{f,\,-} ([^1P_1]_1\,\, [^1P_1]_2) & \approx & 18  \pi\alpha_s
 (X^-_A - 1)\biggl( 2 X^-_A -\frac{17}{3} \biggl)\nonumber\\
 && \times \Bigg[
  a_1^{\parallel, [^1P_1]_2} \frac{m_{[^1P_1]_2}}{m_{[^1P_1]_1}} \
 r_\chi^{[^1P_1]_1}
 +  a_1^{\parallel, [^1P_1]_1} \frac{m_{[^1P_1]_1}}{m_{[^1P_1]_2}} \
 r_\chi^{[^1P_1]_2} \Bigg] ,
 \label{eq:1p11p1-XAA3-2}
 \end{eqnarray}

\begin{eqnarray}
 A_3^{f,\,0}(^3P_1\,\, ^1P_1) & \approx& -18  \pi\alpha_s
  \bigg[ r_\chi^{^1P_1} (X_A^0-2) (2X_A^0-1) \nonumber\\
 && \ \ \ \ \ \ \ \ \ \ +
 a_1^{\parallel,\, ^1P_1}\,a_1^{\perp,\, ^3P_1}  r_\chi^{^3P_1}
  (X_A^0-3)(6X_A^0-11)
 \bigg] ,\label{eq:3p11p1-XAA3-1} \\
A_3^{f,\,-}(^3P_1\,\, ^1P_1) & \approx& -18  \pi\alpha_s
 \Bigg[ \frac{m_{^3P_1}}{m_{^1P_1}} r_\chi^{^1P_1}
 (X^-_A-1) (2 X^-_A -3) \nonumber\\
 && \ \ \ \ \ \ \ \ \ \ \
 +3 a_1^{\parallel,\, ^1P_1} a_1^{\perp,\, ^3P_1}
 \frac{m_{^1P_1}}{m_{^3P_1}}  r_\chi^{^3P_1}
 (X^-_A-2) \biggl(2 X^-_A -\frac{17}{3} \biggl) \Bigg],
 \label{eq:3p11p1-XAA3-2}
 \end{eqnarray}

\begin{eqnarray}
 A_3^{f,\,0}(^1P_1\,\, ^3P_1) & =&  A_3^{f,\,0}(^3P_1\,\, ^1P_1),
 \qquad A_3^{f,\,-}(^1P_1\,\, ^3P_1)  =  -A_3^{f,\,-}(^3P_1\,\, ^1P_1),
 \end{eqnarray}

\begin{eqnarray}
  && A_3^{i,\,0} (V_1\,V_2) \approx  18 \pi\alpha_s
 \bigg( - r_\chi^{V_1} + r_\chi^{V_2}\bigg)
 \bigg({X_A^0}^2-2X_A^0+4-{\pi^2\over 3}\bigg)  ,\label{eq:A3i0VV} \\
 && A_3^{i,\,-} (V_1\, V_2) \approx -18 \pi\alpha_s
 \bigg(
  - \frac{m_{V_2}}{m_{V_1}}r_\chi^{V_1} +
  \frac{m_{V_1}}{m_{V_2}}r_\chi^{V_2}
 \bigg)
 ({X_A^-}^2-2X_A^-+2) , \label{eq:}
 \end{eqnarray}

\begin{eqnarray}
  A_3^{i,\,0} (V\, ^3P_1) & \approx & -18 \pi\alpha_s
   \Bigg[
 a_1^{\perp,\, ^3P_1} r_\chi^{^3P_1} \bigg({X_A^0}^2-2X_A^0-6+{\pi^2\over 3}\bigg) \non \\
 && \qquad\qquad
  +r_\chi^{V} \bigg({X_A^0}^2-2X_A^0+4-{\pi^2\over 3}\bigg)  \Bigg] ,\label{eq:v3p1-XAA3-3}  \\
 A_3^{i,\,-} (V\, ^3P_1) & \approx & 18 \pi\alpha_s
  \Bigg[\frac{m_{^3P_1}}{m_{V}}  r_\chi^{V} ({X_A^-}^2-2X_A^-+2) \non \\ &&  \qquad\quad
 + 3a_1^{\perp, ^3P_1} \frac{m_{V}}{m_{^3P_1}} \
 r_\chi^{^3P_1} (X_A^{-2}-4X^-_A +{2\pi^2\over 3})
  \Bigg],
 \label{eq:v3p1-XAA3-4}
 \end{eqnarray}

\begin{eqnarray}
 A_3^{i,\,0}(V\, ^1P_1) & \approx&  -18 \pi\alpha_s
  \bigg[ r_\chi^{^1P_1} \bigg({X_A^0}^2-2X_A^0+4-{\pi^2\over 3}\bigg) \non \\
   && \qquad\quad
 + 3a_1^{\parallel,\, ^1P_1}\, r_\chi^{V}
({X_A^0}^2-4X_A^0-4+\pi^2)
 \bigg] ,\label{eq:v1p1-XAA3-3} \\
A_3^{i,\,-}(V\, ^1P_1) & \approx& 18 \pi\alpha_s
\nonumber\\
 &&\times
 \Bigg[ \frac{m_{V}}{m_{^1P_1}} r_\chi^{^1P_1}
 ({X_A^-}^2-2X_A^-+2)
 + a_1^{\parallel,\, ^1P_1}
 \frac{m_{^1P_1}}{m_{V}}  r_\chi^{V}
 ({X_A^-}^2-2X_A^--2) \Bigg],
 \label{eq:v1p1-XAA3-4}
 \end{eqnarray}

\be
  A_3^{i,\,0} (^3P_1\, V) & \approx & 18 \pi\alpha_s
   \Bigg[
 a_1^{\perp,\, ^3P_1} r_\chi^{^3P_1} \bigg({X_A^0}^2-2X_A^0-6+{\pi^2\over 3}\bigg) \non \\
 && \qquad\qquad
  -r_\chi^{V} \bigg({X_A^0}^2-2X_A^0+4-{\pi^2\over 3}\bigg)  \Bigg], \\
 A_3^{i,\,0}(^1P_1\, V) & \approx&  18 \pi\alpha_s
  \bigg[ -r_\chi^{^1P_1} \bigg({X_A^0}^2-2X_A^0+4-{\pi^2\over 3}\bigg) \non \\
   && \qquad\quad
 + 3a_1^{\parallel,\, ^1P_1}\, r_\chi^{V}
({X_A^0}^2-4X_A^0-4+\pi^2)
 \bigg] ,
 \en

\begin{eqnarray}
  A_3^{i,\,-}( ^3P_1\, V) = A_3^{i,\,-}(V\, ^3P_1),
\qquad
  A_3^{i,\,-}( ^1P_1\, V) = -A_3^{i,\,-}(V\, ^1P_1),
 \end{eqnarray}

\begin{eqnarray}
  A_3^{i,\,0} ([^3P_1]_1\,\, [^3P_1]_2) & \approx & 18 \pi\alpha_s
 \bigg({X_A^0}^2-2X_A^0-6+{\pi^2\over 3}\bigg)\nonumber\\
 && \times \bigg[
 a_1^{\perp,\, [^3P_1]_1} r_\chi^{[^3P_1]_1}
 - a_1^{\perp,\, [^3P_1]_2} r_\chi^{[^3P_1]_2}  \bigg] ,
 \label{eq:3p13p1-XAA3-3}  \\
 A_3^{i,\,-} ([^3P_1]_1\,\, [^3P_1]_2) & \approx &  -54  \pi\alpha_s
 \bigg({X_A^-}^2-4X_A^-+{2\pi^2\over 3}\bigg)\nonumber\\
 && \times \Bigg[-
  a_1^{\perp, [^3P_1]_1} \frac{m_{[^3P_1]_2}}{m_{[^3P_1]_1}} \
 r_\chi^{[^3P_1]_1}
 + a_1^{\perp, [^3P_1]_2} \frac{m_{[^3P_1]_1}}{m_{[^3P_1]_2}} \
 r_\chi^{[^3P_1]_2} \Bigg] ,
 \label{eq:3p13p1-XAA3-4}
 \end{eqnarray}

\begin{eqnarray}
  A_3^{i,\,0} ([^1P_1]_1\,\, [^1P_1]_2) & \approx & -54 \pi\alpha_s
({X_A^0}^2-4X_A^0-4+\pi^2) \nonumber\\
 && \times \bigg[
 a_1^{\parallel,\, [^1P_1]_2} r_\chi^{[^1P_1]_1}
 + a_1^{\parallel,\, [^1P_1]_1} r_\chi^{[^1P_1]_2}  \bigg] ,
 \label{eq:1p11p1-XAA3-3}  \\
 A_3^{i,\,-} ([^1P_1]_1\,\, [^1P_1]_2) & \approx & -18  \pi\alpha_s
 ({X_A^-}^2-2X_A^--2)\nonumber\\
 && \times \Bigg[
  a_1^{\parallel, [^1P_1]_2} \frac{m_{[^1P_1]_2}}{m_{[^1P_1]_1}} \
 r_\chi^{[^1P_1]_1}
 -  a_1^{\parallel, [^1P_1]_1} \frac{m_{[^1P_1]_1}}{m_{[^1P_1]_2}} \
 r_\chi^{[^1P_1]_2} \Bigg] ,
 \label{eq:1p11p1-XAA3-4}
 \end{eqnarray}

\begin{eqnarray}
 A_3^{i,\,0}(^3P_1\,\, ^1P_1) & \approx& 18  \pi\alpha_s
  \Bigg[  r_\chi^{^1P_1} \bigg({X_A^0}^2-2X_A^0+4-{\pi^2\over 3}\bigg) \nonumber\\
 && \ \ \ \ \ \ \ \ \ \ +
 3a_1^{\parallel,\, ^1P_1}\,a_1^{\perp,\, ^3P_1}  r_\chi^{^3P_1}
  \bigg({X_A^0}^2-4X_A^0+40-{11\pi^2\over 3}\bigg)
 \Bigg] ,\label{eq:3p11p1-XAA3-4} \\
A_3^{i,\,-}(^3P_1\,\, ^1P_1) & \approx& -18  \pi\alpha_s
 \Bigg[ \frac{m_{^3P_1}}{m_{^1P_1}} r_\chi^{^1P_1}
 ({X_A^-}^2-2X_A^-+2) \nonumber\\
 && \ \ \ \ \ \ \ \ \ \ \
 -3 a_1^{\parallel,\, ^1P_1}\, a_1^{\perp,\, ^3P_1}
 \frac{m_{^1P_1}}{m_{^3P_1}}  r_\chi^{^3P_1}
 \bigg({X_A^-}^2-4X_A^- +24-2\pi^2\bigg) \Bigg],
 \\
A_3^{i,\,0}(^1P_1\,\, ^3P_1) & \approx& -18  \pi\alpha_s
  \Bigg[  r_\chi^{^1P_1} \bigg({X_A^0}^2-2X_A^0+4-{\pi^2\over 3}\bigg) \nonumber\\
 && \ \ \ \ \ \ \ \ \ \ +
 3a_1^{\parallel,\, ^1P_1}\,a_1^{\perp,\, ^3P_1}  r_\chi^{^3P_1}
  \bigg({X_A^0}^2-4X_A^0+40-{11\pi^2\over 3}\bigg) \Bigg] ,
 \end{eqnarray}

\begin{eqnarray}
A_3^{i,\,-}(^1P_1\,\, ^3P_1)  =  A_3^{i,\,-}(^3P_1\,\, ^1P_1),
 \end{eqnarray}

\begin{eqnarray}
  && A_{1,2}^{i,\,0} (V_1\,V_2) \approx  18 \pi\alpha_s
 \Bigg[ \bigg(X_A^0-4+{\pi^2\over 3}\bigg)+r_\chi^{V_1}r_\chi^{V_2}(X_A^0-2)^2\Bigg]
 \end{eqnarray}

\begin{eqnarray}
  A_{1}^{i,\,0} (V\, ^3P_1) \approx -A_{2}^{i,\,0} (V\, ^3P_1)& \approx & 18 \pi\alpha_s
   \Bigg[\bigg(X_A^0-4+{\pi^2\over 3}\bigg) \non \\
   && - a_1^{\perp,\, ^3P_1}
  r_\chi^{V}r_\chi^{^3P_1}({X_A^0}^2-5X_A^0+6 ) \Bigg] ,
 \end{eqnarray}

\begin{eqnarray}
 A_1^{i,\,0}(V\, ^1P_1) & \approx&  18 \pi\alpha_s
  \Bigg[  a_1^{\parallel,\, ^1P_1}(3X_A^0+4-\pi^2)
 - r_\chi^{V}r_\chi^{^1P_1}(X_A^0-2)^2
 \Bigg] ,  \\
  A_2^{i,\,0}(V\, ^1P_1) & \approx&  18 \pi\alpha_s
  \Bigg[  -a_1^{\parallel,\, ^1P_1}(X_A^0+29-3\pi^2)
 + r_\chi^{V}r_\chi^{^1P_1}(X_A^0-2)^2
 \Bigg] ,
 \end{eqnarray}

\begin{eqnarray}
  A_1^{i,\,0}(^1P_1\, V) & \approx&  -18 \pi\alpha_s
  \Bigg[  a_1^{\parallel,\, ^1P_1}(X_A^0+29-3\pi^2)
 + r_\chi^{V}r_\chi^{^1P_1}(X_A^0-2)^2
 \Bigg] , \\
  A_2^{i,\,0}(^1P_1\,V) & \approx&  18 \pi\alpha_s
  \Bigg[  a_1^{\parallel,\, ^1P_1}(3X_A^0+4-\pi^2)
 + r_\chi^{V}r_\chi^{^1P_1}(X_A^0-2)^2
 \Bigg] ,
 \end{eqnarray}

\begin{eqnarray}
  A_{1}^{i,\,0} (^3P_1\, V) \approx -A_{2}^{i,\,0} (^3P_1\, V)& \approx & 18 \pi\alpha_s
   \Bigg[\bigg(X_A^0-4+{\pi^2\over 3}\bigg) \non \\
   && + a_1^{\perp,\, ^3P_1}
  r_\chi^{V}r_\chi^{^3P_1}({X_A^0}^2-5X_A^0+6 ) \Bigg] ,
 \end{eqnarray}

\begin{eqnarray}
 A_{1,2}^{i,\,0}( ^1P_1\, V) & =& A_{2,1}^{i,\,0}(V\, ^1P_1),
 \end{eqnarray}

\begin{eqnarray}
  A_{1,2}^{i,\,0} ([^3P_1]_1\,\, [^3P_1]_2) & \approx & 18 \pi\alpha_s
 \Bigg[ \bigg(X_A^0-4+{\pi^2\over 3}\bigg) \non \\
 &-&
 a_1^{\perp,\, [^3P_1]_1}\, a_1^{\perp,\, [^3P_1]_2}\, r_\chi^{[^3P_1]_1}\,
 r_\chi^{[^3P_1]_2}(X_A-3)^2  \Bigg] ,
 \label{eq:3p13p1-XAA3-5}
 \end{eqnarray}

\begin{eqnarray}
  A_{1,2}^{i,\,0} ([^1P_1]_1\,\, [^1P_1]_2) & \approx & 18 \pi\alpha_s\bigg[- 3\,
a_1^{\parallel,\, [^1P_1]_1}\, a_1^{\parallel,\, [^1P_1]_2}(X_A^0-71+7\pi^2) \nonumber\\
 &+&
 r_\chi^{[^1P_1]_1}\, r_\chi^{[^1P_1]_2}(X_A^0-2)^2  \bigg] ,
 \label{eq:1p11p1-XAA3-5}
 \end{eqnarray}

\begin{eqnarray}
 A_1^{i,\,0}(^3P_1\,\, ^1P_1) & \approx& 18  \pi\alpha_s
  \Bigg[  a_1^{\parallel,\, ^1P_1}\bigg(3X_A^0+4-\pi^2\bigg) \nonumber\\
 && \ \ \ \ \ \ \ \ \ \ -
 a_1^{\perp,\, ^3P_1}\,  r_\chi^{^3P_1}\, r_\chi^{^1P_1}
  ({X_A^0}^2-5X_A^0+6\bigg)
 \Bigg] ,  \\
  A_2^{i,\,0}(^3P_1\,\, ^1P_1) & \approx& 18  \pi\alpha_s
  \Bigg[  a_1^{\parallel,\, ^1P_1}\bigg(X_A^0+29-3\pi^2\bigg) \nonumber\\
 && \ \ \ \ \ \ \ \ \ \ -
 a_1^{\perp,\, ^3P_1}\,  r_\chi^{^3P_1}\, r_\chi^{^1P_1}
  ({X_A^0}^2-5X_A^0+6\bigg)
 \Bigg] ,
 \end{eqnarray}

\begin{eqnarray}
 A_{1,2}^{i,\,0}(^1P_1\,\, ^3P_1) & =&  -A_{2,1}^{i,\,0}(^3P_1\,\, ^1P_1),
 \end{eqnarray}
where the logarithmic divergences occurred in weak annihilation
are described by the variable $X_A$
 \be
 \int_0^1 {du\over u}\to X_A, \qquad \int_0^1 {\ln u\over u}\to -{1\over 2}X_A.
 \en
Following \cite{BBNS}, these variables are parameterized as
 \be
 X_A=\ln\left({m_B\over \Lambda_h}\right)(1+\rho_A
 e^{i\phi_A}),
 \en
with the unknown real parameters $\rho_A$ and $\phi_A$. For simplicity,  we shall assume in practical calculations that $X_A^h$ are helicity independent, $X_A^-=X_A^+=X_A^0$.

Note that while our result for
$A_3^{f-}(VV)$ is in agreement with Kagan \cite{Kagan} and Beneke
et al. \cite{BenekeVV} (up to a sign), the relative sign between
$r_\chi^{V_1}$ and $r_\chi^{V_2}$ in  $A_3^{f,0}(VV)$ ($A_3^{i,0}(VV)$) is positive (negative) in
our case [see Eqs. (\ref{eq:vv-XAA3-1}) and (\ref{eq:A3i0VV})] and in \cite{Kagan}, but
negative (positive) in \cite{BenekeVV}.

\section{Explicit expressions for hard spectator  terms}
Using the asymptotic distribution amplitudes, the explicit expressions of the integrals $\int^1_0 dudv$ appearing in the transverse hard spectator interaction amplitudes $H_i^-(M_1M_2)$ and $H_i^+(M_1M_2)$ [see Eqs. (\ref{eq:H1m})-(\ref{eq:H6m})] are summarized in Table \ref{tab:Hterms}.

\begin{table}[t]
\caption{The explicit expressions for the integrals $\int^1_0 dudv\cdots$ appearing in the transverse hard spectator interaction amplitudes $H_i^-(M_1M_2)$ and $H_i^+(M_1M_2)$ described by  Eqs. (\ref{eq:H1m})-(\ref{eq:H6m}), where $\alpha\equiv a_1^{\perp,\, ^3P_1}$, $\beta\equiv a_1^{\parallel,\, ^1P_1}$ and the upper (lower) sign is for $H_1^+$ ($H_5^+$). } \label{tab:Hterms}
\begin{ruledtabular}
\begin{tabular}{l c c c  }
 $M_1M_2$ & $H_{1,5}^-$ &  $H_6^-$ & $H_{1,5}^+$  \\
 \hline
 $V_1\,V_2$ & $9(X_H^--1)$  & 9 &  0 \\
 $V\,^3P_1$ & $9(X_H^--1)$  & 0 &  0 \\
 $^3P_1\,V$ & $27\alpha(X_H^--2)$  & 9 & 0 \\
 $V\,^1P_1$ & $-3\beta(X_H^--1)$  & 9 & $-3\beta$  \\
 $^1P_1\,V$ & $9(X_H^--1)$  & $3\beta$ & $\mp 3\beta$  \\
 $^3P_1\,^1P_1$ & $-9\alpha\beta(X_H^--2)$ & 9 & $-3\beta$  \\
 $^1P_1\,^3P_1$ & $9(X_H^--1)$  & 0 & $\mp3\beta$  \\
 $^3P_1\,^3P_1$ & $27\alpha(X_H^--2)$  & 0 & 0 \\
 $^1P_1\,^1P_1$ & $-3\beta(X_H^--1)$  & $3\beta$ & $-6\beta^2$  \\
\end{tabular}
\end{ruledtabular}
\end{table}

\end{document}